\documentclass[12pt]{article}
\usepackage{amssymb,amsmath, psfrag, epsfig}

\def\ov{\overline}

\font\er = cmr8

\def\nh{\noindent\hangindent=1 true cm \hangafter = 1}  

\def\c{\centerline}

\def\ov{\overline}
\def\be{\begin{equation}}
\def\ee{\end{equation}} 
\def\sy{\scriptstyle} 

\begin{document}
%ca internal conc
\def\cai     {${\rm Ca_i} $}
%ca intern conc no space
\def\cain{${\rm Ca_i}$}

% ca^2+
\def\VR{$V_{rev,L}$}
\def\CA{${\rm Ca^{2+}}$ }
% ca 2+ no space after
\def\CAN{${\rm Ca^{2+}}$}
% ical
\def\IL{${\rm I_{CaL}}$ }
% ical no space after 
\def\ILN{${\rm I_{CaL}}$}
%subtypes with spaces
\def\aa{${\rm Ca_v1.1}$ }
\def\ab{${\rm Ca_v1.2}$ }
\def\ac{${\rm Ca_v1.3}$ }
\def\ad{${\rm Ca_v1.4}$ }
%subtypes sans spaces 
\def\a{${\rm Ca_v1.1}$}
\def\b{${\rm Ca_v1.2}$}
\def\cc{${\rm Ca_v1.3}$}
\def\d{${\rm Ca_v1.4}$}
\def\Vh{$V_{h, \frac{1}{2} } $}
\def\Vhs{$V_{h, \frac{1}{2} } $ }
\def\Vm{$V_{m, \frac{1}{2} } $}
\def\Vms{$V_{m, \frac{1}{2} } $ }
\title{\bf Quantitative aspects  of L-type Ca$^{2+}$ currents} 
\author{ Henry C. Tuckwell$^*$\\
\   \\
Max Planck Institute for Mathematics in the Sciences\\
Inselstr. 22, 04103 Leipzig,  Germany \\
 %\\
%\ \\   
%{Running head: L-type calcium currents}\ \\ 
%\ \\
%{$^*$ 
%tuckwell@mis.mpg.de}\\\\ 
%\   \\}
}
        
\maketitle

\begin{abstract} 

 \CA currents in neurons and muscle cells have been classified as being one of 5 types of which 
four, L, N, P/Q and R were said to be high threshold and one, T, was designated low threshold.
This review focuses on quantitative aspects of L-type currents.   L-type
channels are now distinguished according to their structure as one of four main subtypes
\aa - \d.   
L-type calcium currents play many fundamental roles in cellular dynamical processes including
pacemaking in neurons and cardiac cells,  the activation of transcription factors involved
in synaptic plasticity and in immune cells. 
The half-activation potentials of L-type currents (\ILN)  have been ascribed values 
as low as -50 mV and as high as near 0 mV. The inactivation of \IL   has been found
to be both voltage (VDI) and calcium-dependent (CDI)  and the latter  component may involve
calcium-induced calcium release. CDI is often an important
aspect of dynamical models of cell electrophysiology. We describe the basic 
components in modeling 
\IL  including activation and both voltage and calcium dependent inactivation and the
two main approaches to determining the current. We review, by means of tables
of values from over 65 representative studies,  the various details of the dynamical properties
 associated with \IL 
 that have been found experimentally or employed in the last
25 years in 
deterministic modeling 
in various nervous system and cardiac cells. Distributions and statistics  of several parameters related
to activation and inactivation are obtained. There are few reliable experimental data
on L-type calcium current kinetics for cells at physiological calcium ion concentrations.
Neurons are divided approximately into two groups with experimental  half-activation potentials 
that are high, $\approx$ -18.3 mV, or low,  $\approx$ -36.4 mV, which correspond closely
with those for \ab and \ac channels in physiological solutions. There are very few experimental
data on time constants of activation, those available suggesting values around 0.5 to 1 ms.
In modeling, a wide range of time constants has been employed. A major
problem for quantitative studies due to lack of experimental data has been the use of kinetic parameters
from one cell type for others. 
Inactivation time constants for VDI have been found experimentally with average 65 ms. 
 Examples of calculations of \IL are made 
for linear and constant field methods and the effects of CDI  are  illustrated  for single and double
pulse protocols and the results compared with experiment. 
The review ends with a discussion and analysis of experimental subtype (\aa - \d) properties and their roles 
in normal, including pacemaker activity, and many pathological states.

\end{abstract}  

\rule{80mm}{.5pt}

{\it Keywords:}  L-type calcium currents, neuronal modeling, calcium-dependent inactivation 

*Email address: tuckwell@mis.mpg.de

 \vskip .2 in 
{\bf Abbreviations:}  AHC, auditory hair cells; AM, atrial myocyte; B, Brugada syndrome;
BAPTA, 1,2-bis (2-aminophenoxy) ethane-$N,N,N',N'$-tetra-acetic
acid; BK, big potassium channel; ${\rm Ca_i}$,  ${\rm [Ca^{2+}]_i}$,  internal calcium ion concentration; 
${\rm Ca_o}$,  ${\rm [Ca^{2+}]_o}$,  external calcium ion concentration; 
CDI, calcium-dependent inactivation; CH, chromaffin;  CICR, calcium-induced calcium release;
CORT, cortical;   DA, dopamine; DCN, dorsal cochlear nucleus;  DRG, dorsal root ganglion;
dors, dorsal;   
 DRN, dorsal raphe nucleus; 
 EGTA,  ethylene glycol ($\beta$-amino-ethyl ether)-$N,N,N',N'$-tetra-acetic acid; 
 GABA, gamma-aminobutyric acid; 
 HEK, human embryonic kidney; 
HVA, high-voltage activated; IC, inferior colliculus; KO, knockout; LD, laterodorsal; 
LVA, low-voltage activated; Mag, magnocellular; MR, medullary respiratory; med, medial; MID, midbrain; MN, motoneuron; 
N, neuron; NRT, nucleus reticularis
thalami; 
 PC, pituitary corticotroph; PF, Purkinje fiber; PM, pacemaker; RF, renal failure; SA, sino-atrial; SE, serotonergic;
SH3-GK, src homology 3 - guanylate kinase; SK, small potassium channel;
 SM, skeletal muscle;  SMM, smooth muscle; SN, substantia nigra;
SNc; substantia nigra pars compacta; SON, supraoptic nucleus; SS, disulphide bond; 
SYMP, sympathetic; TR, thalamic relay;
VDI, voltage-dependent inactivation; VM, ventricular myocyte; VWA, Von Willebrand Factor A; WT, wild-type. 
\rule{60mm}{1.5pt}

\tableofcontents

\rule{60mm}{1.5pt}

\section{Introduction}
\subsection{Perspective}
The passage of ions across cell membranes, and within cells, is of fundamental importance in determining 
the electrophysiological responses of nerve and muscle cells. Such responses are manifested ultimately
in the functioning of the nervous and muscular systems, including organs of crucial biological importance
such as the heart and brain. 
In 
the late 19th and early 20th centuries,   key discoveries were made and biophysical theories 
proposed concerning such ionic currents, for example by Nernst (1889),  Planck (1890)
 and Bernstein (1902). 
 With new electrophysiological recording
techniques, many advances were made in the 1940's and 1950's by, amongst others,  in alphabetical
order,  Eccles, Hodgkin, Huxley and Katz - see Huxley (1959) for a summary.
In the 1970's and 1980's, much additional insight
was obtained when recordings were made of currents through single ion channels,
notably by Neher and Sakmann (Hamill et al., 1981).   In the last 20 or so years there has been an enormous
number of discoveries concerning the factors which determine ionic current flows 
in neurons and muscle cells. The present review  concerns modeling 
aspects of the class of calcium currents called L-type, which, as will be seen below, 
 have many consequences beyond
electrophysiology.

For  graphic but brief historical accounts of calcium current 
discoveries see Tsien and Barrett (2005) and Dolphin (2006). 
According to the former review,  {\it``...it is apparent
that \CA  
channels have reached the forefront of the field of ion 
channel research...due
to their vital role in cellular signaling, their diversity, and 
great susceptibility to modulation...''}. Records of the 
first single channel recording of currents identified as being 
L-type were given in (Nowycky et al., 1985). More recent single channel recordings are
in Cens et al. (2006), where a comparison of results for \CA and ${\rm Ba^{2+} }$ 
as charge carrier is shown, and Schr\"oder et al. (1998), where the much
greater magnitude of L-type currents in failing heart are compared with those in normal heart.

The principal motivation for the analysis and quantitative modeling of
L-type calcium currents is that they occur in most nerve and muscle cells. They often 
 play basic roles in pacemaker activity (see Section 5.2)
and more generally in regulating spike frequency by inducing 
 afterhyperpolarization, as for example in the hippocampus
by coupling to SK channels (Tanabe et al.,1998). 
Wu et al. (2008) showed that L-type \CA current in CA1 pyramidal
cells, by coupling to delayed rectifier potassium channels 
(K$_v$7.x), can give rise to long-lasting changes in adaptation.

Comprehensive models of nerve cells may include spatial variations or 
not, but in either case the minimum number of current components
is at least 10 and amongst these there should or will usually be included
several \CA currents. If they are included in a model, 
 L-type currents require a careful  treatment 
 and our aim here is to attempt to summarize several details
of their basic properties and modeling which have been
employed for many kinds of nerve and muscle cell.

\subsection{Ion channels and neurons}
Many protein molecules are embedded  
in the cell membranes of neurons.   Some of these molecules are receptors for the main central 
neurotransmitters glutamate (excitatory) and GABA (inhibitory),  as well as  
transmitter/modulators such as noradrenaline, dopamine, and serotonin all of which are
released from vesicles in response 
to signals arriving at synapses where neighboring cells 
make close contact. (See for example Cooper et al. (2003) for an introduction to basic
neurochemistry and neuropharmacology.) Of particular importance in
determining the way in which a neuron behaves in response to
electrical and chemical stimuli are other protein molecules which serve as entrance and exit
pathways for electrically charged ions. Such molecules are called ion channels.  

If an ion channel is relatively more selective for a certain kind of ion, for example, sodium, ${\rm Na^+}$,
then it is called a sodium channel. The most commonly occurring cation channels in neurons
are sodium, potassium and calcium.  Such ion channels may be open or closed, which means they
may or may not permit the passage of ions through them. 
One of the chief consequences of the ionic currents which flow through such channels when they 
are open
is the alteration of  the electrical potential difference across the cell membrane. 
In the resting state, this potential difference is in the approximate range from -80 mV to
-50 mV. An inward flux of positive ions such as ${\rm Na^+}$ or \CA leads to a diminution
of the potential difference, called depolarization or excitation,
which, if sufficiently strong,  may give rise to an action potential
or spike.  When the membrane is depolarized, ion channels such as those of \CA undergo
conformational changes which lead to their opening for certain time intervals. Hence they are
called voltage-gated ion channels.  
\subsubsection{Activation and inactivation} 

 The process of opening the channel is called activation. However, channels may be
 in several different states because often there is also a process of inactivation which
 is not simply the cessation of activation.  For a channel to be conducting, the inactivation
 component must be switched off and the activation component switched on. If we denote the
 probability of activation being on by $m$ (for example) and the probability of inactivation being off as $h$,
 then Table 1 gives the various channel states and their probabilities. It will be seen that
 the sum of the probabilities in the third column of the table add to unity, as they must. 
 For more details,
 see, for example, texts such as Levitan and Kaczmarek (1997) or Koch (1999).

 \begin{center}
\begin{table}[!h]
\caption{Channel configurations with activation and inactivation}
    \begin{tabular}{llll}
  \hline
  {\er  Activation process}   & {\er Inactivation process}  & {\er  Probability} &  {\er Conducting state} \\
  \hline
 {\er Off (deactivated) }& {\er  On (inactivated) }  &  $(1-m)(1-h)$  &{\er  Non-conducting}  \\
 {\er On (activated)} & {\er On (inactivated) } &  $m(1-h)$  & {\er Non-conducting}  \\
{\er   On (activated) }& {\er Off (de-inactivated)}  &  $mh$  & {\er Conducting}  \\
   {\er  Off (deactivated)} & {\er Off (de-inactivated) } &  $(1-m)h$  & {\er Non-conducting  }  \\
  \hline 
   \end{tabular}
\end{table} 
\end{center} 
 
In the case of voltage-gated ion channels, the activation and inactivation probabilities 
usually depend on voltage and time. Hence if $V$ 
denotes membrane potential and $t$ denotes time, then $m=m(V,t)$ and $h=h(V,t)$. As in the
pioneering work of Hodgkin and Huxley (1952) on squid axon, the activation and inactivation
variables satisfy approximately differential equations whose solution enables one to
predict reasonably accurately  the ion currents flowing across nerve or other membrane. This is further
elaborated on in Section 2.3.

\subsection{Voltage-gated calcium channels}

Calcium currents, which are found in all excitable cells, were divided into the two main 
groups of low-threshold or low-voltage activated (LVA)  and high-threshold or
high-voltage activated (HVA).  The former group contains the T-type (T for transient)
and the latter group
consists of the types L, N, P/Q and R (L for so called long-lasting, N, either for neither T nor L, or
neuronal,
P for Purkinje, and R for resistant).  For interesting reviews of the history of the discoveries
of these various types and the experiments that preceded them, back to 1953, see Tsien and
Barrett (2005) and Dolphin (2006).

 Although L-type currents,  the main topic
of the present article, were originally
designated as belonging to the HVA group, their properties are diverse 
(Avery and Johnston, 1996; Lipscombe, 2002), varying
greatly with cell-type
and ionic environment - see the data below  and in particular the review of Lipscombe et al. (2004).
They have been implicated as playing an important
role in pacemaker activity  in some neurons and cardiac cells 
(see for example, Kamp and Hell, 2000; Mangoni et al., 2003; Brown and Piggins, 2007;
Marcantoni et al., 2007; Putzier et al., 2009a, b; Vandael et al., 2010)
and being involved in the amplification of certain 
synaptic inputs (Bui et al., 2006). 

               \begin{figure}[!h]
\begin{center}
\centerline\leavevmode\epsfig{file=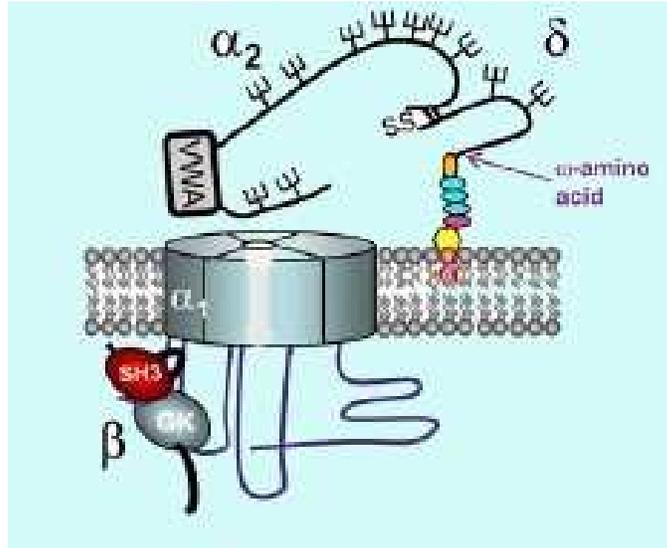,width=3.5 in}
\end{center}
\caption{Recent schematic of a voltage-gated calcium channel,
taken with permission from Davies et al. (2010). The main pore containing
subunit which straddles the membrane is $\alpha_1$ which has 4 juxtaposed components.
The $\beta$ subunit is intracellular and the $\alpha_2-\delta$ subunit
projects into the extracellular compartment.  In this example there is no $\gamma$ subunit. }
\label{}
\end{figure}

               \begin{figure}[!h]
\begin{center}
\centerline\leavevmode\epsfig{file=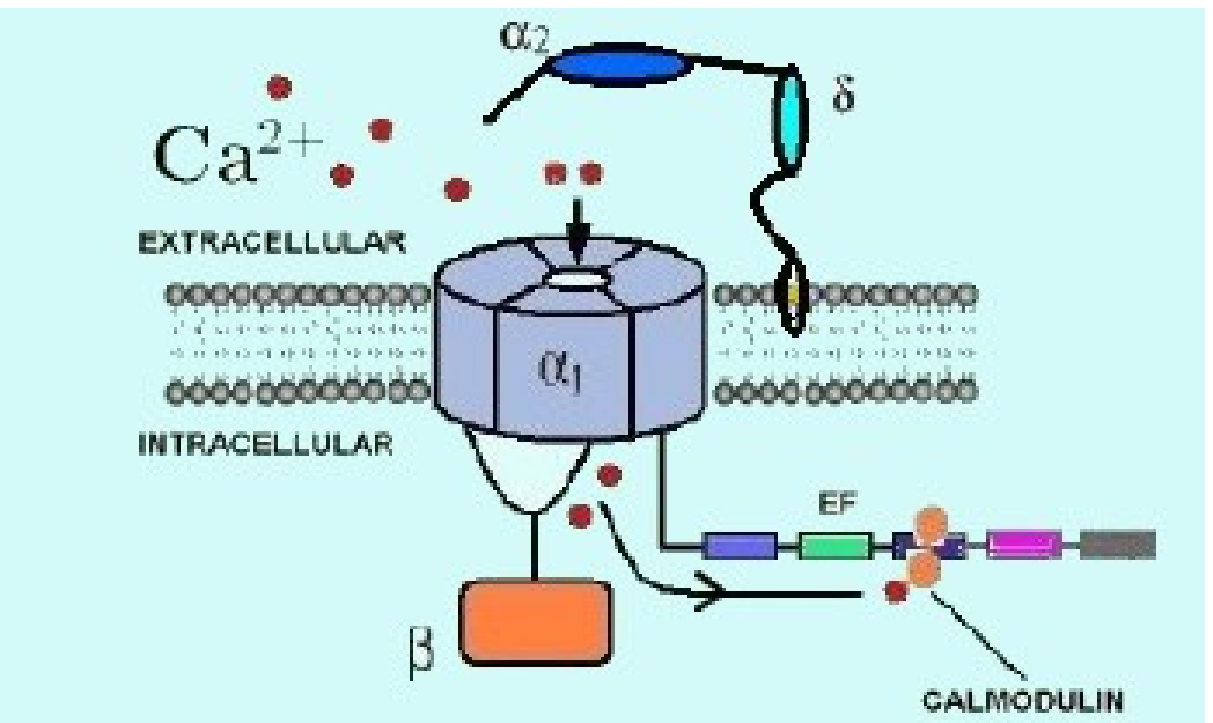,width=3.5 in}
\end{center}
\caption{Sketch representation of an L-type \CA channel depicting
the mechanism of CDI.  Calcium ions are depicted as
filled red circles. 
The calcium binding protein calmodulin 
is shown tethered to one of the intracellular motifs alongside the EF hand.  For
the role of calmodulin in inactivation, see Section 2.1.  This figure has been composed 
from figures in Bodi et al. (2005), Dolphin (2009), Davies et al. (2010) and Doan (2010).}
\label{}
\end{figure}

Calcium channel molecules have up to four subunits, $\alpha_1$, $\alpha_2$-$\delta$, $\beta$ and
$\gamma$, which may exist 
in different forms, and which modulate the conductance and dynamical properties of the
channel (Dolphin, 2006, 2009; Davies et al., 2010). The ten forms of the conducting pore subunit, $\alpha_1$, 
lead to an expansion of the above groups (Catterall et al., 2005; Dolphin, 2009)
to 10 main subtypes. According  to the accepted 
nomenclature, L-type channels consist of the subtypes  ${\rm Ca_v1.1-Ca_v1.4} $. The remaining
``high-threshold'' currents, P/Q, N and R are respectively  ${\rm Ca_v2.1}$ -${\rm Ca_v2.3}$ and the
T-current subtypes are ${\rm Ca_v3.1-Ca_v3.3}$. 
Of the L-channel subtypes,  sub-type \aa  is mainly found in skeletal muscle and the subtype \ad   is
found mainly in retinal cells (Lipscombe et al., 2004; Baumann et al., 2004;  
Catterall et al. 2005; Lacinov\'a, 2005; Dolphin, 2009; Weiergr\"aber et al., 2010). 
In cardiac myocytes and most central nervous system cells the subtypes are mainly
\ab  or  \cc.  See subsection 5.1 for further discussion.

 In Figure 1 is shown  a recent schematic representation of the molecular
structure of a voltage-gated calcium channel,
showing the subunits $\alpha_1$, $\beta$ and $\alpha_2-\delta$.  Note that until recently
 the $\delta$ subunit had been thought
to straddle the membrane rather than project into the
extracellular compartment
(Bauer et al., 2010; Dolphin, 2010).    Figure 2 schematically depicts some 
of the structures involved in the
important phenomenon of calcium-dependent inactivation, which
is elaborated upon in Section 2. 

Since there are several different types of calcium currrent, and in particular
high threshold ones, whose current-voltage relations are not
completely distinguishing, in order to
identify them, pharmacological methods are used. 
Each channel type/subtype, acts as a receptor for specific molecules,
most often toxins, leading to reduced calcium current. In some cells,  R-type currents,
once thought not to have a specific blocker and which
are borderline between HVA and LVA, 
have been more recently found to be blocked by the tarantula toxin SNX-482 (Li et al.,1997; 
Newcomb et al.,1998). Table 2, adapted from Striessnig and Koschak (2008) and Catterall et
al. (2005), 
 summarizes the main agents used for blocking or partially blocking the various
types/subtypes. Note, however,  that dihydropyridines, phenylalkylamines and 
benzothiazepines block all L-type channels to some degree.

\begin{center}
\begin{table}[h]
\caption{Summary of pharmacological agents used }
\c{to identify various types/subtypes}
\smallskip
\begin{center}
    \begin{tabular}{lcl}
  \hline
Type      &   Subtype by  & Agent   (Example)  \\
     &   $\alpha_1$ subunit &    \\
  \hline
L  &  \aa    & Dihydropyridines (Nifedipine)    \\

  &   \ab   & Phenylalkylamines (Verapamil)   \\
 
&  \ac & Benzothiazepines (Diltiazem) \\
& \ad  &     \\
P/Q  &  ${\rm Ca_v2.1}$  &  $\omega$-Agatoxin IVA     \\

N  &   ${\rm Ca_v2.2}$   &  $\omega$-Conotoxin-GVIA \\
R  &   ${\rm Ca_v2.3}$    &  SNX-482   \\

T &    ${\rm Ca_v3.1}$   & Kurtoxin    \\
 
 &   ${\rm Ca_v3.2}$ &   \\
&  ${\rm Ca_v3.3}$ &     \\
  \hline 
   \end{tabular}
\end{center} 
\end{table} 
\end{center}

It is noteworthy that one of the most important areas in which the quantitative
study of L-type (and other) \CA currents plays a fundamental role is in
the electrophysiology of the heart.  
In medicine, L-type calcium currents are the prime target  of important  
calcium channel blockers such the benzothiazapine diltiazem,
 the phenylalkylamine verapamil  and the dihydropyridines such as
nifedipine, which are used to treat, {\it inter alia}, hypertension, angina and cardiac
arrhythmias.

Within the above subtypes,  
various configurations of other subunits lead to channels, with quite 
different properties (Dolphin, 2009; Catterall, 2010). Thus,  L-type ${\rm Ca_v1.3}$ variants may have differing
current magnitudes and pharmacology (Andrade et al.,  2009). One cannot therefore ascribe definite parameters
in the dynamical description of L-type calcium currents  
based on the subtypes  ${\rm Ca_v1.1-Ca_v1.4} $.   
For example, the $\beta$ subunit regulates current density (defined, for example,  as
current per unit area) 
and the properties of activation and inactivation.
There are four forms of the $\beta$ subunit,  
$\beta_3$ and $\beta_4$  being highly expressed in the brain (Dolphin, 2003).

\subsubsection{A brief background and useful references}

It has become apparent that quantitative descriptions of L-type calcium currents
are often a fundamental component in computational modeling of neuronal and muscle cell dynamics. 
The first descriptions of L-type current dynamics in such a context appear to be 
those of Belluzzi and Sacchi (1991) and McCormick and Huguenard (1992), although there had been
many models containing calcium currents,  prior to the division into the above types, 
for example for cardiac cells. In fact Belluzzi and Sacchi's (1991) model for a sympathetic neuron in
the rat superior cervical ganglion described a high-threshold current which had an inactivation
similar to the L-type of Fox et al. (1987) but an activation similar to the N-type current. 

For an introduction to fundamentals about ion channels see, for example,   Levitan and Kaczmarek (1997).
Additional reviews on voltage-dependent \CAN-channels pertaining to their molecular
structure, nomenclature and function, including modulation and regulation are contained in, 
for example, in chronological order,
Tsien et al. (1988), Bean (1989), Tsien and Tsien (1990),  Anwyl (1991), Catterall (1995),
De Waard et al. (1996),  
Jones (1998),  Hofmann et al. (1999), Catterall (2000), Ertel et al. (2000), 
Kamp and Hell (2000),  Lacinov\'a (2005),  Zamponi (2005) and Catterall (2010).   Reviews addressing specific related topics
include those on synaptic transmission (Meir et al., 1999; Neher and Sakaba, 2008), 
T-type currents (Perez-Reyes, 2003;
Cueni et al., 2009),
calcium-dependent inactivation in neurons (Budde et al., 2002), 
dynamics of calcium signaling in neurons (Augustine et al., 2003), 
models of calcium sparks and waves (Coombes et al, 2004), L-type currents in the heart (Bodi et al., 2005),
the role of calcium currents
in circadian rhythms (Brown and Piggins, 2007), calcium release and the roles of ryanodine receptors 
in heart and skeletal muscle diseases
    (Zalk, 2007), \CA channels in chromaffin cells (Marcantoni et al., 2008)  and calcium dynamics
    in relation to absence epilepsy (Weiergr\"aber et al. 2010). Calcium ion influx 
    through L-type channels leads via various
signaling pathways to the activation of transcription factors such as CREB
and hence the expression of genes that are essential for synaptic plasticity and other
important cellular processes (Dolmetsch et al., 2001;
Hardingham et al., 2001; Mori et al., 2004; Power and  Sah, 2005;
  Satin et al., 2011). 
The role of calcium ion influx through  L-type
  channels in immune cells has also been recently reviewed (Suzuki et al. 2010). 
It was also recently demonstrated that 
calcium currents of
L-type, together with R-type, are involved in the activation of SK
channels, which attenuate excitatory synaptic transmission in
 pyramidal cells of the medial prefrontal cortex (Faber, 2010).

Mathematical modeling of the electrophysiology of cardiac cells has a history 
spanning nearly the last 50 years -  see for example DiFrancesco and Noble (1985) -  and 
reviews by Noble (1995),  Wilders (2007), Fink et al.  (2011) and Williams et al. (2010).
Brette et al. (2006) reviews background physiology and biophysics 
of calcium currents in cardiac cells. The modeling  
has taken into account details of 
the dynamics of several ionic currents, particularly calcium currents (as for example in Luo and Rudy, 1994). 
For a discussion of the dynamics underlying  the large difference between the relatively short action potential 
duration in neurons and
heart cells, see Boyett et al. (1997). 
 In  ventricular cardiomyocytes  L-type calcium currents play a pivotal role (Benitah et al., 2010)
 although T-type currents, important in development and in some pathologies (Ono and Iijima, 2010)
are also present (Bean, 1989) and play a role in pacemaker activity (Bers, 2008). 
 L-type channels,  located in the sarcolemma, are involved in arrythmias, and
 are activated by depolarization and inactivated by both voltage-dependent
 and intracellular \CA mechanisms. The inactivation of L-channels, which is elaborated 
 upon below, is complex, not only in cardiac cells,
 and not completely understood (Imredy and Yue, 1994; Findlay et al. 2008; Grandi et al., 2010).  
 See also Lacinov\'a and Hofmann (2005) and 
 the review of mechanisms of calcium and
 voltage-dependent inactivation in Cens et al. (2006). 
 The website of David Yue (Calcium signals lab) http://web1.johnshopkins.edu/csl/
 contains a wealth of information about calcium dynamics.

 In the electrophysiology of neurons and muscle cells, calcium currents usually have 
 crucial roles. Hence there is  the need to represent  as accurately as possible, in mathematical or
 computational modeling,  calcium entry into cells along with related 
 processes,  such as buffering and pumping.  The latter 
 two topics are not explored in depth here, but see for example Standen and Stanfield (1982),  Blaustein (1988),
 Tsien and Tsien (1990), McCormick and
 Huguenard (1992), 
 Tank et al. (1995), Lindblad et al. (1996),  Blaustein and Lederer (1999), Stokes and Green (2003),  Shiferaw et al. (2003), 
 Rhodes and Llin\'as (2005), Roussel et al. (2006),  Higgins et al. (2007), Friel and Chiel (2008) 
and Brasen et al. (2010).

\section{Quantitative description of L-type Ca$^{2+}$ currents}

 We are here concerned with  deterministic approaches that have been
employed in 
 the  quantitative description of  
L-type  \CA  currents. Such  descriptions are probably adequate in many
preparations although it has not been found to be the case for high
rates of action potentials in cardiac myocytes, where a many-state Markov chain model,
used in conjunction with the
constant field method (Mahajan et al,  2008), was found to be more accurate.  See also Jafri et al. (1998), Sun et al. (2000), 
Fink et al. (2011), Grandi et al. (2010) and Williams et al. (2010) for more details on the
Markov chain approach for determining L-type channel-open probabilities.
  Destexhe and Sejnowski (2001) contains 
 a brief introduction to such models, especially for voltage-dependent sodium channels. As far as can be discerned,
until now there have been no Markov models of L-type calcium channels in non-cardiac cells. 
 
 \subsection{General description: VDI and CDI}
Whereas activation kinetics of L-type channels are  voltage-dependent (Budde et al., 2002;
 Lacinov\'a, 2005), inactivation has often been found  to have dependence on both voltage and the
 internal   ``concentration'' of calcium ions, as discovered by Brehm and Eckert (1978) in
 {\it Paramecium} and Tillotson (1979)  in molluscan neurons.  The two components are called
 CDI (calcium-dependent inactivation) and VDI (voltage-dependent inactivation).
 A hallmark of CDI is that it has a maximal effect where the relevant calcium currents
 are themselves maximal.  Other markers are the reduction in CDI if ${\rm Ba^{2+}}$ is the charge carrier or if
 the buffers EGTA or BAPTA are present in the pipette (see Budde et al., 2002, for a discussion). 
 
 The relative contributions
 of CDI and VDI vary widely amongst cell types and modeling these contributions
 is not always as straightforward as in the 
 simplified scheme presented below. 
Sometimes CDI may be absent, especially in the case of \ad channels (Koschak et al., 2003;
Baumann et al. 2004, Singh et al., 2006; Wahl-Scott et al., 2006; Liu et al, 2010). 
 Indeed, often CDI is not included in modeling and often there is no
 inactivation at all, neither voltage nor calcium dependent,  but it is not
 clear if this renders some calculations inaccurate.  

In  Tables A1.1-A1.4
 of the Appendix, there are
given data for about 65 studies, which involved  
 L-type (or undefined high threshold calcium) 
currents. This survey is not exhaustive but representative of works 
from 1987  to the present. 
 About half of these 
are experimental and about half concern mathematical modeling. 
Some salient aspects of the experimental studies are discussed in Section 3. 
Here we note that for the 24 neuronal modeling studies, 16 included no
inactivation,  3 included VDI only, 3 included CDI only and a further 2 included
both VDI and CDI. It is not clear if inactivation was omitted
in several cases simply because the models employed 
data extrapolated from other cell types or the inactivation was
considered to be unimportant or extremely slow. 
For the 7 models for cardiac cell L-type calcium currents,
2 studies included VDI only and 5 included both CDI and VDI.
It is claimed (De Waard et al., 1996) that
 inactivation of L-type currents is generally slow in neurons and secretory cells but more rapid
 in cardiac cells. The data in Table A3 indicate that this may not always be the case, but the
 definition of the term rapid is no doubt flexible. 
 
 Generally,
 CDI has been cited as being more rapid than VDI (Budde et al., 2002; Lacinov\'a and Hoffman, 2005;
 Grandi et al., 2010).  The molecular basis of the 
 dynamics of CDI has been an active and fascinating research area in the last two decades
 (Imredy and Yue, 1992,1994;  Lee et al., 1999;  Peterson et al., 1999; Qin et al., 1999;
 Z\"uhlke et al., 1999; Erickson et al., 2003; Soldatov, 2003; Bazzazi et al., 2010). It was found that 
calmodulin, which is tethered to the channel, 
 must bind \CAN, as depicted in Figure 2, 
 whereupon a configurational change takes place resulting in inactivation.  
Crump et al. (2011) have reported that calmodulin and
\CA can compete to limit CDI in \b. 
 The roles of the $\beta$ and $\alpha_2\delta$ subunits in 
determining the properties of \ab channels have been discussed 
in Ravindran et al. (2008) and Ravindran et al. (2009).

 There have been developed,
 since the original discovery of CDI,  two main modeling ideas concerning the spatial distribution 
 of calcium ions which participate in  the inactivation process.  One, called shell theory and posited
 originally by Standen and Stanfield (1982),  is that there is a region
 of some 100 nm depth inside the cell membrane where calcium ions may accumulate, giving a concentration ${\rm Ca_i^*}$ 
 much higher than in the remaining cytoplasm and that it is ${\rm Ca_i^*}$ which should be used to determine the
 rate of CDI. This approach is widely used in neuronal modeling, if indeed CDI is included. The second approach, called domain theory
 (Sherman et al., 1990),
 takes account of the calcium ions just inside the pore where they have entered the cell.  In fact, according to 
 Imredy and Yue (1994) and Cens et al. (2006), inactivation (CDI) can be induced in a section of membrane which contains a single
 L-type channel.   However, in some cells, in a region with many channels, averaging over all channels should 
 make the often used and simpler 
 shell approach a reasonable approximation.

 Special consideration is made for some cardiac and other muscle cells 
 where calcium entry, principally through L-type channels, leads to \CAN-induced calcium release (CICR) of stored calcium 
 via ryanodine receptors and  results in rapid CDI.  Many articles have addressed the modeling of CICR  and
 the geometrical details and 
 the formation of localized
 increases in \cai, called calcium sparks,  and their triggering of a large local increase in \cai \hskip .05in  (see for example 
 and references therein, Shiferaw et al., 2003; Soeller and Cannell, 2004; Koh et al., 2006; Bers, 2008;
 and Groff and Smith, 2008).   
 Many of these approaches employ Markov models (Hinch et al., 2004; Shannon et al., 2004;
 Greenstein et al. 2006; Faber et al., 2007). 
 Scriven et al. (2010) have provided detailed information on the
geometry and numbers of \ab channels and ryanodine receptor clusters in rat VM which are
useful in modeling \CA dynamics.

 Ryanodine receptors may also be coupled to L-type 
 calcium channels in some neurons so that CICR plays a role.
A comprehensive review containing many examples of CICR in neurons, 
including that involving L-type currents in the hippocampus, was 
compiled by Verkhratsky (2005). 
Coulon et al. (2009) report the occurrence of both low threshold and
high threshold calcium currents in connection with CICR in the
thalamus. In bullfrog sympathetic neurons, Albrecht et al. (2001) showed
that small elevations of  ${\rm Ca_i}$ evoked
by weak depolarization lead to ${\rm Ca_i}$ accumulation by
the  endoplasmic reticulum, and that ${\rm Ca_i}$ accumulation became stronger
after inhibiting CICR with ryanodine. A mathematical model was 
presented in support of these results. In another interesting related study,
 Hoesch et al. (2001)
concluded that caffeine is a reliable agonist for CICR in rabbit vagal
sensory neurons, but that
caffeine-activated rises in ${\rm Ca_i}$ in nerve cells could not 
be attributed solely to release from intracellular stores.
See Tsien and Tsien (1990),  Friel and Tsien (1992), 
Chavis et al. (1996) and Ouyang et al. (2005) for further
examples and discussions of neuronal CICR.

\subsection{The basic model for L-type \CA current} 
In the following, the membrane potential is $V$,  the internal calcium concentration is \cain, 
 the external calcium concentration is ${\rm Ca_o}$ and 
 $t$ is time.  All the deterministic formulations of the L-type calcium current employed
 in modeling to date are included in the general form
 \begin{equation}
 {\rm I_{CaL}} = m^{p_1}(V,t)h^{p_2}(V,t)f({\rm Ca_i},t){\rm  F}(V, {\rm Ca_i, Ca_o}),
 \end{equation} 
 where $m(V,t)$ is the voltage-dependent activation variable,  $h(V,t)$ is the voltage-dependent 
 inactivation variable and $f({\rm Ca_i},t)$ is the (internal) calcium-dependent inactivation variable. 
 The factor ${\rm F}$ contains  membrane biophysical parameters
 and, as described below, is  of the Ohmic (or linear) form (as in (15)) used in the original Hodgkin-Huxley model,
 or the constant-field form (as in (17) or (19)), often called the  Goldman-Hodgkin-Katz form. 
 The values of $p_1$ and $p_2$ are ideally 
 dictated by best fits of current-voltage relations to experimental data.
 The power $p_1$  to which $m$ is raised, is about equally frequently
 $p_1=1$ or $p_1=2$, with invariably $p_1=1$ for cardiac cells.  For skeletal muscle, before the L-type 
 was distinguished, the value $p_1=3$ was employed (Standen and Stanfield, 1982). 
 Details are given in Tables  A1.1- A1.4 in the Appendix. 
 The value of $p_2$, if indeed VDI is included, is invariably $p_2=1$. 
 (Putting $p_2=0$ implies no VDI.)  
 The notation $f_{Ca}$ is often used for $f$ and $f$ often used for $h$, but the present notation
 avoids excessive subscripts in subsequent formulas.  If other calcium currents were under consideration,
 the variables for L-type current might be usefully written as $m_L$, $h_L$ and $f_L$.  In some reports,  
 L-type \CA channels are described as being 
 also permeable
  to ${\rm Na^+}$ and ${\rm K^+}$, the relative permeabilities being given for ventricular
  myocytes as 2800:3.5:1 (Luo and Rudy, 1994; Faber et al., 2007) and 3600:18:1 (Shannon et al., 2004).
  The contributions from ${\rm Na^+}$ and ${\rm K^+}$ are evidently sufficient to bring the reversal potential for \IL
 from its value around the Nernst potential for \CA (about 120-150 mV), which it would be close to 
  if it was through a purely \CAN-conducting channel, 
 to the observed 
  values of about 70 mV. However, in what follows there is no focus on the ${\rm Na^+}$ and ${\rm K^+}$
  components of the L-type current.
  In modeling neuronal or other cell-type dynamics,  there are of course many other 
  components, one of which will be the intracellular calcium concentration ${\rm Ca_i}(t)$ whose value will directly influence
  \ILN,  ${\rm Ca_o(t)}$ usually being regarded as constant.
 
 Each of the variables $m$, $h$ and $f$ takes values between 0 and 1, inclusively, and the
 product $m^{p_1}h^{p_2}f$ is interpretable as the probability that the channel is open or conducting or
 equivalently gives the expected fraction of such channels in a large sample. 
 As can also be seen from Tables A1.1-A1.4, there have been many forms other than what might be called the full
 description as in Equ. (1) in which there there is both time-varying voltage-dependent inactivation and time-varying
 calcium-dependent inactivation.  Thus, as mentioned in Section 2.1,  in some computations,
 there is no inactivation whatsoever,
 or there may be just one or the other of voltage-dependent or calcium-dependent inactivation, either with
 or without time dependence. Sometimes, if a time constant for an activation or inactivation variable
 is very small, the steady state value of the variable is assumed to hold instantaneously with no time dependence. 
 With a few exceptions, this has been the case in modeling calcium-dependent inactivation. 
 
 \subsection{Activation and inactivation variables}
 Most of the material in this subsection is standard but is repeated here for
 notational purposes and to make the account self-contained. According to the above basic model
 there are three dynamical variables determining the magnitude of the current. These are the
 activation variable, always assumed to be purely voltage-dependent, the voltage-dependent inactivation
 variable, and the calcium-dependent inactivation variable. 

\subsubsection{Activation variable $m$ and voltage-dependent inactivation variable $h$}

 The activation variable  $m(V,t)$ is usually assumed to satisfy the differential equation
 \begin{equation}
 \frac{dm}{dt}=  \frac{ m_{\infty}-m} { \tau_m}, 
 \end{equation} 
 where $m_\infty(V)$ is the steady state value $m(V,\infty)$ and  $\tau_m$ is a time constant 
which may depend on $V$.   Similarly for the voltage-dependent inactivation variable 
  \begin{equation}
 \frac{dh}{dt}= \frac{ h_{\infty}-h} { \tau_h}.
 \end{equation} 

In the classical approach,  
the voltage dependence of the steady state activation variable
is written in the Boltzmann form
\begin{equation}
m_{\infty}(V) =  \frac{ 1} { 1 + e^{ - (V-V_{m, \frac{1}{2} } )/k_m } }
\end{equation}
 where $V$ is the membrane potential. The half-activation potential is given by 
 $m_{\infty}(V_{m, \frac{1}{2} } )=0.5$, and if the derivative of $m_{\infty}$  is denoted by $m'_{\infty}$ then
the slope factor is $k_m= \frac{1}{4  m'_{\infty}(V_{m, \frac{1}{2} }  ) }$. 
 Similarly, if a voltage-dependent inactivation is included then its steady state form may be written as
 \begin{equation}
 h_{\infty}(V) =  \frac{ 1} { 1 + e^{ (V-V_{h, \frac{1}{2} } )/k_h } }
 \end{equation}
 where 
 $h_{\infty}(V_{h, \frac{1}{2} } )=0.5$ and the slope factor is  
 $k_h= -\frac{1}{4  h'_{\infty}(V_{h, \frac{1}{2} }  )}$.

 In many instances the Boltzmann forms are not used but rather forward and backward
 rate coefficients $\alpha(V)$ and $\beta(V)$ respectively, are introduced,  so that
 \begin{equation}
  \frac{dm}{dt} =  \alpha_m(V) (1-m) - \beta_m(V) m
 \end{equation}
 with a similar equation for $h$. 
It is usually the case that $m_{\infty}(V)= \frac{\alpha_m(V)}{\alpha_m(V) + \beta_m(V)}$ and
$\tau_m(V)= \frac{1}{ \alpha_m(V)+ \beta_m(V)}$ and similarly for the inactivation variable, but
as will be seen from Table A2, variations have been employed.

The Boltzmann forms for $m_{\infty}$ and $h_{\infty}$ have the advantage of making it
immediately transparent at which voltages these functions have their half-maximal values
and over what ranges of $V$ they are significantly different from zero or one.  If  $\alpha$ and
$\beta$ are given, and  $m_{\infty}(V)= \frac{\alpha_m(V)}{\alpha_m(V) + \beta_m(V)}$, then 
one may determine an approximating Boltzmann curve
by graphical inspection or calculation to ascertain the $V _{\frac{1}{2}}$ and $k$-values. 
 
 \subsubsection{Calcium-dependent inactivation variable $f({\rm Ca_i},t)$}

 It will be seen in the summary of various models presented below that the
 inactivation variable $f({\rm Ca_i},t)$ has entered the modeling in many different forms.
 The most general, which contains all those employed, can be written as
 \begin{equation}
  \frac{df}{dt}= \frac{ f_{\infty}-f} { \tau_f}, 
  \end{equation}
   where $f_\infty({\rm Ca_i})$ is the steady state $f({\rm Ca_i},\infty)$ and  $\tau_f$ is a time constant which may depend on ${\rm Ca_i}$.
  Here $f_{\infty}$ and $\tau_f$ are defined through
  \begin{equation}
  f_{\infty}= \frac{1} { 1 + \big( \frac{{\rm Ca_i}}{K_f} \big)^n}
  \end{equation} 
  \begin{equation} 
  \tau_f= \frac  {1}       { \alpha_f   \big[ 1 + \big(     \frac{{\rm Ca_i}}{K_f}    \big)^n \big]  }
  \end{equation} 
  where $\alpha_f$, with units time$^{-1}$,   is a constant 
  and $K_f$ in mM is the value of ${\rm Ca_i}$ at which the steady state inactivation has half-maximal
  value. The index $n$ has been given the values 1 (Standen and Stanfield, 1982), 2 (Luo and Rudy, 1994)
  and 3 (Fox et al., 2002).  It can be argued that $n$, which is often referred to as a Hill coefficient,
  is the approximate number of calcium ions which bind to a channel
  or associated molecule to give inactivation.  However this reasoning is
  open to considerable doubt (Weiss, 1997). H\"ofer et al. (1997) reported that the Hill coefficient for binding of \CA
  to the site mediating CDI  was close to 1 with an ``inhibition constant '' of 4$\mu$M.  
In the original
  derivation of Standen and Stanfield (1982) it was assumed that the underlying reaction 
  was in fact simply
    \begin{eqnarray}
  {\rm Ca+R}  \begin{matrix}    {\sy \beta_f} \\
\noalign{\vskip -1true mm}  \rightleftharpoons \\ 
\noalign{\vskip -2 true mm}  {\sy \alpha_f}   \end{matrix} {\rm CaR},         
\end{eqnarray}
with $K_f=\alpha_f/\beta_f$. Using standard reaction rate theory (e.g. McConigle and Molinoff, 1989)
this scheme gives the ratio of
inactivated to activated as
\begin{equation}
1-f= \frac{{\rm Ca_i}}{{\rm Ca_i} + K_f}.
\end{equation}
 (Recall that $f$ is the interpretable as the  probability that a channel is not inactivated
 by a \CAN-dependent mechanism.)
 The alternative formulation of the time-dependent behaviour in the case of $n$ 
 calcium ions binding to give an inactivated channel is through the differential equation
 \begin{equation}
 \frac{df}{dt}=\alpha_f(1-f) - \beta_f^*{\rm Ca_i}^n f
 \end{equation} 
  where  $\beta_f^*=\alpha_f/K_f^n$ which is consistent with (7)-(9). 
  However, refinements of this basic model would be needed to incorporate
  other findings such as  the dependence of CDI on voltage in cardiac myocytes (Faber et al., 2007).

 \subsection{Determining the current}
 
 Although they are well-known, we give here for completeness
 the two most generally used methods for calculating membrane ion currents (see e.g., Tuckwell, 1988; Koch, 1999).
 These are the linear and constant field methods, of which the former is simpler. In the 22 works on 
 non-cardiac cells summarized
 in Table A1.1, which state the method of determining the current, 15 use the linear method
 and 7 use the constant field method. All cardiac cell studies in Table A1.4 use the constant field method.
 Although the constant field method is considered more appropriate, if the voltage doesn't spend
 much time above 0 then the linear method may be sufficiently accurate, 
 especially when there
 is voltage-dependent inactivation, which is supported by the calculations in Section 4.1.

 \subsubsection{Linear method}
 This method,  employed
 by Hodgkin and Huxley (1952) in their fundamental modeling of the action
 potential in squid giant axon, consists of multiplying the membrane conductance 
 $G_L$ by the driving potential $V-V_{rev,L}$, where $V_{rev,L}$ is a reversal potential,  to give 
 \begin{equation}
 {\rm I}_{\rm {CaL, Lin}}= G_L(V-V_{rev,L}).
 \end{equation}
 This method is called linear (or Ohmic), as the term $V-V_{rev,L}$  is linear in $V$, but not of course the
 conductance.  
 The conductance $G_L$ is the product of a maximal value $G_{L,max}$ and one to three
 of the factors $m(V,t)^{p_1}$, $h(V,t)^{p_2}$ and $f({\rm Ca_i},t)$,  
 all of  which are dimensionless and take values between 0 and 1 inclusively. 
 That is,
 \begin{equation}
  G_L= m(V,t)^{p_1}h(V,t)^{p_2} f({\rm Ca_i},t)G_{L,max}.
  \end{equation}
  Assuming a membrane area $A$ sq cm, a channel density $\rho_L$ per sq cm
  and a single channel open conductance of $\ov{g}_L$, the maximal conductance
  is   $G_{L,max}=A\rho_L \ov{g}_L$.
  Then in terms of fundamental quantities
  \begin{equation}
 {\rm I}_{\rm {CaL, Lin}}=  A\rho_L \ov{g}_Lm(V,t)^{p_1}h(V,t)^{p_2} f({\rm Ca_i},t)(V-V_{rev,L}).
  \end{equation} 
As can be seen from Tables A1.1-1.4,  $p_1$ and $p_2$ are usually small non-negative integers between 0 and 2 for L-type currents.
If  $\ov{g}_L$  is in mS and voltages are in mV, then  $ {\rm I}_{\rm {CaL, Lin}}$ given by (15) is in $\mu$A. 

 The reversal potential for an ion type is often taken to be its Nernst potential which for \CA  is 
 \begin{equation}
 V_{Ca}=  \frac {RT}{2F}  \ln \bigg(\frac{ {\rm Ca_o}  }  {    {\rm Ca_i}}  \bigg)
 \end{equation}
  where ${\rm Ca_o}$ and  ${\rm Ca_i}$ are external and internal concentrations, $R$, $T$ and $F$ have their
 usual meanings, $RT/F$ having units of volts.  Although $V_{Ca}$ is sometimes taken as the value of $V_{rev,L}$,
 it is usually of magnitude about 120 mV, which is much higher than experimentally determined
 values of $V_{rev,L}$ which are around 60-70 mV. In modeling, using $V_{Ca}$  thus gives substantially larger
 calcium currents than expected.

 \subsubsection{Constant field method}
  From the basic expression for the current density under the constant field assumption (see for example Tuckwell, 
  1988, Chapter 2), the L-type calcium
  current, in $\mu$A,  through a membranous area of $A$ square cm is given by the
Goldman-Hodgkin-Katz flux equation
 \begin{equation}
 {\rm I}_{{\rm CaL,CF}} =  AP_Lm^{p_1}(V,t)h^{p_2}(V,t)f({\rm Ca_i,t})\frac{4FV}{ V_0 } 
 .  \frac {\big[ {\rm Ca_i} - {\rm Ca_o} e^{ -       \frac{2V}{ V_0  }  }      \big] } 
             {    1- e^{ - \frac{2V}{ V_0 } }  },   
\end{equation}  
where for convenience of expression we have defined the temperature-dependent voltage 
\begin{equation}
V_0= \frac{RT}{F},
\end{equation}
$F$ is Faraday's constant (96500 coulombs/mole), $P_L$ is the membrane permeability coefficient  
in cm/sec,  and ${\rm Ca_i}$ and ${\rm Ca_o}$ are the intracellular and extracellular concentrations of \CA in mM.
Note that the partition coefficient, here assumed the same for the intracellular and
extracellular boundaries, is absorbed into the permeability coefficient. However, if the intracellular 
and extracellular partition 
coefficients are different, then additional multiplicative factors will appear with ion concentrations, as in, for example,
Sun et al. (2000) and Fink et al. (2011). 
In terms of single channel properties, the current in $\mu$A through an area of $A$ cm$^2$ is
 \begin{equation}
 {\rm I}_{{\rm CaL,CF}} =  A\rho_LP^*_Lm^{p_1}(V,t)h^{p_2}(V,t)f({\rm Ca_i}, t)\frac{4FV}{ V_0  } 
 .  \frac {\big[ {\rm Ca_i} - {\rm Ca_o} e^{ -       \frac{2V}{V_0  }  }      \big] } 
             {    1- e^{ - \frac{2V}{ V_0} }  }
\end{equation}  
where $\rho_L$ is the density of channels in cm$^{-2}$  and $P^*_L$ is the 
single channel permeability in cm$^3$/sec. 
 
\section{Data on L-type activation and inactivation}
 A large number of sources of data for L-type activation and inactivation, including both VDI and CDI,  have been
 considered, back to the first uses of the term  ``L-type'' (Nowycky et al., 1985; Fox et al., 1987).
 Data on the half activation potentials
 $V_{m, \frac{1}{2} }$ and slopes $k_m$ are given in Tables A1.1-A1.4.   In these and all subsequent tables
 the authorship of articles is given only by first author and date to minimize column width. 
In the first column of Tables A1.1-1.4 are given carrier concentrations. M denotes a purely 
mathematical modeling
study, the absence of M can denote either a purely experimental work, or experimental work
in conjunction with modeling. The characterization of currents as L-type
by various authors has been assumed to be correct. 
 Voltage-dependent inactivation parameters  $V_{h, \frac{1}{2} }$ and $k_h$ are also given if available.
 Table A1.1 contains data for neurons and secretory cells, whereas Table A1.2 is restricted to
 cardiac cells. 
 In some cases, marked with 
 asterisks ($^{\ast}$), parameters have been estimated approximately from data presented graphically,
 but the original articles should be consulted for details. 
Table A1.3 has data for neurons and secretory cells with both VDI and CDI
 and Table A1.4 presents these data for cardiac cells.  In Table A1.1  the entry for Joux et al. (2001) for 
  $V_{m, \frac{1}{2} }$ has two values as fits were to a double Boltzmann function. 
  Note that  the concentrations ${\rm Ca_o}$ may differ in the various 
preparations, (and occasionally be ${\rm Ba_o}$) as described in the first column of Tables A1.1-1.4. 
 For example, the value in the first entry (Fox et al., 1987) is  ${\rm Ca_o}$=10mM, 
which may explain why the value of  \Vms  seems rather high.
 In general, much of the variability
in the data presented is attributable to various ionic carrier concentrations and
differing subunit isoforms. In order to keep the Tables relatively
simple, many such details are  omitted,  but these can be obtained from
 the original sources in most cases.

\subsection{Steady state activation $m_{\infty}(V)$ and inactivation $h_{\infty}(V)$}

 In the top part of Figure 3 the steady-state activation $m_{\infty}$  is plotted against $V$ for the two lowest and two highest
 values of the half-activation potential found in 
 a comprehensive, but not exhaustive, search of the literature, including
all modeling and experimental results in the Tables A1.1-1.4.  The two lowest are for midbrain dopamine neurons, 
 with half-activation
 potentials of -50 mV (Amini et al., 1999; 
Komendantov et al., 2004).

               \begin{figure}[!h]
\begin{center}
\centerline\leavevmode\epsfig{file=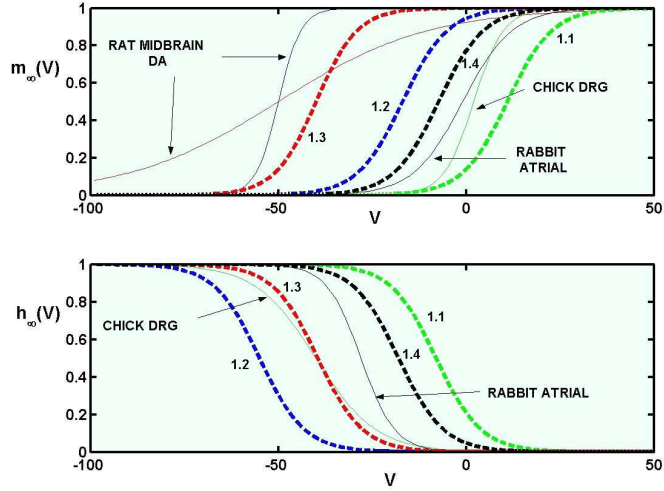,width=3.8in}
\end{center}
\caption{Steady state activation $m_{\infty}$  and inactivation $h_{\infty}$  versus membrane
potential for L-type currents in 
neurons for extreme cases.  
{\it Top panel.}  Activation curves $m_{\infty}(V)$ are given for L-type currents in midbrain DA neurons, 
(blue solid curve, Amini et al., 1999; red solid curve,  Komendantov et al., 2004)
chick DRG (green solid curve, Fox et al., 1987) and rabbit atrial myocytes (black solid curve, Lindblad et al., 1996). 
These are the cases with the two highest and two lowest half-activation potentials found
  in the literature. The two lowest (midbrain DA)  have  \Vm=-50 mV, 
  and the highest two are for rabbit atrial cell, with \Vm=-0.95 mV and chick DRG 
 with \Vm=2 mV.  Also given are examples of activation curves (dashed) for the four subtypes,
 \aa (green, from Catterall, 2005),
  \ab  (blue, average from Lipscombe et al., 2004 and Catterall, 2005), \ac (red, average from
  Lipscombe et al., 2004 and Helton et al., 2005) and \ad (black, mean of lowest and highest, from
  Catterall, 2005).  
{\it Bottom panel.}   Steady state inactivation  $h_{\infty}(V)$ for cases shown in the top panel with the
  same color coding.  For the two cells with the lowest two half-activation potentials, inactivation was 
  given only as \CAN-dependent - see Tables A1.1-A1.4.}
\label{fig:wedge}
\end{figure}

 The two highest are +5 mV 
 for ventricular myocytes (Shiferaw et al., 2003) and  +2 mV for chick dorsal root ganglion (Fox et al. 1987). 
 Also shown in the top part of the figure are examples of the $m_{\infty}$ curves found for the channel subtypes 
 \a-\d.  It can be seen that the curves for the subtypes cover or span those for the lowest and highest activation 
 curves for cells except for the one for midbrain dopamine neurons.

 In the lower part of Figure 3 are shown  the corresponding  steady state V-dependent inactivation $h_{\infty}$ 
 curves for the
 two highest half-activation potentials. For the other two cases with \Vm =-50,  there was only CDI.
 Also shown in the lower part of Figure 3 are the $h_{\infty}$ curves for each channel subtype.  The curve for 
 \ab seems more extreme than that for any cell.  
  In some cases Boltzmann functions for $m_{\infty}(V)$ and $h_{\infty}(V)$ were not given but rather explicit 
 expressions for the forward and backward rate functions $\alpha_m(V)$, $\beta_m(V)$, $\alpha_h(V)$ and
 $\beta_h(V)$.  Such formulas when available are given for all cell types in Table A3. 
 The parameters  of fitted Boltzmann functions for the
 corresponding $m_{\infty}$ and $h_{\infty}$ were given in Tables A1.1-A1.4. When explicit expressions were given for
 the activation time constants $\tau_m(V)$, these are given in Table A4.

 \subsubsection{Distribution of half-activation potentials}
It can be seen from Tables A1.1-A1.4 that there is great variability in the parameters 
for the activation and inactivation properties of L-type currents in various cells.
The underlying reasons for variability include, (a) the occurrence of
different subtypes of L-channels, some examples of whose properties are listed in
Table A6; (b) the various locations of the channels 
over the soma-dendritic surface; 
(c) the interaction of L-type channels with
other neighboring ionic channels; and
 (d) the ionic compositions of the intracellular
and extracellular compartments of the cell under consideration. 
One other source of variability is that there are different methods, some 
more accurate than others, for calculating Boltzmann curves.

    \begin{figure}[!ht]
\begin{center}
\centerline\leavevmode\epsfig{file=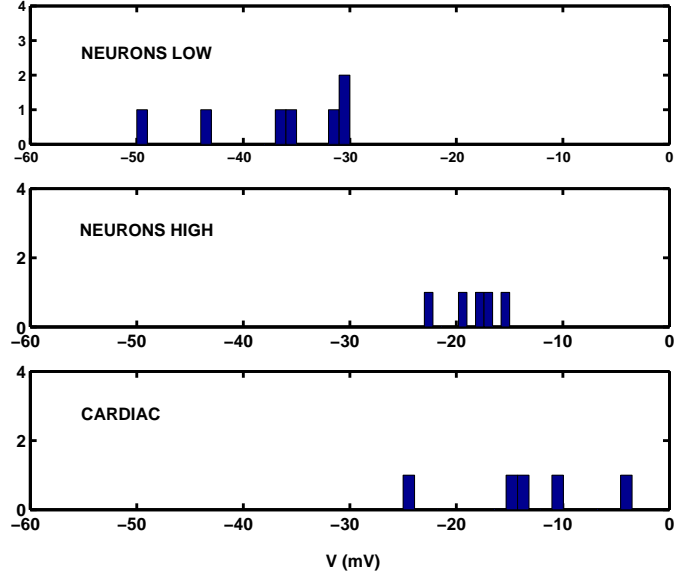,width=3.5in}
\end{center}
\caption 
{Histograms of the  half-activation membrane potentials $V_{m, \frac{1}{2} }$ for L-type \CA currents from
the data of Tables A1.1-A1.4.  The data, which
are based on selected experiments (see text) are divided into 3 groups. Top, low neuron group. 
Middle, high neuron group. Bottom, cardiac cells. 
Note that the mean values for \ab and \cc, using Table A6 values with physiological calcium ion concentrations,
are -18.7 mV and -38.9 mV, respectively.}
\label{fig:wedge}
\end{figure}

The majority of data in tables A1.1-A1.4 concerns modeling studies and it is useful to
separate out the experimental values, and to subdivide these into  those for neurons and cardiac cells.
In particular, the experimental  results considered here are only those 
for calcium concentrations in the physiological range
because  high concentrations of either barium or calcium ions can shift
the half-activation potential in the depolarizing direction by 10 or more mV
(Jaffe et al., 1994; Muinuddin et al., 2004). With these restrictions
there are 13 half-activation potentials given for neurons and 5 for cardiac cells.
Histograms for the  $V_{m, \frac{1}{2} }$ values are shown in Figure 4. 
For neurons, the
data are divided into a high ($>-30$ mV) and a low group, the topmost panel
 being for the low group. 
The mean for the neuron high group  
is $<V_{m, \frac{1}{2} }>= -18.3$ mV whereas that for the low group
 is $<V_{m, \frac{1}{2} }>= -36.4$ mV. 
The mean slope factors for these two groups of neurons
are $<k_m>= 8.4$  mV for the high group and  $<k_m>= 5.8$ mV for the low group.
These average slope factors are obtained on omitting 
the outliers 2.45 in the high group and 20.0 in the low group.
 The mean value
 of $V_{m, \frac{1}{2} }$ for the experiments with 
cardiac cells is -13.6 mV with a standard deviation of 7.8 mV,
and the corresponding average $k_m$ is  6.84 mV with standard deviation 1.61 mV. 
It is then of interest to note that the average values of the half 
activation potentials
for \ab and \cc, using Table A6 values with physiological calcium ion concentrations,
are -18.7 mV and -38.9 mV, respectively.  These values correspond well with
the mean values of the high and low groups of neurons.  The
values for cardiac cells tend to be higher than those in even the
high neuron group, but the sample sizes are too small to draw
any statistically significant conclusions.

For all the 25 neuronal modeling studies the mean value of 
$V_{m, \frac{1}{2} }$ is -21.0 mV with a standard deviation
of 12.21 mV, and for $k_m$ the mean is 7.26 with a standard
deviation of 2.33. 
The mean modeling study parameters correspond 
roughly to the experimental values for the high neuron group.
  For the 7 cardiac cell (all kinds) modeling studies the
mean half-activation potential $V_{m, \frac{1}{2} }$ is -8.8 mV with a standard
deviation of 9.1 mV and the corresponding mean for $k_m$ is
6.4 mV with a standard deviation of 0.52.  

For the parameters of the Boltzmann function for the steady state
inactivation,  $V_{h, \frac{1}{2} }$ and  $k_h$, there are, unfortunately
little appropriate experimental  data, and none which
could be considered reliable,  though many modeling studies have
employed estimates (see Tables A1.1-A1.4) and see below. 
In order to
provide crude estimates based on experiment, one may use the results of
Koschak et al. (2001) as given in Table A6 for the differences
in voltage between $V_{m, \frac{1}{2} }$ and $V_{h, \frac{1}{2} }$ 
for \ab and \ac in high
barium concentration. This procedure yields values which are summarized
in Table 3 for the activation and inactivation  parameters.
\begin{center}
\begin{table}[h]
\caption{Parameters for L-type steady state }
\c{activation and inactivation in neurons}
\smallskip
\begin{center}
    \begin{tabular}{lccc}
  \hline
      &   High neuron  &  Low neuron & All models  \\
     &   group (expt) &   group (expt) &   \\
  \hline
 $V_{m, \frac{1}{2} }$  &  -18.7     &   -38.9 & -21.0  \\

$k_m$  &   8.4 &   5.8 & 7.26  \\
 
$V_{h, \frac{1}{2} }$  & -42.0$^*$  & -61.9$^*$ & -38.95   \\
    
$k_h$  & 13.8$^*$  &  6.6$^*$ & 6.82   \\
  \hline 
   \end{tabular}
\vskip .06 in
\c{$^*$ Estimated - see text. }
\end{center} 
\end{table} 
\end{center} 
For cardiac cells, based on the few experimental studies
available, the mean  $V_{h, \frac{1}{2}   }$ is -40.0 mV with standard
deviation 12.8, the corresponding mean value of $k_h$ being 6.85 mV 
with a standard deviation of 2.04 mV.

Of the 25 neuronal modeling studies in the tables,  only 6 included VDI and 
these yielded a mean value
of -38.95 mV for $V_{h, \frac{1}{2}   }$ with standard
deviation 14.29. For the corresponding values of the slope factor
$k_h$, the mean is 6.82 with a standard deviation of 2.16,
where one value is omitted as an extreme outlier. 
For the cardiac modeling, the mean value of $V_{h, \frac{1}{2}   }$ is
-30.1 mV with a standard deviation of 10.9 mV and the
corresponding mean for $k_h$ is 6.1 mV with a standard deviation
of 2.00 mV.

%
%
%
%The distribution of  half-activation values, $V_{m, \frac{1}{2}   }$    is depicted in Figure 4 
%by means of a histogram of the values in Tables A1.1-A1.4. The extreme values are -50 mV
%and +5 mV. The grand mean for all results is -19.26 mV and the standard deviation
%is 12.98 mV.  Downward pointing arrows indicate the $V_{m, \frac{1}{2}   }$-values for examples of the two
%channel subtypes \ab and \ac which constitute nearly all the L-type channels in these cells. 
%There seems to be a peak in the distribution around  -12 mV which is not far from the value for
%\b.  Apart from that apparent peak, the distribution is fairly uniform across the interval from -40 mV 
%to +5 mV.  However, in using data on activation functions it is important
%to note that high concentrations of either barium or calcium ions can shift
%the half-activation potential in the depolarizing direction by 10 or more mV
%(Jaffe et al., 1994; Muinuddin et al., 2004).  Furthermore,  
%increasing temperature may also cause large hyperpolarizing shifts (Peloquin et al., 2008). 
%Some results presented in the tables are from single channel experiments such as those of
%Fisher et al. (1990) and Magee and Johnston (1995). 

  \subsection{Time constants $\tau_m(V)$ for activation}
Some values of $\tau_m(V)$ for activation, with
corresponding inactivation, if available,   are reported
in Table A3. This table has been divided into a part
with experimental results and a part on values of
which have been employed in modeling studies. 
As seen in Table A3-Experimental, 
there are not many explicit experimental results available
for the time constants of activation (or inactivation) 
of L-type \CA currents.
In neurons values of  $\tau_m(V)$ range from about 0.5 ms
in sympathetic neurons (Belluzzi and Sacchi, 1991)
 to 2.3 ms in hippocampal pyramidal
cells (Avery and Johnston, 1996),  a maximum value of 1.541 ms being reported for
thalamic relay cells (McCormick and
Huguenard, 1992). These values seem compatible with
those given for the appropriate subtypes in Table 6. 
For skeletal muscle, a value of 19.82 ms (Standen and
Stanfield, 1982; here assuming L-type)  is also compatible with
the relatively high value for the \aa subtype.  The experimental values
reported for smooth muscle are around 4.0 ms 
(Muinuddin et al., 2004). Some available data for
 $\tau_m$ for subtypes are given in Section 5.

In Table A3-Modeling, there are reported 23 values employed
for   $\tau_m$, with a large range of values, many of which
are much larger than the values for subtypes reported in Table 6.
Many of the values employed for a given neuron type are
not based on experiments for that neuron type but adapted
from those for other, sometimes quite different, neurons. 
Often quite disparate values appear for what are possibly
the same kind of cell. Discrepancies may arise
because it is difficult to categorize a specific
channel type in a neuron without a detailed voltage-clamp
and pharmacological analysis. 
We consider cells in the following categories.
\smallskip  \\
\noindent {\it Dopaminergic neurons.} The value of about
0.134 ms was employed by
(Li et al., 1996)  which contrasts greatly with the value 19.5 ms used 
by both Amini et al. (1999) and Komendantov et al. (2004).
The approach of Li et al. (1996) is critically discussed by
Amini et al. (1999).
\smallskip  \\
\noindent {\it Cardiac cells.} Concerning the listed values,  in the  three modeling studies
on ventricular myocytes,  two (Luo and Rudy, 1994; Shannon et al., 
2004) have $\tau_m$-values with a maximum around 2.25 ms whereas
 a third 
puts the maximum value much higher at about 19.3 ms  (Fox et al., 2002) and
at a considerably more hyperpolarized level. 
In their analysis of the firing properties of rabbit sino-atrial
node pacemakers, Zhang et al. (2000) chose a value of about 3.8 ms
for the maximum activation time constant. Thus in most
of these studies the magnitudes of $\tau_m$ are comparable
with or near  the values for  \ab and \ac  channels (see Table 6)  which are 
plentiful in the heart.  
\smallskip  \\
\noindent {\it Hippocampal cells.}  
The range of activation time constants for hippocampal
pyramidal cell models is from 0.83 ms for CA1 (Poirazi et al., 2003) 
to 4.56 ms for CA3 (Migliore et al., 1995). For the latter
cell-type, Jaffe et al. (1994) used a value with a maximum of 1.5 ms.
According to Table 6 and
Figure 12 all of these values are 
within the ranges of the ones given for 
\ab and \ac channels, both of which are prevalent in the hippocampus.
\smallskip  \\
\noindent {\it Thalamic relay cells.}  
The experimental findings of McCormick and Huguenard (1992)
  for $\tau_m$, with a maximum of 1.541 ms at $V=-15$ mV, 
were implemented in a model by the same authors.  The same 
values were later employed by Hutcheon et al. (1994). Rhodes and
Llinas (2005) put the maximum value of the time constant somewhat larger at
2.1 ms at -19 mV, but in the table of parameter values the current is
referred as a generic $I_{Ca}$ rather than specifically as an 
L-type current.  All of these values are compatible with the expected
subtype values.
\smallskip  \\
\noindent {\it Pituitary corticotrophs.}  The two modeling studies
of these cells listed in the table have fairly large values of $\tau_m$.
In the first study, LeBeau et al. (1997) set the value at 27 ms
for all $V$, whereas Shorten and Wall (2000) had a maximum
for $\tau_m$ of value 11.32 occurring at $V=-12$ mV. 
There seems to be a fairly large difference between these
values and the sub-type values, but there are many sources
of variability as discussed elsewhere in this review. 
\smallskip\\

\noindent {\it Cortical pyramidal cells.}  In the one parameter
set available (Rhodes and Llinas, 2001), the maximal time constant for 
L-type activation was 
set at 2.1 ms, occurring at -19 mV. This value is well within the
range of the available data for the appropriate subtypes as given 
in Table 6 and Figure 12.
\smallskip  \\

               \begin{figure}[!hb]
\begin{center}
\centerline\leavevmode\epsfig{file=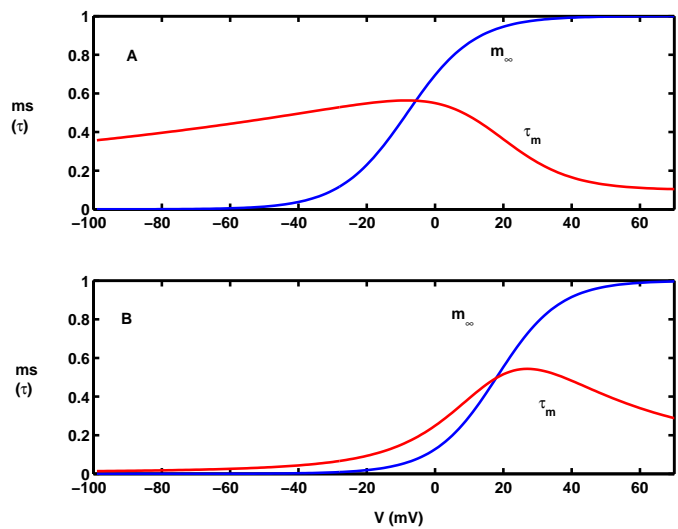,width=3.5in}
\end{center}
\caption 
{Activation time constants for high threshold \CA currents.  Two sets of experimental
 results for sympathetic neurons. In both cases the corresponding steady state
activation function is shown, using the same scale. A. From Belluzzi and Sacchi (1991)
with 5mM \CAN. 
B. From Sala (1991), with 4mM \CAN.}
\label{fig:wedge}
\end{figure}

\noindent {\it Spinal motoneurons.} The L-type activation listed
in the 3 listed modeling studies of these neurons (Booth et al, 1997;
Carlin et al. 2000; Bui et al, 2006) 
is much slower than 
the values available  for either of the subtypes \ab  \ac which are
candidates for being present in CNS neurons.    The first of these
studies set $\tau_m=40$ ms, whereas the two subsequent ones put
$\tau_m=20$ ms whereas the largest value in Figure 12 is about 4 to 5 ms.
\smallskip  \\
\noindent {\it Smooth muscle.} In modeling
a rat mesenteric smooth muscle cell, Kapela et al. (2008) for $\tau_m(V)$
used a Gaussian-like function with a maximum value of 3.65 ms at $V=-40$ mV. 
Such a value is close to the experimentally reported time constant in
(Muinuddin et al., 2004).

    \begin{figure}[!ht]
\begin{center}
\centerline\leavevmode\epsfig{file=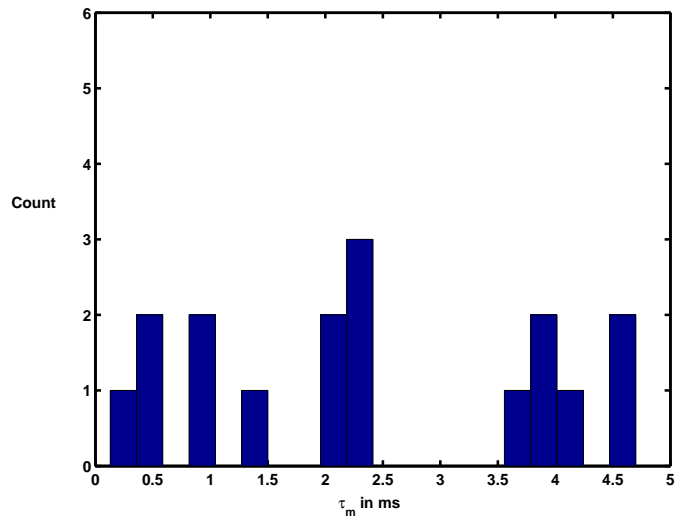,width=3.5in}
\end{center}
\caption 
{Histogram of the maximum values of  time constants $\tau_m$, in ms, for activation
 for L-type \CA currents using data from Table A3. Includes both experimental and
modeling studies less than 10 ms.}
\label{fig:wedge}
\end{figure}

Figure 5 shows two plots of $\tau_m$ versus $V$ for two of the few available
experimental data sets (but see Section 5 for results for L-subtypes).These 
are both for rat sympathetic neurons (A, Belluzzi and Sacchi, 1991; B,  Sala, 1991)
and are, in the absence of reliable L-type data in physiological or near
physiological solutions, taken from HVA studies which may contain L-type
and other high threshold components.  These data sets are very 
different in magnitude and form from those used in many modeling studies,
where sometimes the time constant for activation is taken to be a
large value, constant and independent of voltage. They are also considerably 
less than the values reported for \ab and \ac in 2mM \CA in Section 5.
Molitor and Manis (1999) report experimental values of  $\tau_m$ with a maximum
of about 1.2 ms at about V=-20 mV in  guinea pig dorsal cochlear nucleus neurons. 
In Figure 5 are
also drawn the steady state activation functions, also based on experiment,  for comparison.

The data of Table A3 on maximum values of activation time constants are collected
into a histogram in Figure 6, but this includes only values less than 10 ms.
The mean for the experimental results on neurons is
1.34 ms. 
 For all the data shown (n=17) the mean
is 2.38 ms  and the standard deviation is 1.51.

For time constants  $\tau_h$ for voltage dependent inactivation
in neurons  there are very few
data. The average of the two available experimental maximum
values is 65 ms and the average of 5 modeling maximum values
is 140.9 ms.

\subsection{Magnitudes of \IL or related quantities}
Data on the magnitudes of \IL in various cells is collected in Table A5.
Such data are presented in several forms, viz, (a) the actual
current density, usually in pA/pF, which may be for the whole cell or a part thereof (b) 
the permeability  (c) the  conductance per unit area $g_L$  (d) the whole cell
conductance $G_L$ (d) the fraction of total calcium current that is \ILN. 
 It can be seen that the proportion of \IL across cell types is extremely variable from nearly
all of the calcium current in some muscle and cardiac cells (Kapela et al., 2008; Benitah et al., 2010)
to just a few percent, for example in serotonergic (dissociated) cells of the
dorsal raphe nucleus (Penington et al., 1991).   The roles of \IL are presumably very different in these examples. 
From Table A5, the largest \IL current density is 19.3 pA/pF in supraoptic nucleus (Joux et al., 2001),
but almost as high magnitudes are found in rat CA1 pyramidal cells (Xiang et al., 2008)  and ventricular
cardiomyocytes in rats with renal failure (Donohoe et al, 2000).   The highest reported conductance density
is 7 mS per square cm in rat CA1 pyramidal cells (Xiang et al., 2008). 
Note that current densities can differ significantly in different preparations such as those with different ages, 
or with different times in culture.

\section{Examples of calculations of \IL}
In this Section we apply the basic model (1) to determine the L-type calcium current in
three cases. In the first, only VDI is considered, whereas in the remaining two examples, 
VDI and CDI are both operative.  

\subsection{Steady state currents obtained by the
linear and constant field methods with voltage dependence only}

 The activation and inactivation variables, which are dimensionless and take values in [0,1],
 may depend on membrane potential or ion concentrations, or both, but in this section 
 only voltage dependence is considered. Here we wish to compare 
 current-voltage relations obtained by the two methods outlined in Section 2.4. 
In order to compare the steady state currents ${\rm I_{CaL,Lin}}$ and ${\rm I_{CaL,CF}}$ calculated 
by the two methods, an average 
maximal conductance density $g_L$  (from the values given in Table A5) of 1.3925 mS/cm$^2$ was
assumed for the determination of ${\rm I_{CaL,Lin}}$, although this value varies greatly
from cell to cell.  For the steady state activation variable $m_{\infty}$ used in the calculation of ${\rm I_{CaL,Lin}}$, two values of
$V_{m,\frac{1}{2}}$ were employed, being  a relatively low value, $V_{m,\frac{1}{2}}=-29.6$ mV,
and an intermediate value $V_{m,\frac{1}{2}}=-14.4$ mV.  The corresponding values of
$k_m$ were 6.97 mV (low) and 7.78 (intermediate). When inactivation was included,
the values for $V_{h,\frac{1}{2}}$ were $V_{h,\frac{1}{2}}=-43.5$ mV
(low) and $V_{h,\frac{1}{2}}=-36.2$ mV (intermediate), with corresponding values for $k_h$ of
9.55 mV (low) and 6.18 (intermediate).

Four sets of values for $p_1$ and $p_2$ were employed, giving the forms 
$m, m^2, mh$ and $m^2h$.
The calculated results are shown in the four panels of Figure 7.  
  \begin{figure}[!hb]
\begin{center}
\centerline\leavevmode\epsfig{file=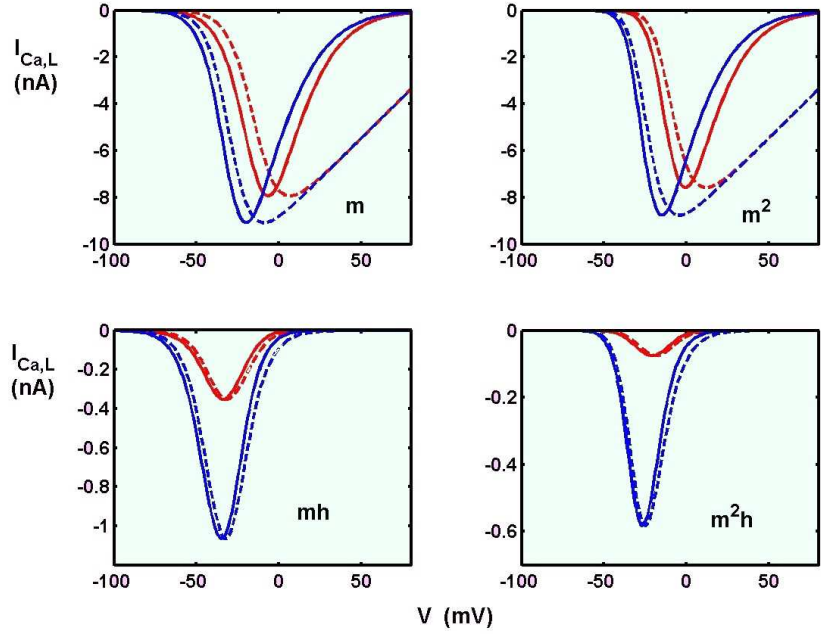,width=4.75in}
\end{center}
\caption 
{Calculated asymptotic steady state  ($t=\infty$) currents for L-type \CA currents
with average activation and inactivation properties for a relatively  low voltage activation
 ($V_{m, \frac{1}{2} } = -29.7$ mV), blue curves, and an intermediate value
($V_{m, \frac{1}{2}} = -14.4 $ mV), red curves.  The solid curves are for the constant field
current (formula (17)) and dashed lines are for the linear current case (formula (15)). The top two panels 
are for activation only ($m$ and $m^2$) and the bottom two panels are for activation with inactivation ($mh$ and $m^2h$).
In each of the 8 sets of curves, the maximum amplitudes of the constant field and linear currents are
scaled to be equal, for a conductance density of 1.3925 mS/cm$^2$ in the linear case.
Temperature 25$^o$C.}
\label{fig:wedge}
\end{figure}
In each case the current ${\rm I_{CaL,Lin}}$ (linear method) was calculated first,  the temperature 
being set at 25$^o$ C, the internal calcium concentration
${\rm Ca_i}$  fixed at 100 nM and the external concentration ${\rm Ca_o}$ fixed at 2 mM, giving $V_{Ca}=128.4$ mV
which was used as the reversal potential.  The constant field current ${\rm I_{CaL,CF}}$ was then calculated
and the permeability adjusted to make the peak (inward, negative) current the same as that in the linear case.
This gave 8 values of the permeability which are listed in Table 4 which are close to those reported for
L-type calcium currents (Hutcheon et al., 1994; Luo and Rudy, 1994). 
  
\begin{table}[!h]
      \begin{center}    
\caption{ Estimated $P_L$ from ${\rm I_{CaL, CF}}$
in 10$^{-4}$ cm/sec}
    \begin{tabular}{lllll}
  \hline
     $V_{m,\frac{1}{2}}$   &  $m$ &   $m^2$  & $mh$  &   $m^2h$      \\
  \hline
 Low (-29.6 mV) & 3.01 & 3.40  & 2.08 & 2.41 \\
 
    Intermediate (-14.39 mV)  & 4.41  & 5.26&   2.11 & 2.76 \\
                             \hline
  \end{tabular}
\end{center} 
\end{table}

The following can be seen from the results in Figure 7.  Details of the values of $V$
at which minima in \IL  (maximal current amplitude) occur are given in Table 5. Firstly,
{\it without inactivation} (top two
panels), as expected,  the constant field and linear calculations are similar in
magnitude until
a value of $V$ just greater than that at which the minimum of \IL (maximum amplitude)
occurs. However, the minima in the constant field case
occur  at voltages about 11-13 mV more negative than for the linear case, both for 
open probabilities given by $m$ or $m^2$ and whether \Vms is ascribed the low or high value. 
For larger values of $V$ the two calculations give strikingly different results,
as the constant field currents are almost symmetric about their minima and return to
almost zero at values of  $V$ around 70 mV whereas the linear calculation
has the current returning to zero very gently by comparison, and of course
reaching zero at the reversal potential of \CA (about 128 mV).  Whether such differences
are significant for any given cell depend on how long the voltage stays at values greater than those
at the minima (maximum amplitude), which are given in Table 5. 

Changing from an $m$ to an $m^2$ dependence
for \IL shifts the voltage at which the greatest current amplitude occurs by about 5 or 6 mV in
the depolarizing direction. Furthermore, there is not a large difference in the
magnitude of the current between the low and high \Vms cases. 
\begin{table}[!h] 
\begin{center}
\caption{Membrane potentials in mV  at maximal current amplitude}
\begin{tabular}{lllll}
\hline
& \multicolumn{2}{c}{\Vm=-29.7 mV }& \multicolumn{2}{c}{\Vm=-14.4 mV}\\
\hline
& Linear & Const Field & Linear & Const Field\\
\hline
$m$ &-9 & -20& 6.5 &-7 \\

$m^2$ &-4 & -15& 12 &-1 \\

$mh$ &-32 & -35& -32 &-34 \\

$m^2h$ &-25 & -27&-18 &-21 \\
\hline
\end{tabular} 
\end{center}  
\end{table}  
Secondly, {\it with inactivation} (bottom two panels) the current/voltage relations are
practically symmetric for all combinations of \Vms and  the factors $mh$ and $m^2h$.
Thus, the constant field calculations are practically the same as the linear calculations
for all values of $V$.  Inactivation, either as $mh$ or $m^2h$ shifts the voltage of
maximal current amplitude by very large amounts in the hyperpolarizing direction.
The smallest shift is  -12 mV with $m^2h$ and the constant field calculation and the 
largest is -30 mV with $m^2h$ and the linear calculation. 
The magnitudes of the currents are reduced by an order of magnitude
by the inactivation and the reductions are much greater for the higher value of \Vm (red curves).
The difference for the linear and
constant field results for the voltage of maximal current amplitude is only a few mV.

\subsection{Time dependent  \IL for high and low group neurons with VDI: $m$ or $m^2$?}
The question arises as to when the time-dependent 
results for an $mh$ form of current differ significantly from
those for an $m^2h$ form. This was investigated using
the parameters for steady state activation and inactivation
for the two sets called high and low summarized in Table 3.

   \begin{figure}[!h]
\begin{center}
\centerline\leavevmode\epsfig{file=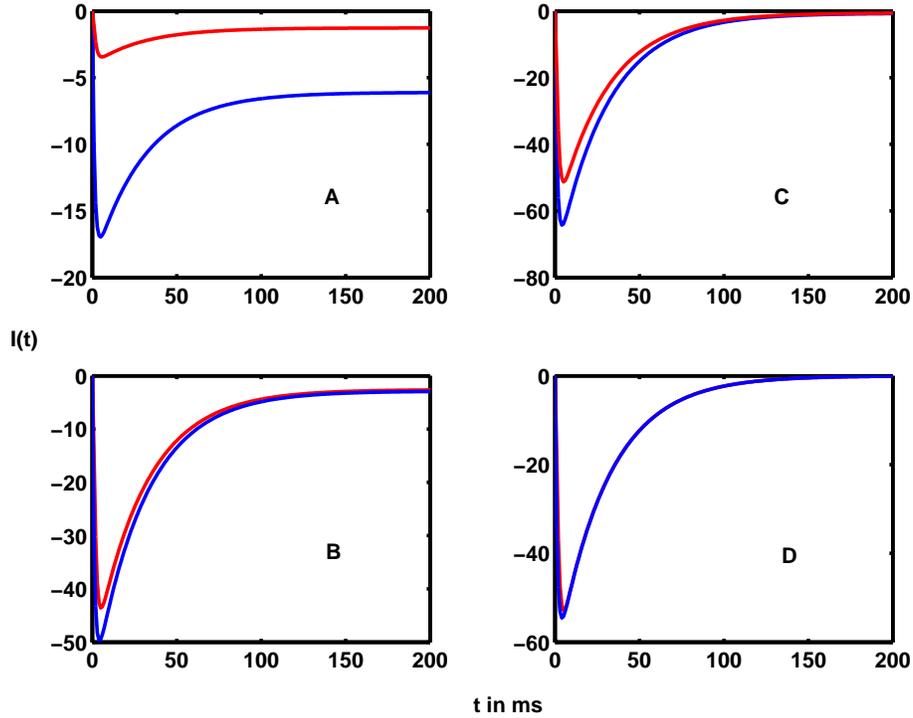,width=4.75in}
\end{center}
\caption 
{In all plots, current in arbitrary units 
versus time in ms. Blue curves are for $p=1$ and red curves are for $p=2$. A. High neuron
group parameters (see Table 3) with a step from -80 mV to - 30 mV.  B. High neuron group
parameters with a voltage step from -80 mV to 0 mV. C. Low neuron group parameters
with a step from -80 mV to -30 mV. D. Low neuron group parameters with a voltage
step from -80 mV to 0 mV.}
\label{fig:wedge}
\end{figure}
 The time constant of activation
was set at the average maximal experimental value of 1.34 ms
 for neurons and the time constant for inactivation was set at
29.9 ms as found in Belluzzi and Sacchi (1991), this being
one of the only two experimental values available. 
Current flows were computed for voltage steps
from an initial value $V_0=-80$ mV to final values
$V_1=-30$ mV and $V_1=0$ mV. The following Hodgkin-Huxley (1952)
formula may then be applied
\begin{eqnarray*}  I(t; V_0, V_1) &=&\ov{ g}_L (V_1-V_{rev,L})
[m_1 - (m_1-m_0) e^{-t/\tau_m}] ^p \\
&\times&  [h_1 - (h_1-h_0) e^{-t/\tau_h}], \end{eqnarray*}
where we have employed the abbreviations
$m_0= m_{\infty}(V_0),  m_1= m_{\infty}(V_1)$ 
for the activation,  $h_0= h_{\infty}(V_0),  h_1= h_{\infty}(V_1)$
for the inactivation, 
$ \tau_m$ and  $\tau_h$ being the
time constants.  For these calculations $\ov{ g}_L$ was set at unity
as only relative magnitudes are of interest.

The results are shown in Figure 8.
It can be seen that in most of the examples shown, there are only
minor differences in the currents computed by the 
$mh$ versus $m^2h$ form.  The case where the difference is very
significant (plot 8A) is for the high neuron group with a voltage step
from - 80 mV to -30 mV.  However, only a detailed neuron
model, which is beyond the scope of this article,  could  ascertain whether
significant differences arise in the ongoing spiking activity
when L-type currents are one of an array of current types. Nevertheless, 
it has been demonstrated that 
the choice of the form of the open probability contribution from voltage
dependent terms, $m^{p_1}h^{p_2}$, can  
have significant consequences for the current/voltage
relations or for current versus time.  In the ideal situation
voltage-clamp data for both activation and inactivation
would be used to find the best fitting theoretical form for a
component current, as in the original Hodgkin-Huxley (1952) analysis.
Care is required in reporting parameters for activation
and functions so that it is clear whether they are
for $m$ or $m^2$.  It is also important that the experimental data have been
analyzed correctly.

\subsection{Two time constants in  \IL inactivation}  
For L-type calcium current inactivation, most reports indicate that there are two time constants, 
one fast and one slow (Luo and Rudy, 1994; Romanin et al., 2000; Meuth et al., 2001; Budde et al., 2002; 
Lacinova and Hoffman, 2005; Faber et al., 2007;
Findlay et al., 2008).   According to Lacinova and Hoffman (2005), such is the case for smooth muscle, ventricular myocytes,
several neurons and the \ab channel. 
Furthermore,  in cells, fast time constants range from 7 to 50 ms and slow ones from 65 to 400 ms whereas for the
\ab channel, the ranges are 20 to 100 (fast) and 160 to 2000 (slow).  See Table A3 for several data although no distinction
is made between fast and slow inactivation time constants in the cited works. 
In an earlier study of calcium currents (not
specifically stated as L-type) in CA1 pyramidal cells,
Kay (1991) found, with 5mM \CAN, a fast time constant with a maximum value of about 240 ms at V=-10 mV, and a slow time constant
with a maximum of about 2200 ms at V=-11 mV. It was pointed out that typically (in cells without CDI)  the amount of inactivation 
increases with increasing V, and that the rate of inactivation tends to be Gaussian-like as a function of V, with a
maximum around the voltage \Vhs at which the steady state inactivation is half-maximal.

The relative contributions of CDI and VDI vary.  Lacinova and Hoffman (2005) state that in general
for a brief pulse, CDI is responsible for about 80\% of inactivation and VDI responsible for the remaining 20\%.
The faster time constant is usually taken to be associated with CDI. However, in some cases
there are taken to be two
time constants for VDI of L-type Ca currents, such as in ventricular myocytes
(Faber et al., 2007; Findlay et al., 2008) where the
dichotomy is attributed to the existence of two voltage-dependent inactivation states. 
Two time constants were also demonstrated by Morad and Soldatov (2005) for barium currents
with no CDI for \b.
     \begin{figure}[!h]
\begin{center}
\centerline\leavevmode\epsfig{file=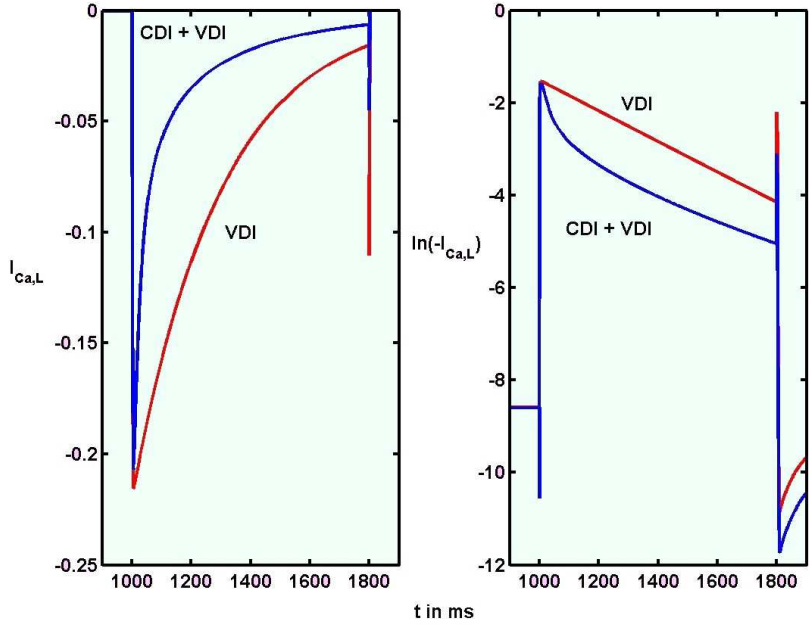,width=4.75in}
\end{center}
\caption 
{Showing two time constants for calculated L-type \CA 
currents evoked by a test pulse of duration 800 ms at 10 mV after
a holding potential of -60 mV. The blue curves are  for CDI and VDI, whereas
the red curves are for no CDI. The figure on right shows clearly the change in 
time constant from early to late (blue curve) compared with the single
time constant for VDI only (red curve).}
\label{fig:wedge}
\end{figure}

To illustrate the occurrence of two time constants we show results obtained for
\IL for the basic model of Equation (1) using the constant field expression for the 
current.  These are designed to be comparable to the experimental results on thalamic relay
(LGN) neurons in Budde et al. (2002), where a step to a test voltage of 10 mV is made for 800 ms after
holding at -60 mV.  
The parameters used were as follows. For activation,
\Vm  was set at  -37.5 mV with slope factor 5 mV, and for voltage-dependent
inactivation,  \Vh was set at -59.5 mV with a slope factor of 10 mV.
The time constant for activation was given by
\begin{equation}
\tau_m = 1 + 2 e^{ -  [(V - V_{m, \frac{1}{2} })/15]^2} 
\end{equation} 
and that for VDI was set at $\tau_h=300$ ms for all $V$. 
In addition, a calcium pump term was added of the form
\begin{equation}
f_{pump}({\rm Ca_i}) = K_c \frac{{\rm Ca_i}}{{\rm Ca_i} + K_p}
\end{equation}
where $K_p=0.0005$ mM (Rhodes and Llin\'as, 2005) and $K_c=0.000002$. 
Initial internal calcium concentration was set at 70 nM and extracellular 
at 2mM.  The standard value of the multiplicative factor $A\rho_LP^*_L$ in (19) was chosen to
make the maximum amplitude of \IL about 0.2 nA as in Budde et al. Figure 2.
For the CDI term we use the standard model (7) and (8) 
with $n=1$, $\tau_f=10$ ms (constant) and  $K_f=0.0002$ mM.

Results are shown for \IL in Figure 9, the contribution from the 
pump being small and not included.  During the pulse the internal calcium
concentration rose to a maximum of about 300 nM.  In the left panel, \IL is 
plotted against time in the cases of CDI + VDI (blue curve) and 
for VDI only (red curve).   With CDI the initial decline of current is 
very rapid and merges into  a slower decay 
after about 100 ms; with VDI only, the decline in current
is slow and relatively uniform as can be seen in the right hand panel
where $ {\rm ln (-I_{Ca,L})}$ is plotted against time to reveal changes in time constant.
The curve for VDI is a straight line, indicating a single time constant. 
The early (fast)  time constant for CDI + VDI was approximately 49 ms and 
the slow time constant about 264 ms. Both of these values are in the ranges 
given by Lacinova and Hoffman (2005) quoted above.  They are also similar
in magnitudes to the values reported in Budde et al. (2002). 
\subsubsection{Effects of varying parameters}
The effects of varying several parameters on the inactivation of \IL  according
to the above test pulse procedure are shown in Figure 10. Here both CDI and VDI are
present, but the graphical results are given only for the first 200 ms.
The parameters used here are as given by Rhodes and Llin\'as (2005) as in Table A1.3, using the same
$f_{\infty}$, 
but with  time dependent $f({\rm Ca_i},t)$.
The standard set of parameters is taken to be as for Figure 9, with the following variations. 
In the {\it top left} panel,  the current is calculated for four values of the CDI time constant 
$\tau_f$, being 5, 10, 30 and 100 ms.  Note that most modeling
of L-type currents effectively takes $\tau_f=0$  (see Table A1.3), as the steady state is
assumed to be attained immediately.    When $\tau_f$ is 5 or 10 ms (blue and red curves), the 
current has almost the
same time course as it does when the immediate steady state assumption is made.
For $\tau_f=30$ ms (green curve), which is the value used by Standen and Stanfield (1982), the decay of
current is much slower for the first 100 ms, very different from that for the immediate steady state assumption.
For $\tau_f=100$ ms (black curve), the decline does not reach values comparable with the remaining three
cases for at least 200 ms. In the {\it top right} panel  are shown \IL for four values of the 
time constant $\tau_h$ for voltage-dependent inactivation,  no dependence on $V$ being assumed. The value
of $\tau_f$ is set at 10 ms. 
             \begin{figure}[!h]
\begin{center}
\centerline\leavevmode\epsfig{file=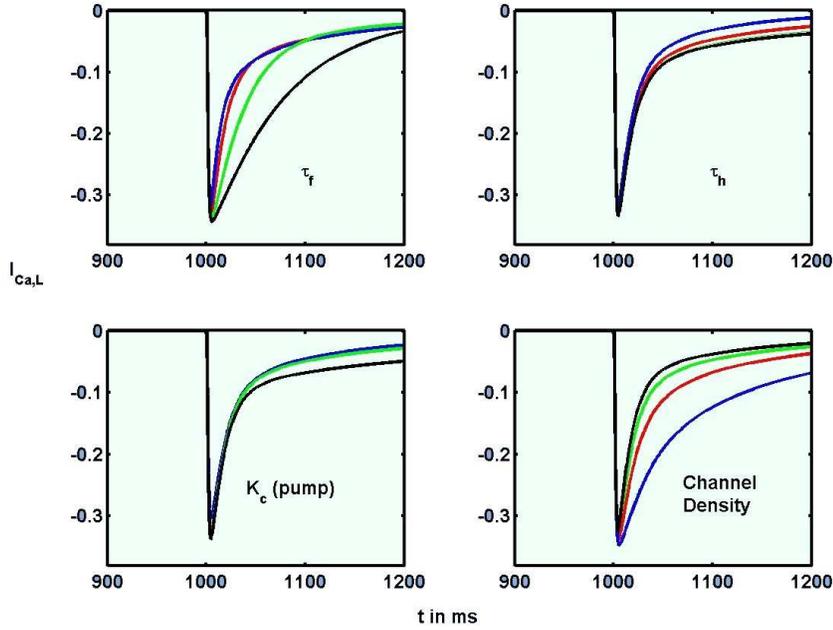,width=4.75in}
\end{center}
\caption 
{The effects of changing 4 parameters on the time course of the L-type calcium current
when there are both CDI and VDI. {\it Top left}: the time constant $\tau_f$ for CDI is varied.
Values in ms are 5 (blue), 10 (red), 30 (green) and 100 (black).  {\it Top right}: the time 
constant $\tau_h$ of VDI is varied. Values in ms are 100 (blue), 200 (red), 300 (green)
and 400 (black).  {\it Bottom left}: The effects of various calcium pump strengths $K_c$ 
on \IL - see Equation (21). Values of $K_c$ are 1 (blue), 2 (red), 3 (green) and 10 (black) times 
a base rate of 0.000001. {\it Bottom right}: How the channel density $\rho_L$,  through the magnitude
of the current,  affects \ILN.  The multiplicative factor $A\rho_LP^*_L$  has values 0.25 (blue), 1 (red), 2 (green) and 3 (black)
times the base rate, where the base rate is the standard value described for the previous figure.}
\label{fig:wedge}
\end{figure}
Results are shown for time constants $\tau_h$  (in ms) of 100 (blue), 200 (red), 300 (green) and 400 (black),  which values
encompass approximately  the values employed (Table A3). As could be anticipated,
increasing $\tau_h$ from 100 to 400 has a negligible effect for the first 20 ms but quite noticeable effects
on the late current.  

In the {\it bottom left} panel are shown results of varying the parameter $K_c$ which determines the rate of
calcium pumping as in (21).  The values employed are 1,2,3 and 10 times a base rate with blue, red, green and
black curves respectively.   When the pump rate is higher,  \CA is less available for CDI so the
current has a larger magnitude. However, in the situation modeled here,  it can be seen that a significant
difference in \IL occurs when the
pump rate is increased tenfold, but that increases of 100\% and 200\% have only a small effect. 
Finally, in the {\it bottom right} panel are shown the time courses of \IL for various channel
densities, being 1 (blue), 4 (red), 8 (green)  and 12 (black)  times a base rate.  The current amplitude is assumed 
proportional to the channel density so the higher channel densities result in more effective CDI. 
These results indicate that these four quantities have a significant influence on the amounts of CDI and VDI
and thus the magnitude and time course of \ILN.

\subsubsection{CICR}
If there is a substantial amount of CICR, as occurs in cardiac and other muscle cells, then there
may be two phases of CDI,  consisting of an early rapid component due to release of \CA from the 
sarcoplasmic reticulum and then a slower phase in response to calcium ion entry through L-type channels
(or other calcium channels) 
(Budde et al., 2002; Bodi et al., 2005; Cueni et al., 2009; Empson et al., 2010). This raises the possibility of three time constants in the decay of \ILN, two
for the two phases of CDI and one for VDI, or even four time constants if as mentioned above there are
two for VDI. 
The effect of \CA released from sarcoplasmic reticulum  is graphically illustrated in Hinch et al. (2004). 
The basic model (1) can incorporate the influence of  CICR  on CDI by taking account of the time-course of the 
\CA  concentration,  as stated in Section 2.1. This quantity will usually be obtained from 
either ordinary  (for ${\rm Ca_i}(t)$) or partial differential equations (for ${\rm Ca_i}(x,t)$, where $x$ is from 1 to 3 space 
variables) which incorporate calcium ion concentration changes due to pumping and buffering as well as 
fluxes through various channels. See Luo and Rudy (1994) for such
a deterministic example.   A detailed stochastic model for CICR in cardiac myocytes has been
analyzed by Williams et al. (2007).

\subsection{Calculations for double-pulse protocols}

According to Budde et al. (2002), CDI tends to be rapid and results in a ``U-shaped''  inactivation 
curve  when  investigated using a double pulse protocol (Tillotson, 1979; Kay, 1991; Meuth et al., 2001).  
We show corresponding results for the standard model for
L-type calcium currents as described in section 2. The voltage is held at a ``conditioning''
 value, $V_{COND}$  followed by a pause and then a ``test''
pulse at $V_{TEST}$ is delivered. Of interest is the relationship between the first and second responses.
As in Figure 1 of  Budde et al. (2002), 
mathematically we have
\newcommand{\q}{\quad}
\begin{equation}
V(t) =
\left\{\begin{array}{ll}
V_1,   \q & \hbox{ } 0 \leq t < t_1, \\
                  V_{COND}, \q &\hbox{$t_1 \leq t < t_2$,  } \\
                  V_1 \q &\hbox{$t_2 \leq t < t_3$,  } \\
                  V_{TEST} \q &\hbox{$t_3 \leq t < t_4$,  }\\
                   V_1 \q &\hbox{$t \geq t_4$.    }
\end{array}
\label{eq:diff}
\right.
\end{equation}

   \begin{figure}[!hp]
\begin{center}
\centerline\leavevmode\epsfig{file=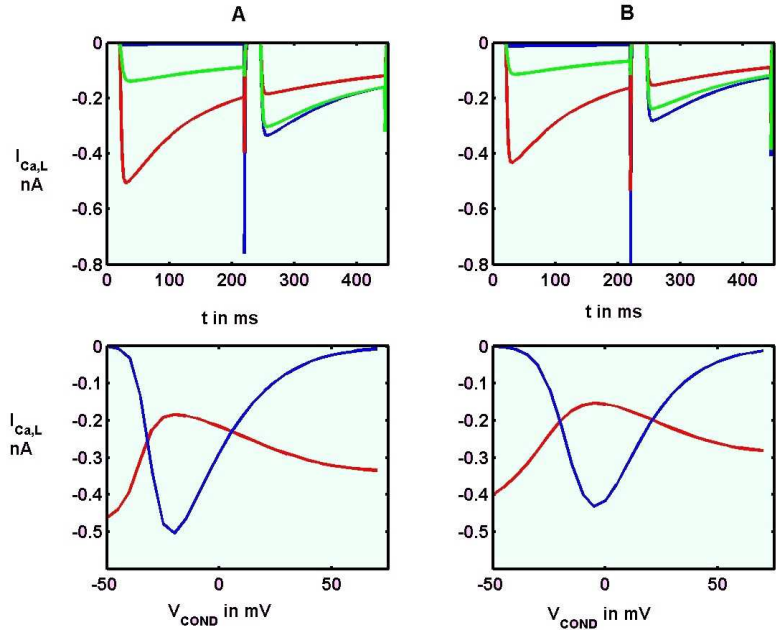,width=4.5in}
\end{center}
\caption 
{Calculated L-type calcium currents in the two-pulse protocol described in the text using the standard model given by (1) with VDI and 
CDI. Time is in ms, voltages in mV and currents in nA.  In the left column are results with data for activation and VDI 
of Pospischil et al. (2008), and in the right column of Rhodes and Llin\'as (2005). The voltage is held at $V_{COND}$ from
$t=20$ to $t=220$  and at $V_{TEST}$ from $t=225$ to $t=445$. In the bottom graphs the amplitude (minimum)
of \IL is plotted against the conditioning pulse voltage for the first pulse (blue curves) and for the second pulse (red curves).
$V_{TEST}$ is at the minmum of the blue curves which occur at -20 mV for the left set of results and -5 mV for the right set. 
In the top graphs, responses are shown for three particular values of $V_{COND}$. The red curves are for $V_{COND}$ at the value
of $V_{TEST}$, which gives a large reduction in \IL  on the second pulse due to CDI. The blue curves are for $V_{COND}=70$
where the current on the first pulse is very small. The green curves are for $V_{COND}=-35$ (left) and $V_{COND}=-25$ (right). 
The results may be compared with those of Figure 1 in Budde et al. (2002). }
\label{fig:wedge}
\end{figure}
We use the constant field method and assume,  as the results of Section
4.2 indicate a relative insensitivity to the power of $m$, 
\begin{equation}
{\rm I_{CaL}} = {\rm F}(V,{\rm Ca_i, Ca_o}) m^2(V,t)h(V,t)f({\rm Ca_i},t)
\end{equation}
where ${\rm F}$ contains physical constants and the constant field factor as given by (17). 
For the CDI term we use (7) and (8) with $n=1$, $\tau_f=30$ mS (constant) and  $K_f=0.0005$ mM. 
            
For the voltage-dependent activation and inactivation parameters we use the two sets
for thalamic relay cells as given by Pospischil et al. (2008)  (see Table A1.1) and Rhodes and Llin\'as (2005)  (see Table A1.3).
The time constant for activation was chosen as in Equation (20), 
and the inactivation time constant was set at $\tau_h=500$ ms for all $V$.

The conditioning pulse is on from 20 to 220 ms and the test pulse is on from 225 to 445 ms. 
The voltage at the test pulse is taken as that at which the response to the first pulse has the
largest (negative) amplitude. 
Results of these calculations are shown in Figure 11. 
The curves in the bottom part of the figure show the magnitudes of the currents in 
response to the first (blue) and second (red) pulses. The reduction in the amplitude of
in the second pulse is most prominent when the response to the first pulse is the greatest,
because  ${\rm Ca_i}$ has increased the most, resulting in the greatest amount of CDI. 
In the top parts of the Figure are shown responses to two-pulse clamps for three particular
values of $V_{COND}$, as explained in the Figure caption.
Although such results indicate that the standard
model for \IL   with the usual voltage-dependent activation, coupled with VDI and CDI
may be adequate for describing
certain experimental paradigms, it is
uncertain whether this is the case in applications to single cell models, mainly due to
the lack of concrete data.    The relative contributions of CDI
and VDI depend on the particular cell type, mainly through the various 
combinations of subtypes of L-channels as well as modulating
factors such as sub-unit configurations and various ligands.

\subsection{Difficulties in the accurate modeling of \IL}
Mathematical modelers of nerve cell physiology are
invariably faced with the dilemma of insufficient
accurate data concerning the many ion channels
influencing a neuron or muscle cell's activity.
It is hard to think of a single neuron for which
such data is available for all known channel types. 
  As an example, 
L-type current data, for a current denoted by
$I_L$,  were not available for thalamic
relay neurons when they were first comprehensively
modeled (McCormick and Huguenard, 1992). 
Instead, the parameters were 
obtained from the calcium current results of Kay and Wong (1987)
for CA1 pyramidal cells, where in fact an L-type current
had not been pharmacologically identified.
Kay and Wong hypothesized that the channels they had
analyzed were from a homogeneous population. 
The same or similar data 
were employed by Traub et al. (1991) for $I_{Ca}$ in CA1 pyramids, 
 Hutcheon et al. (1994) for $I_L$ in thalamic relay cells,
Traub et al. (1994) for for $I_{Ca}$ in CA3 pyramids, 
Wallenstein (1994) for $I_L$ in 
nucleus reticularis
thalami neurons and Wang (1998) for $I_{Ca}$ in cortical
pyramidal cells. In none of these models was calcium-dependent
inactivation included.  

It is hoped that in the future, 
experiments can be performed to remedy the paucity
of accurate data required for L-type (and other) current modeling.
There are other limitations apart from the use of data from one cell
type for another. The fact that the majority of experiments are
performed on dissociated cells means that existing data must be biased
towards those for cell somas, so that guesses have to be made
for dendrites.  To make choice of parameters
even more difficult is the use of various different
concentrations of either calcium or barium ions which
may give results for such quantities as activation potentials
which are very different from those for physiological concentrations.
Temperature is another key factor which may alter the dynamic
properties considerably. Furthermore, results for heterologous 
ion channels may not be the same as for the corresponding
channels in native cells. 

Thus, given the many uncertainties
in the parameters describing the kinetics of the various ion
channels, it is hard to know whether model-based 
predictions are truly descriptive of cell electrophysiology
even when they agree with experimental
observations.  The hope is then that  the modeling
 results are robust enough with respect to relatively small changes in the 
parameters.  To construct a model
of a nerve cell is a very large undertaking, and 
fortunately many researchers have had the courage
to proceed with such endeavours in the light of uncertainties 
concerning the underlying channel data.

\section{L-channel subtype properties}

  L-type currents are not only important for neuronal and
cardiac cell dynamics. Although
L-type channels play a limited role in the process of synaptic transmission, they are crucial
 for activity-dependent gene expression and for regulating
plasticity at certain synapses (Hardingham et al. 2001; Dolmetsch et al., 2001; Helton et al. 2005). 
Thus they have been found to play an essential role  
in long-term alterations in synaptic efficacy underlying
learning and memory in the hippocampus (Kapur et al., 1998;  Graef et al., 1999; Leitch  et al., 2009)
where  both \ab and \ac channel subtypes
are predominantly located in
postsynaptic dendritic processes and somata. 
Furthermore,  current through  \ac channels has been demonstrated at the output 
synapses of mice AII amacrine cells (Habermann et al., 2003).

\subsection{Distribution}
The \aa subtype is mainly found in skeletal muscle and \ad is found in the
 retina, but also in human T lymphocytes (Kotturi and Jefferies, 2005).
  The remaining subtypes \ab and \ac constitute the main
 L-type channels in neurons, endocrine cells and heart cells. 
 The \ab subtype is widely expressed in heart, brain, smooth muscle and
 endocrine systems (Ertel et al., 2000).
In the brain it is found in cortex,
 hippocampus, thalamus, hypothalamus, caudate putamen and 
 amygdala (Splawski et al., 2004). In neuroendocrine
cells, \ab and \ac channels are involved in action potential generation, bursting activity
and hormone secretion (Lipscombe et al., 2004; Marcantoni et al., 2007). 
According to Vignali et al. (2006),  60\% of calcium current in both mouse
 pancreatic A- and B-cells
is L-type, with mainly \ab in B-cells and both \ab and \ac in A-cells. 
 The \ac subtype is found in
 sinoatrial node, cochlear hair cells, and dendritic neuronal processes (Dolphin, 2009).
Schlick et al. (2010) reported distributions of subtypes in many brain
regions, including during development,  and found that patterns tended to be intrinsic
 rather than dependent on neural activity. Using barium as charge carrier,
Navedo et al. (2007) investigated isoform contributions in mouse
arterial smooth muscle. Both \ab and \ac were involved in
calcium sparklet formation but  \ab was primarily
responsible (see Table A6).

 Immunofluorescence studies have determined  that both \ab
and \ac have mainly proximal locations on neurons,
particularly on the dendrites, and they may  make a
substantial contribution to the  inward current and
the action potential (Holmgaard et al., 2008).  
 According to Martinez-Gomez  and Lopez-Garcia (2007), in mouse
 and rat spinal neurons \ab 
 channels tend to localize in soma and proximal
dendrites whereas \ac channels are also found in distal dendrites.
However, Zhang et al. (2006) found that in the cat,  \ac channels were dense in ventral
horn motoneurons, occurring mainly on somata and proximal dendrites and being responsible for 
plateau potentials in these cells. Forti and Pietrobon (1993) 
found 
that functionally different L-type \CA channels coexist
in rat cerebellar granule cells.
Although \ab and \ac
are often found in the same neuronal processes,
particularly dendrites, their subcellular distributions are sometimes 
distinct (Hell et al. 1993). An interesting graphical representation
of the distribution of various calcium channels over the 
surface of hippocampal
pyramidal cells was given in Magee and Johnston (1995).

Differential effects of corticosterone have been found for \ab and \ac
in the CA1 region of hippocampus and the basolateral amygdala (Liebmann et al., 2008). 
In the former, corticosterone increases the amplitude of the slow afterhyperpolarization
whereas in the latter no such effect is observed.  The different  responses are thought to reflect 
the expression of  \ac  in hippocampus and the absence of this subtype
 in the basolateral amygdala.

\subsection{Role in pacemaking}

 In the heart, \ac  is found in atrial tissue where it  contributes to pacemaking
but not in ventricular muscle that expresses
\ab (Lipscombe et al. 2004). 
According to Torrente et al. (2011), SA node pacemaker activity
controls heart rate and local \CA release is tightly
controlled by \ac channels. In \ac KO mice, the 
frequency of \CA transients in pacemaker cells was reduced by 45\%
 relative to WT.  In another study with KO mice,  Zhang et al. (2011) found that \ac channels are 
highly expressed in atrioventricular node cells and played  
a critical role in their firing.   
Mesirca et al. (2010)  pointed out that
atrioventricular node cells may contribute to pacemaker activity in the case
of SA node failure.

 The properties of  \ac channels  
which make them more suitable  for pacemaking activity
than \ab channels  were discussed
in Vandael et al. (2010) with particular attention to
dopaminergic neurons, cardiac pacemaker cells and adrenal
chromaffin cells. 
 \ac 
activates with steep voltage dependence at voltages considerably
more negative than  \ab
and has faster activation but slower and
less complete voltage-dependent inactivation.
 According to Striessnig (2007), such activation
at low voltages enables \ac
\CA currents  to support pacemaking
activity in sinoatrial node and
some neurons (for example Putzier et al., 2009a).  For adrenal
chromaffin cells, Marcantoni et al. (2010) found that
although \ab and \ac occurred with about equal frequency
in WT cells, most of which fired at about 1.5 Hz,  \cc-deficient cells had
 significantly less firing.  The essential role of \ac and their coupling
to BK channels in the pacemaker activity of these cells was established.
Alternative splicing  of the C terminus of the
 $\alpha_1$ subunit of \ac has been shown
to give two forms, of which one (short, 42A) is expected to be 
more supportive of pacemaker activity (Singh et al., 2008)
with 73.6\% inactivation of $I_{Ca}$ in 30 ms. See also Table A6.

\subsection{Biophysical properties: activation and VDI}

In Table A6 are listed some of the available data on voltage-dependent
activation and inactivation for examples of the four main subtypes. 
As pointed out in the Introduction, there are no definitive results for each subtype because
of the modulatory effects of other subunits and varying experimental conditions such as
the use of
different hosts and different ionic concentrations. 
That is, one can expect considerable 
variability in channel properties for a given subtype from different preparations. 
As seen in the examples shown in Figure 3, \ac channels can have quite  low activation
thresholds compared to the \ab subtype and which are, in the data given here, similar to that of T-type. 
The latter have strong
VDI whereas \ac currents are reported to have have weak VDI and strong CDI (Lipscombe, 2002).
Very few modeling studies have distinguished subtypes, an exception being Shapiro and Lee (2007). 
It is also apparent from the steady state inactivation function for \ab shown in Figure 3,
based on Hu and Marban (1998), 
that virtually no \ab channels would be available to open around resting membrane 
potential (about -60 mV). 
 
In a pioneering study using tsA201 cells, Koschak et al. (2001) determined electrophysiological
properties for \ab and \ac (see Table A6) with current densities of 24.5 and 24.3 pA/pF, respectively,
in 15 mM Ba$^{2+}$. 
Generally in the cases reported 
\ac has  faster kinetics than \ab (Helton et al., 2005) which is seen in the data of
 Table 6 and Figure 12.
Inspection of Table A6 reveals surprisingly that \Vms is lower in the example for  \ac than for \b, but that
\Vhs is higher for \ac than for \ab.  
Unfortunately the incompleteness of the data for all subtypes
prevents a quantitative comparison of their relative current amplitudes, either
singly or in combinations. Such a program will be useful
for future work when more data are available.

\begin{table}[!h]
      \begin{center}    
\caption{Examples of data on activation time constants for L-current subtypes}
    \begin{tabular}{lllll}
  \hline
       Reference   & $ \rm{ Ca_ v1.1}$   &   $ \rm{ Ca_ v1.2}$ &  $ \rm{ Ca_ v1.3}$   &  $ \rm{ Ca_ v1.4}$  \\
  \hline
   {\er  Koschak (2001)} & & 2.3 ms & 1.2 ms &   \\
     &   &  15mM Ba$^{++}$ &  15mM Ba$^{++}$   &     \\

 {\er  Hildebrand (2004)} &  {\er Control} & 1.34 ms &  &   \\
    {\er  2mM Ba$^{++}$} &  {\er  10 $\mu$M allethrin}  & 1.23 ms &   &     \\

   {\er  Catterall (2005)} &  $>$ 50 ms &  1 ms & $<$ 1 ms & $<$1 ms   \\
     & 10 mV &  10 mV &  10 mV &     \\

   {\er  Helton  (2005)} & &  $2.43e^{-0.02V}$ & $1.26e^{-.02V}$  &   \\
      &  &  $V\in[-20,20]$ &  $V \in [-55,10]$  &  \\

   {\er  Shapiro (2007)} &  {\er  Model, MN} & 30 ms & 24 ms &   \\

  {\er Luin (2008)}  &  {\er  young}  53.6*   &   & &    \\
   & {\er old}  62.5*  &   & &   \\

     \hline
   \end{tabular}
\end{center} 
\end{table}

There are only limited data on activation time constants for 
L-channel subtypes, some being  given in Table 6. Those for \ab and \ac
are graphed in Figure 12.  Data from Koschak et al.  (2001) are for tsA201 cells. The 
data of Helton et al. (2005) were given over the membrane
potential ranges $-55 \leq V_m \leq 10$ for \ac and $-20 \leq V_m \leq 20 $ for \b. 
The following expression was employed by Putzier et al. (2009a) for the activation time
constant of  $ \rm{ Ca_ v1.3}$:
\begin{equation}
\tau_m = \bigg(    \frac  { 0.020876(V + 39.726)   }  { 1- e^{    -(V+ 39.726)/4.711}  }  
+ 0.19444e^{ - (V+15.338)/224.21}    \bigg)^{-1}.
\end{equation}

\begin{figure}[!hbp]
\begin{center}
\centerline\leavevmode\epsfig{file=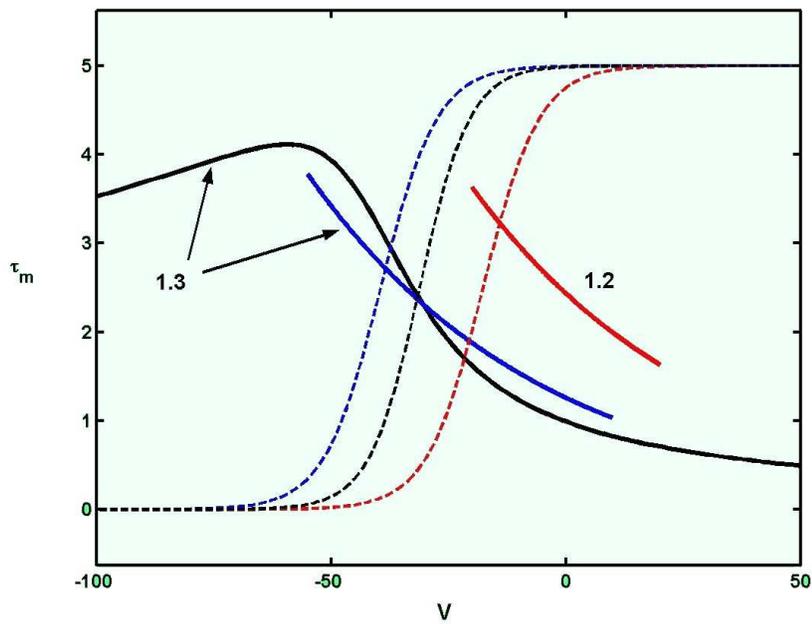,width=4.75in}
\end{center}
\caption 
{Solid lines: time constants for activation, in ms, versus membrane potential in mV, 
 found for examples of L-channel subtypes  \ab and  \cc. Red and blue curves
from Helton et al. (2005) (see also Table 6), black curve from Putzier et al. (2009a) (see
also Equation (24)).   Dashed lines: corresponding activation
functions $m_{\infty}$, scaled upwards.}
\label{fig:wedge}
\end{figure}

This function is plotted in Figure 12 and can be seen to be approximately the same
as the function $1.26e^{-.02V}$ given by Helton et al. (2005) over the range $[-55,10]$. 
Also shown for comparison in Figure 12 are the steady state activations $m_{\infty}(V)$
corresponding to the three graphs of $\tau_m$ versus $V$.  

\subsection{Pathologies linked to \ab and \ac channels}
Several pathologies have been linked to \ab and \ac channels.
Examples include 
autism, bipolar disorder, and Parkinson's  disease (Andrade et al., 2009). On the other
hand, \ab and \ac were shown to not be directly related to absence epileptogenesis
(Weiergr\"aber et al., 2010).  Mutations of \ab have been found to
result in Timothy syndrome (Splawski et al., 2004), which is a multi-organ disease
manifesting in cardiac arrythmias,  metabolic and immune system deficiencies and autism. 
It has been posited (Barrett and Tsien, 2008; Yarotskyy  et al., 2009)  that strongly decreased 
voltage-dependent inactivation, but not CDI,  associated with
the mutant channel subtype is a  
possible underlying cause.    Alterations in L-channel 
density and activity have been found
 in aged rat hippocampal neurons
with the possibility of a role in cognitive decline and the occurrence of Alzheimer's
 disease (Thibault and
Landfield, 1996; Thibault et al.,  2001, 2007;
Norris et al., 2010). 
The expression of amyloid precursor protein 
was shown to increase L-type \CA currents, thus leading to the inhibition of 
 calcium oscillations (Santos et al., 2009).  Further, performance of rats in a maze task was found to be
 strikingly negatively correlated with
L-channel density (Thibault and Landfield, 1996). 
 
Altered properties of L-type calcium channels  have also been 
implicated in muscular dystrophy (Collet et al., 2003) and epilepsy
(Fuller-Bicer et al., 2009). N'Gouemo et al. (2009) found upregulation
of various high threshold calcium currents including ${\rm I_{CaL}}$
in rat models of epilepsy. 

 The role of \ac in pacemaking in sino-atrial node and  inner hair cells of the cochlea  has been
demonstrated  in KO mice which exhibit significant sinus bradycardia and hearing
 loss (see Lipscombe et al.,
2004).   Striessnig et al. (2004)  reviewed  the role of L-type calcium channels in human disease
caused by genetic channel defects (channelopathies)
in $\alpha_1$ subunits. Recently it has been demonstrated in humans that a  
mutation in a gene which encodes the $\alpha_1$-subunit of \ac channels is associated
with deafness and sino-atrial node dysfunction (Baig et al., 2010).  The possibility
and usefulness of subtype-specific therapeutic drugs has been discussed in 
Striessnig et al. (2006) and more recently in Zuccotti et al. (2011). 
A  comprehensive review of L-subtype properties (and other calcium currents)
and associated channelopathies was compiled by  Piedras-Renter\'ia et al. (2007).

\section{Concluding remarks} 
L-type \CA currents have many fundamental roles, including providing basic depolarization
in action potential generation, amplification of synaptic inputs (established in spinal motoneurons) and 
involvement in signaling pathways for activating
transcription factors such as CREB and hence the
expression of genes that are essential for synaptic plasticity and other
important cellular processes.  \ad  L-type
channels are also involved in \CAN-signaling in immune cells

Quantitative descriptions of L-type \CAN currents   thus play an important role in understanding 
the dynamics of many processes in neurons and other cells, particularly cardiac myocytes.
We have here reviewed the basic deterministic methods for a mathematical description of these
currents. Whereas activation is usually taken to be voltage dependent, inactivation has
varying degrees of CDI and VDI and each cell type requires a 
separate analysis of their relative contributions, which requires
experimental investigation in addition to fundamental current-voltage relations.
Incorporating L-type \CAN currents in a neuronal model
is often hampered by lack of knowledge as to the location of the corresponding
ion channels, which is important to ascertain their 
contributions  to cell dynamics. An additional complication is that some cells
have more than 1 subtype of L-type channels with separate 
properties and locations.  

The use of Markov chain theory, not described here,  
for  transitions between various
channel states has proven useful for cardiac cells but as far as can be discerned  has not yet been
applied to L-type channels in neurons. 
The deterministic, or averaging approach, outlined
in this review has the advantage of simplicity
and is thus likely to continue to be employed in modeling L-type (and other) calcium currents
in multi-component nerve cell models. It is nevertheless not capable of capturing the fine details of the 
gating schemes for the activation and inactivation processes (Peterson et al., 1999;
Erickson et al., 2003; Tanskanen et al., 2005; Tadross and Yue, 2010; Williams et al., 2010)
 which are made
complicated by the  roles of subunits and the varying degrees of voltage-dependent
and calcium-dependent dynamics.  See also Section 4.4 
for a discussion of some of the general difficulties encountered in
constructing accurate mathematical models for neurons.

\section{Acknowledgements}
I am grateful to Professor Nicholas Penington, Physiology and
Pharmacology, SUNY Downstate Medical Center, Brooklyn,
for useful comments on the manuscript and for valuable references, 
and to Professor Annette Dolphin, Department of Neuroscience, 
 University College London,  for
permission to reprint Figure 1. The referees' remarks were an
invaluable source of improvement and gave rise to a more complete content.  
Support from the Max Planck Institute is appreciated.

\section{Appendix: Tables of data on L-type currents}
The following tables are representative of experimental
and modeling studies involving L-type calcium currents
or calcium currents with properties similar to (or claimed to
be similar) to L-type. In the years following the introduction of the 
term L-type, the term was often taken to imply a high threshold
current, although later some L-type currents were found to be
activated at quite low membrane potentials. The pharmacological
distinction between the various types is desirable but has
not always been possible. Except in the case of
experiments on subtypes which are clearly identified, usually in transfected cells, 
original sources may be consulted to determine the
likelihood that  the ion channels under
consideration belong to the L-type category. 
Table A1.1 contains activation and inactivation data and 
modeling studies as opposed to experimental determinations
have been marked M. Table A3 on time constants has been divided into an 
experimental part and a modeling part. 
\newpage
\hoffset  .21  in

{\small
\begin{center}

{Table A1.1}
\centerline{Parameters for L-type Ca$^{2+}$ activation and, if included, VDI}
\centerline{Non cardiac cells  \hskip .1 in  *Denotes estimated \hskip .1 in  M denotes model}
\smallskip

\begin{tabular}{llllll}
  \hline
     Source, cell type    & Form for current  &  $V_{m,\frac{1}{2}} $  & $k_m$  &      $V_{h,\frac{1}{2}} $        & $k_h$ \\
 &  if given  &  &&& \\
  \hline
 {\er  Fox (1987)} &  & 2  & 4 & -40 & 8\\
 {\er Chick DRG, 10mM \CA} &  & & & & \\

    {\er Kay (1987)} & $m^2$  &-19  & 7.73 &  & \\
  {\er CA1 pyramidal, 2mM \CA} & & & & & \\

 {\er  Fisher (1990)} &  & 17  & 4.7 &  & \\
 {\er CA3 pyramidal, 100mM Ba$^{2+}$} &  & & & & \\

    {\er Belluzzi (1991)} & $m^2h$, const. field &-8.1 & 9.85 &-19.91 &4.5 \\
  {\er SYMP N, 5mM \CA} & & & & & \\

 {\er Sala (1991)} & $m^2$ &-4 & 9.7 & & \\
  {\er SYMP N, 4mM \CA} & & & & & \\

    {\er Thompson (1991)} & &  & &-78 & 9.9 \\
  {\er CA1 pyramidal, 3-10mM Ba$^{2+}$ } & & & & & \\

   {\er Traub (1991), M} & $m^2h(V-V_{Ca})$ &-19 & 7.73 & & \\
  {\er CA3 pyramidal, $V_{Ca}$=140} & & & & & \\

  {\er McCormick (1992), M} & $m^2$, const. field  &-14.1*& 8.7*& &\\
  {\er TR cell, 2mM \CA} & & & & & \\
   
  {\er Schild (1993), M$^{\dagger}$} & $m^2h(V-V_{Ca})$ & -2.8 & 9.85 & -14.61 & 4.5 \\
{\er Rat med, 2.4mM \CA } & & & & & \\
  
   {\er Hutcheon  (1994), M} & $m^2$, const. field & -14 & 9.1&& \\
  {\er TR med dors, 2mM \CA} & & & & &\\
  
   {\er Jaffe (1994), M} & $m^2$, const. field & 4 & 4.7  && \\
  {\er CA3 pyramidal} & & & & &\\

   {\er Traub (1994), M} & $m^2(V-V_{Ca})$ &-19 & 7.73 & & \\
  {\er CA3 pyramidal, 2mM \CA} & & & & & \\

   {\er Wallenstein  (1994), M} & $m^2$, const. field & -14.1 & 8.7 && \\
  {\er NRT N, 2mM \CA} & & & & &\\

  {\er Johnston (1995)} & $m^2h$, const. field & -15 & & & \\
  {\er General} & & & & &\\
  
 {\er  Kuryshev  (1995) }&& -12.3 & 7.8 & &\\
  {\er Rat corticotropes, 10mM Ba$^{2+}$} & & & & &\\
  
    {\er Migliore (1995), M}  & $m^2h(V-V_{Ca})$ &  -11.3 & 5.7 &&\\
  {\er CA3 pyramidal, 2mM \CA} & & & & &\\
   \hline
\end{tabular}
\newpage

  \c{Table A1.1 continued} 
\vskip .1 in 
    \begin{tabular}{llllll}
  \hline
     Source, cell type    &Form for current  &  $V_{m,\frac{1}{2}} $  & $k_m$  &    $V_{h,\frac{1}{2}}$       & $k_h$ \\
         &if given  &  &  &      &  \\
  \hline 
{\er Magee (1995)}  &   &  9  &  6  &&\\
  {\er CA1 pyramidal, 20mM Ba$^{2+}$} & & & & &\\

    {\er Avery (1996) } &  & -30& 6  &  &    \\
 {\er CA3 pyramidal, 2 mM \CA} & & & & & \\

    {\er Li (1996), M}  & $m^2(V-V_{Ca})$ &  -20.0 & 5.3 & &\\
  {\er DA SN, $V_{rev,L}$=120} & & & & &\\

  {\er LeBeau   (1997), M}  &  $m^2$, const. field &  -12& 12& &\\
  {\er PC} & & & & &\\

    {\er Booth  (1997), M}  & $m(V-V_{Ca})$ &  -40 & 7 & &\\
  {\er Spinal, \VR=80} & & & & &\\

 {\er Wang  (1998), M}  & $m(V-V_{Ca})$ &  -20 & 9 & &\\
  {\er CORT pyramid, $V_{Ca}$=120} & & & & &\\

 {\er Molitor  (1999)}  &    &  -10 & 8   & &\\
  {\er DCN N,  10mM Ba$^{2+}$} & & & & &\\

{\er Athanasiades  (2000), M} & $m^2h(V-V_{Ca})$ & -22.8 & 9.85 & -34.61  &  4.5  \\
 {\er  MR, 2.4 mM \CA  } & & & & & \\

  {\er Guyot(2000) }& & -15 & 6 & -33.5 &  4.5 \\
  {\er Bovine adrenal, 20mM Ba$^{2+}$   } & &&& & \\

    {\er Shorten (2000), M} &  $m^2$, const. field  & -18 & 12 &  &    \\
 {\er  PC, 20 mM \CA} & & & & & \\
                             
    {\er Carlin (2000), M (HVA)}  &  $m(V-V_{Ca})$ & -10 & 6 &  &    \\
      {\er Carlin   (2000), M (LVA)}  &  $m(V-V_{Ca})$ & -30 & 6 & &   \\
 {\er Spinal MN, \VR=60} & & & & & \\
  
      {\er Joux (2001)} &  &  -27.3  &  &  &  \\
 {\er  SON, 5mM \CA (Double Bolt)} & &-12.4 & & & \\
  
   {\er Zhuravleva (2001)} &  &  -43.4  & 5.0  & -78.2   &  11.5  \\
 {\er LD thalamic N, 10mM Ba$^{2+}$ } & & & & & \\
   
    {\er Collet (2003)} &  $m(V-V_{Ca})$   &  -1.1  & 6.7  & control &  \\
 {\er Mouse, SM, 2.5 mM \CA } & &-6.4 & 6.2 & mdx & \\

    {\er Schnee (2003)} &    &  -35  & 4.7  &  {\er high frequ}  &  \\
 {\er Turtle AHC, 2.8 mM \CA } & &-43 & 4.2 & {\er low frequ}  & \\

    {\er Durante  (2004)  } &    & -31.1& 5.5$^*$ &  &  \\
 {\er Rat DA SNc, 2mM \CA} & & & & & \\

    {\er Muinuddin  (2004)  } &   {\er distal}    & -7.8 & 16.8 & -33.4   &  11.6  \\
 {\er Cat SMM, 20mM Ba$^{2+}$} &  {\er proximal}  &-6.1  & 17.3 & -36.5 & 13.2 \\

        {\er  Jackson  (2004)}  &  & -36 & 5.1 &  &  \\
 {\er Dorsomedial SCN, 1.2 mM \CA} & & & & & \\
 
\end{tabular}
\newpage

  \c{Table A1.1 continued} 
\vskip .1 in 
    \begin{tabular}{llllll}
  \hline
     Source, cell type    &Form for current  &  $V_{m,\frac{1}{2}} $  & $k_m$  &    $V_{h,\frac{1}{2}}$       & $k_h$ \\
         &if given  &  &  &      &  \\
  \hline 
         {\er  Liu  (2005)}  &  & -16.6  &2.45 & -16.8  &  5.95  \\
 {\er Rat IC, 2 mM \CA } & & & & & \\
  
          {\er Bui   (2006), M } & $m$ & -33 & 6 &  &   \\
{\er  Spinal MN, \VR=60} & & & & & \\
  
   {\er  Marcantoni (2007)}  &  & -30* & 9* &   &  \\
 {\er Adrenal CH cells, 2 mM \CA} & & & & & \\
  
  {\er  Kager (2007), M}  & $m^2h(V-V_{Ca})$  &  -11.3 & 5.7 && \\
 {\er CA3 pyramidal, 1.5 mM \CA} & & & & & \\

   {\er  Kapela  (2008), M}  & $mh$ const. field & -40$^{\dagger}$ & 8.3 & -42   & 9.1 \\
{\er Mesenteric SMM, 2.5 mM \CA} & & {\er Printed as 0}& & & \\

  {\er Luin (2008)}  &  {\er  young} &8.73     &7.72 & -5.77  &  7.30  \\

    {\er Cultured human SM, 10 mM \CA  }  & {\er old}    & 14.47  &7.74 & 2.38  &  10.1  \\

   {\er  Pospischil (2008), M}  & $m^2h(V-V_{Ca})$ & -33.4  & 4.45& -58.5  & 20.28 \\
{\er TR, CORT, \VR=120} & & & & & \\

  {\er  Xiang  (2008)}  &  & -6 & 16.7  & -30.5    & 11.2 \\
{\er CA1 pyramidal, 5 mM \CA} & & & & & \\
  
{\er  Marcantoni (2009)}  &  & -15 & 11$^*$  &   &  \\
{\er Adrenal CH cells, 2 mM \CA} & & & & & \\

{\er  Marcantoni (2010)}  &  & -23 (WT)& 6.9 &   &  \\
{\er Adrenal CH cells, 2 mM \CA}  &  & -18.1(KO) &  7.9 &   &  \\
{\er Adrenal CH cells, 2 mM \CA} & & & & & \\
  \hline 
  
  \end{tabular}
  
  \vskip .05   in

\newpage
 {Table A1.2}
  
  {Parameters for L-type Ca$^{2+}$  activation and, if included, VDI}
\smallskip
\c{Cardiac cells}
    \begin{tabular}{llllll}
  \hline
     Source, cell type    &Form for current  &  $V_{m,\frac{1}{2}} $  & $k_m$  &      $V_{h,\frac{1}{2}} $        & $k_h$ \\
   &if given  &   &   &           &  \\
  \hline
    {\er Lindblad  (1996), M}  & ($m_1h + m_2)\times$ & -0.95(1)  & 6.6(1)     & -28 & 4.52\\
  {\er Rabbit AM, 2.5 mM \CA} &$ \times(V-V_{Ca})$ &   -33 (2) & 12(2)& &\\
  
      {\er Handrock (1998)}  & & -3.5 & 4.9  & -60 & 7.5\\
  {\er Failing human VM, 1 mM \CA} & &    && &\\
  
    {\er Van Wagoner (1999) }& & -14* & 7.75* & -31.7 &  4.6 \\
  {\er Human AM, 1 mM \CA} & &&& & \\

      {\er Donohoe  (2000) }  & & & &-25.9&  \\
      {\er Rat VM, 1 mM \CA}  & & &    & -25.1 {\er RF} &  \\
                             
   {\er Zhang (2000), M} & $\big(m_1h+m_2\big)\times$ &  -23.1 (1) & 6 & -45 &  5\\
 {\er  Rabbit SA PM, 2 mM \CA } &$\times (V-V_{Ca,L})$ & -14.1 (2)  & 6 & & \\

      {\er Arikawa  (2002) }  &  {\er (Control) }  & -10.7 & 6.2 &-28.2 & 6.2  \\
      {\er Rat VM }  &  {\er (Diabetic) } & -11.5  &  6.0  & -27.2 & 6.4  \\

       {\er Mangoni (2003)} & & -25 &  & -45 &  10\\
 {\er  Mouse SA PM, 4 mM \CA } & & & & & \\

   {\er Mangoni  (2003)} &  & -3&   & -36 &  4.6\\
 {\er  Mouse SA PM 1.3KO} && & & &\\

  {\er Benitah (2010)}  & &-15 & 8.5$^*$   & -35  & 5.95 $^*$  \\
{\er VM, 1.5 mM \CA }  & & & & & \\
\hline
  \end{tabular}
%%\end{center} 
% %\footnote{**  Not able to be well- fitted to a Boltzmann function}
 \newpage

     {Table A1.3}
     
{Parameters for L-type Ca$^{2+}$ current activation, VDI and/or CDI.}
\smallskip

Neurons and secretory cells 
    \begin{tabular}{llllll}
  \hline
     Source, cell type    & Form for current  &  $V_{m,\frac{1}{2}} $  & $k_m$  &      $V_{h,\frac{1}{2}}$    {\er and/or}    & $k_h$ \\
            &     &       &    & {\er equation for $f({\rm Ca_i},t)$}       &  \\
  \hline
  {\er Amini (1999), M }& $mf_{\infty}(V-V_{Ca})$ & -50 & 3 & &  \\
  {\er MID  DA, 2.4 mM \CA} & & & & $f_{\infty}=\frac{ 0.00045} {(0.00045 + {\rm Ca_i})} $   & \\

   {\er Rhodes (2001), M}& $m^2hf_{\infty}(V-V_{Ca})$  & -19.3 $^\dagger$  & 7.5 & -42  & 8  \\
 {\er CORT pyramidal} & & & & $f_{\infty}= \frac{0.0005}{0.0005+ {\rm Ca_i}} $& \\

     {\er Poirazi (2003),M} & $mf_{\infty}$ const. field & -30.1 & 4.8  & & \\
 {\er  CA1 pyramidal soma} && & & $f_{\infty}=\frac {.001}{.001 + {\rm Ca_i}}$ &\\

  {\er Komendantov  (2004)  } & $mf_{\infty}(V-V_{Ca})$ & -50& 20 &   &  \\
 {\er MID DA, 2.4 mM \CA} & & & & $f_{\infty}= \frac{ 0.00045} {(0.00045 + {\rm Ca_i})} $     & \\

   {\er Rhodes (2005), M}& $m^2hf_{\infty}(V-V_{Ca})$  & -19$^\dagger$  & 8 & -42  & 8  \\
 {\er TR, \VR=60} & & & & $f_{\infty}= \frac{0.0001}{0.0001+ {\rm Ca_i}} $& \\

   {\er  Komendantov (2007), M}  & $m(f_{1,\infty} +f_{2, \infty} )\times$ & -27 & 4.5 & $f_{1,\infty}
 = \frac{0.2\times 0.0001^4}{0.0001^4 + {\rm Ca_i}^4}$   &  \\
 {\er Mag Endocrine, 2.4mM \CA} & $\times(V-V_{Ca})$  & & & $f_{2,\infty}
 = \frac{0.8\times 0.002}{0.002 + {\rm Ca_i}}$  & \\
  \hline
  \end{tabular}

\newpage

     {Table A1.4}
     
{Parameters for L-type Ca$^{2+}$ current activation, VDI and/or CDI}
\smallskip

{  Cardiac and skeletal muscle cells} 
\vskip .05 in 
 \begin{tabular}{llllll}
  \hline
     Source, cell type    & Form for &  $V_{m,\frac{1}{2}} $  & $k_m$  &      $V_{h,\frac{1}{2}}$  {\er and/or}        & $k_h$ \\
        &  current   &       &    & {\er equation for $f({\rm Ca_i},t)$}       &  \\
              \hline
 {\er Standen (1982), M}  & $m^3f$ & -21.1 & 4.69 & $df/dt=(f_{\infty}-f)/\tau_f $  &   \\
   {\er Insect SM, ${\rm I_{Ca}}$ }  & const. field& & &$ f_{\infty}= \frac{ 1} {    1 + ({\rm Ca_c}/K_f) } $& \\
             & & & &    ${\rm Ca_c}$= shell conc \CA     & \\
                  & & & &      $\tau_f=\frac{1}{\alpha_f(1 + \frac{{\rm Ca_c}}{K_f}) } $     & \\
                        & & & & $\alpha_f=0.1ms^{-1}$  & \\ 
                        & & & &  $ K_f=1.0\mu M$   & \\

    {\er Luo  (1994), M}&  $mhf_{\infty}$   & -10 & 6.24 & $\approx -35^*$ & $\approx 8.6$  \\
  {\er VM, 1.8 mM \CA} & const. field & & &  $ f_{\infty}= \frac {1}{ 1 +  \big({\rm Ca_i}/0.0006\big)^2 }$    &   \\

    {\er Fox  (2002), M} & $mf$ & -10 & 6.24&    -12.5 &  5 \\
  {\er VM, 2 mM \CA} & const. field& & &  $df/dt=(f_{\infty}-f)/\tau_f $ & \\
             & & & &  $ f_{\infty}= \frac{ 1} {    1 + ({\rm Ca_i}/K_f)^3 } $& \\
                        & & & & $\tau_f=30, K_f =  0.18\mu M$   & \\                    

            {\er Shiferaw  (2003), M} & $mhf_{\infty}$  & 5& 6.24 &  -35 & 8.6 \\
 {\er  VM, 1.8 mM \CA} & const. field & & & $f_{\infty} = \frac {.0005}{.0005+ {\rm Ca_i}}$&\\

    {\er Shannon (2004), M} & $mhf$ & -14.5 & 6  &-34.9* & 3.7  \\
 {\er VM, 1.8 mM \CA} & const. field & & & $df/dt=11.9(1-f) -$     & \\
 &  & & & $ -1.7{\rm Ca_i}f$     & \\
  {\er Stewart (2009), M} & $mh_1h_2f$ & -8 & 7.5  & -20  & 7  \\
 {\er PF, 2 mM \CA} & const. field & & &     $ f_{\infty}=0.4 +  \frac {0.6}{ 1 +  \big({\rm Ca_i}/0.05 \big)^2 }$  & \\
 &    {\er See source for details}&  & &     $ \tau_f=2+  \frac {80}{ 1 +  \big({\rm Ca_i}/0.05 \big)^2 }$  & \\
  \hline
  \end{tabular}

\newpage

\hoffset 0 in 
{Table A2}

{ Forward and backward rate functions, $\alpha$ and $\beta$ for $m$,  and, if given,  $h$}
\smallskip

{ All cell types}
   \begin{tabular}{llll}
  \hline
     Source, cell type    & $\alpha_{m,h} (V)$, $\beta_{m,h} (V)$    & $\tau_m $ & $ \tau_h $\\
     \hline 
            {\er Standen (1982)}  & $\alpha_m= \frac{0.013 (V+20) }  { 1- e^{ -(V + 20)/3   }  }  $       &  $\frac{1} {\alpha_m + \beta_m}  $ &    \\
 {\er Insect SM, ${\rm I_{Ca}}$ } & $\beta_m= 0.31e^{  -(V + 20)/25  }    $   & &\\

  {\er Kay (1987)}  & $\alpha_m= \frac{1.6 }  { 1- e^{ -0.072(V -5)   }  }  $       & &    \\
 {\er CA1 pyramidal} & $\beta_m=  \frac{0.02 (V+8.69) }  {e^{ (V + 8.69)/5.36}-1  }    $   & &\\

  {\er Sala (1991)} & $\alpha_m= \frac{0.058(11.3-V) }  {e^{(11.3-V)/13.7 } -1  }  $       &  $\frac{1} {\alpha_m + \beta_m}  $  &    \\
   {\er SYMP N, 4mM \CA} & $\beta_m=  \frac{0.085 (V+15.4) }  {e^{ (V + 15.4)/9.9}-1  }    $   & &\\

       {\er McCormick (1992)} & $\alpha_m=\frac{1.6} { 1 + e^{ -0.072(V-5)}   }   $   & & \\
              {\er TR N} &  $ \beta_m =\frac{0.02(V-1.31)  } {e^{ (V-1.3)/5.36)    } -1 }  $      & $\frac{1} {\alpha_m + \beta_m}  $  &  \\
  {\er Hutcheon  (1994) }& & & \\
       {\er TR N med dors} &   & $\frac{0.33} {\alpha_m + \beta_m}  $   & \\
        {\er  Rybak  (1997)} &   & & \\
       {\er Respiratory N} &  & $\frac{1} {\alpha_m + \beta_m}  $  &  \\
         &  &  &    \\

        {\er Lindblad  (1996)}  &$\alpha_m=   \frac{ 16.720(V+35.0) }  {1-   e^{    -(V + 35.0)/2.5  }  }    +$        
         & $\frac{1} {\alpha_m + \beta_m}  $   &$a + be^{ - (\frac{V -c} {d} ) ^2 }$   \\
               & $+ \frac{  50 }   { 1-e^{ -V/4.808  }    } $  &  & {\er a=1.5, b= 211}   \\
         {\er Rabbit AM}  & $ \beta_m=\frac{ 4.480(V-5.0) }  { e^{  (V-5.0)/2.500      }   - 1   }    $   & & {\er c=-37.427}\\
 &  $ \alpha_h= \frac{ 8.490(V + 28.0)  }  {e^{ (V+28.0)/4.000}  -1  }    $     &  & {\er d=20.213}  \\
                        &  $\beta_h= \frac{67.922 } {1 + e^{-(V+28.0)/4.000      } }            $   & &\\

         {\er Athanasiades  (2000)}  & $\alpha_m^*=0.346e^{0.0925(V+20)}   $  &    $0.1+\frac{1}{\alpha_m^* + \beta_m^*} $&  $17+\frac{1}{\alpha_h^* + \beta_h^*} $   \\
 {\er  MR }  & $ \beta_m^*= 1.8818e^{-0.00732(V+20)} $   &  & \\
  & $\alpha_h^*= 0.2419e^{0.145(V+20)} $ & &\\
   & $\beta_h^*= 0.0434e^{-0.02013(V+20)    }$ & &\\

       {\er Rhodes (2001)} & $\alpha_m=\frac{1600} { 1 + e^{ -(V-5)/13.9}   }   $   & & \\
              {\er CORT pyramid} &  $ \beta_m =\frac{20(V+8.9)  } {e^{ (V+8-9)/5)    } -1 }  $      & &  \\

         {\er Poirazi (2003)}  & $\alpha_m= \frac{0.055 (V+27.01) }  { 1- e^{ -(V + 27.01)/3.8    }  }  $      &  $\frac{0.2} {\alpha_m + \beta_m}  $ &   \\
 {\er  CA1 pyramidal, soma} & $\beta_m= 0.94e^{  -(V + 63.01)/17  }    $ & &\\

            {\er  Kager (2007)} & $\alpha_m= \frac{5.23 (V-71.5) }  { 1- e^{ -(V +?? 71.5)/10    }  }  $    &  $\frac{1} {\alpha_m + \beta_m}  $&    \\
 {\er CA3 pyramidal} & $\beta_m= 0.097e^{  -(V +?? 10)/10.86  }    $    & &\\

  {\er  Pospischil  (2008)} & $\alpha_m= \frac{0.055(V+27)}  { 1- e^{ -(V+27)/3.8    }  }  $       &  $\frac{1} {\alpha_m + \beta_m}  $&  $\frac{1} {\alpha_h + \beta_h} $    \\
 {\er TR, CORT} &  $\beta_m= 0.94e^{  -(V +75)/17 }    $   &  &   \\
   & $\alpha_h= 0.000457e^{  -(V +13)/50 }    $     &    &  \\
     & $\beta_h= \frac{0.0065}  { 1+e^{ -(V+15)/28    }  }  $  &    &  \\
 \hline
   \end{tabular}

   \newpage 
   \hoffset .1 in 
   {Table A3 - Experimental}

{ Time constants for L-type \CA activation and, if included, VDI}
\smallskip

{ All cell types}
    \begin{tabular}{lllll}
\hline
     Source, cell type    & $ \tau_{m, max} $  &  $V_{\tau_m,max} $  &      $ \tau_{h,max} $        &  $V_{\tau_h, max} $  \\
  \hline
      {\er Standen (1982)} & 19.82 &  -28 &  & \\
  {\er Insect SM, ${\rm I_{Ca}}$ } & & & &  \\

      {\er Kay (1987)} & 1.5-2.07 &  -18 &  & \\
  {\er Insect SM, ${\rm I_{Ca}}$ } & & & &  \\

    {\er Belluzzi (1991)} & 0.564&-8.0 & 29.90 &-22  \\
  {\er SYMP N} & & & &  \\

    {\er Sala (1991)} & 1.16 & 1 &  &  \\
  {\er SYMP N} & & & &  \\ 

    {\er Thompson (1991)} & &   & 100  & \\
  {\er CA1 pyramidal} & & &  & \\

  {\er Avery  (1996)} & 2.3 &   &  &  \\
  {\er CA3 pyramidal} & & &  & \\

    {\er Molitor (1999)} & 1.5 & -10  &  & \\
  {\er  DCN N} & & &  & \\

         {\er Donohoe  (2000) }  & & & $\tau_{fast}=17$, $\tau_{slow}=57$  & at 0 mV  \\
      {\er Rat VM}  &  &  &  $\tau_{fast}=15$, $\tau_{slow}=46$   & RF \\

         {\er Muinuddin (2000) }  &  4.0 ms        {\er distal}  &  & $\tau_1=4402$, $\tau_2=255$ &   \\

 {\er Cat SMM, 20mM Ba$^{2+}$} &  4.2 ms {\er proximal}  & &  $\tau_{1}=3905$, $\tau_2=222$   &  \\         

    {\er Schnee (2003)} &  $\approx$ 0.5  &   &    &  \\
 {\er Turtle AHC} &  &   & &   \\

        {\er  Jackson  (2004)}  & $\approx$ 2  &  &   &  \\
 {\er Dorsomedial SCN} & & &  & \\

  {\er Luin (2008)}  &  {\er  young}  53.6*   &   & &    \\
 {\er Cultured human SM}  & {\er old}  62.5*  &   & &   \\

   \hline
        \end{tabular}
%  
%
% %\bb ** Minimum not maximum 
%
%
%
% 
%  
%  
     \newpage 
   \hoffset .1 in 
   {Table A3 - Modeling}

{ Time constants for L-type \CA activation and, if included, VDI}
\smallskip

{ All cell types}
    \begin{tabular}{lllll}
\hline
     Source, cell type    & $ \tau_{m, max} $  &  $V_{\tau_m,max} $  &      $ \tau_{h,max} $        &  $V_{\tau_h, max} $  \\
  \hline

  {\er Traub (1991, 1994) } & 2.11&  -18 & &  \\
{\er CA3 pyramid} & & & &  \\

  {\er McCormick (1992)} & 1.541 &-15   &  &  \\
  {\er TR neuron} & & &  & \\

   {\er Luo (1994) } & 2.23&  -10 & 25.2  & -10   \\
{\er VM} & & & Min not max &  \\

   {\er Hutcheon  (1994)} & 1.541 & -15 & &  \\
  {\er TR med dors } & & & & \\

   {\er Jaffe  (1994)} & 1.5 &   & &  \\
  {\er CA3 pyramidal } & & & & \\

   {\er Wallenstein  (1994)} & 1.541 & -15   & &  \\
  {\er NRT N} & & & & \\

    {\er Migliore (1995)}  & 4.56  &  -11 &  &\\
  {\er CA3 pyramidal} & & & & \\

    {\er Li  (1996)}  & 0.134 & 4 &  &\\
  {\er DA  SN} & & & & \\

    {\er Lindblad   (1996)}  & 4.7  & -39  & 226 & -37.4\\
  {\er Rabbit AM } & & & & \\
 
    {\er LeBeau (1997)}  & 27  (all $V$)  &  &  & \\
  {\er PC} & & & & \\

    {\er Booth (1997)}  & 40  (all $V$)  &  &  & \\
  {\er Spinal MN } & & & & \\
 
    {\er Amini   (1999) }&  19.5 & -45 &  CDI only & \\
  {\er MID DA} & & & &  \\

  {\er Athanasiades (2000)} & 0.563 & -28 & 29.9 & -42  \\
 {\er  MR  } & & & &  \\
 
    {\er Zhang  (2000)} & 3.8  & -37 & 40 & -28  \\
 {\er  Rabbit SA PM } & & &  $\sim {\rm const}, V>-28$ &  \\
  
    {\er Shorten (2000)} & 11.32  &-12   & &   \\
 {\er  PC } & & & &  \\

                               {\er Carlin (2000) (HVA)}  & 20, all V& -10 & &   \\
      {\er Carlin  (2000) (LVA)}  & 20, all V & -30 &  & \\
 {\er Spinal MN} & & & &  \\

    {\er Rhodes (2001)} & 2.1 &-18   &   20.0 &   \\
 {\er  CORT pyramid } & & & &  \\

       {\er Fox  (2002)}  & 19.3 & -33 & 230  & $\approx <-60$ \\
 {\er VM} & & & &  \\
   \hline
        \end{tabular}
%  
%
% %\bb ** Minimum not maximum 
%
%
%
\newpage     
  {Table A3 - Modeling, continued} 

  \smallskip 
    \begin{tabular}{lllll}
    \hline
     Source, cell type    & $ \tau_{m, max} $  &  $V_{\tau_m,max} $  &      $ \tau_{h,max} $        &  $V_{\tau_h, max} $  \\
\hline
      {\er Poirazi  (2003)} & 0.826 & -35 & 4.8  &   \\
 {\er  CA1 pyramidal soma} && & & \\

   {\er Komendantov (2004)  } & 19.5 & -45 &CDI only  & \\
 {\er MID DA} & & & &  \\
 
     {\er Shannon (2004)  } & 2.38 & -15 &   25.2 & -15  \\
{\er VM} & & &Min not max &  \\

     {\er Rhodes  (2005)  }&2.1  &-19   & 200, all V &  \\
 {\er TR N} & & & &  \\

    {\er Bui  (2006) } &20, all V  &  &  &  \\
{\er  Spinal MN} & & &  & \\

  {\er  Komendantov (2007)}  &0.9, all V &  & &  \\
 {\er Mag Endocrine} & & & & \\
  
     {\er  Pospischil  (2008)}  &  7.04 &-38 & 449.8 & -67 \\
{\er TR, CORT} & & & &  \\
  
   {\er  Kapela (2008)}  &  3.65 &-40 & 110 & -35 \\
{\er Mesenteric SMM } & & & &  \\
  \hline
  \end{tabular}
\newpage 

{Table A4}

{Time constants if given by explicit formulas}
\smallskip

{All cell types} 
 \begin{tabular}{lll}
  \hline
     Source, cell type    &  $\tau_m$   &  $\tau_h$ \\
  \hline

  {\er Belluzzi (1991)}   &   $  0.1 +   \bigg(  0.343e^{0.925V} *+ 1.88e^{-0.00732V}   \bigg)^{-1} $  & \\
{\er SYMP N, 5mM \CA}& & \\

      {\er Luo (1994)$^\dagger$}  &  $\frac { 1 - e^{-(V+10)/6.4} } { 0.035(V+10)(1 + e^{ - (V+10)/6.4} )} $  & $\frac{1} { a +  be^{-(0.0337(V+10))^2} }$ \\
  {\er VM} & &$a=0.02, b=0.0197$ \\

    {\er Li  (1996)}  & $ \frac{0.4}        {          5e^{\theta}  +    \frac{ \theta}  {      e^{\theta} -1       }                  }$,     & \\
  {\er DA SN} &  $\theta=-(V+11)/8.3$ & \\

    {\er Amini (1999)} & $   1.5 + 18e^{ - \big( \frac{V+45}{20}  \big)^2} $ &  \\
   {\er Komendantov (2004)} & &  \\
    {\er MID DA} & &  \\

{\er Shorten (2000), PC} &  $   \frac { 27} {e^{(V+60)/22} + 2e^{-2(V+60)/22}      } $  &  \\

    {\er Fox   (2002), VM  }&  $  \bigg(  \frac{0.25e^{ -0.01V}  } { 1 + e^{-0.07V}   }  +  \frac{ 0.07e^{  -0.05(V+40) } } {1+ e^{ 0.05(V+40)}    }   \bigg)^{-1}   $ &   $30 + \frac{200}{  1 + e^{(V+20)/9.5} } $   \\
 %{\er VM} & &  \\

        {\er Shannon   (2004)}  &  $\frac { 1 - e^{-(V+14.5)/6} } { 0.035(V+14.5)(1 + e^{ - (V+14.5)/6} )} $  & $\frac{1} { a +  be^{-(0.0337(V+14.5))^2} }$ \\
  {\er VM} & &{\er a=0.02, b=0.0197}  \\

   {\er Rhodes  (2005)  }& $0.6 + \frac{6} { \cosh[( V + 19)/24]) }$ & 200  \\
 {\er TR N} & &  \\

   {\er  Kapela  (2008)}  & $   1.15 + e^{ - \big( \frac{V+40}{30}  \big)^2 }$ & $45+  65e^{ - \big( \frac{V+35}{25}  \big)^2} $   \\
{\er Mesenteric SMM }  & & \\
  \hline
\end{tabular}
%
%
%%  $^\dagger$ Compares several sets of inactivation parameters.  
%
%
%
\newpage  

{Table A5}\\
{Magnitude of L-type \CA currents or related quantities}\\
{All cell types}   
 \begin{tabular}{lll}
  \hline
     Source, cell type    & L-type magnitude & Remarks \\
  \hline
   {\er Penington (1991)} &  $\sim$ 4\% of total ${\rm I_{Ca}}$  &  Soma, dissociated cells\\
  {\er DRN SE  } & &    \\
 
  {\er McCormick (1992)} &$\sim$ 1pA/pF $^*$,  & Dissociated cells\\
  {\er TR  N} & $p_L$= 80.10$^{-6}$ cm$^3$/sec &    \\
  
   {\er Schild (1993)} & $g_L=1.44$ mS/cm$^2$  & Cell, G$_L$=0.075$\mu$S,   \\
{\er Rat med} & & C=0.052 nF  \\

  {\er Hutcheon   (1994)} & 1.3mS/cm$^2$  &  $p_L$ as in McCormick(1992)\\
  {\er TR med dors} & & \\

 {\er Kuryshev  (1995) } & 52\% of HVA is L-type & $\sim$70pA \\
  {\er Rat corticotropes} & &  10mM Ba$^{2+}$ \\

    {\er Li (1996)}  & 0.19 mS/ cm$^2$ &\\
  {\er DA SN} & (dendrites)  & \\

      {\er Booth (1997)}  & 0.33 mS / cm$^2$ &\\
  {\er Spinal MN} & (dendrites)  &   \\

        {\er Schr\"oder (1998)}  & 13.2 fA (control)& Ensemble averages \\
  {\er Human VM} & 38.2 fA (failing heart) & single channel (${\rm Ba^{2+}}$) \\
  
    {\er Amini  (1999) }& G$_L$=0.0005$\mu$S & $\sim$ 30pA; g$_L$=0.018 mS/ cm$^2$   \\
  {\er MID DA} & &  (sphere, rad 15$\mu$ )  \\

    {\er Van Wagoner  (1999) }& 9.13 pA/pF  & 3.35 pA/pF \\
  {\er Human AM} & (control)  &  (atrial fibrillators)\\
 
  {\er Athanasiades (2000)} & G$_L$=0.075$\mu$S & g$_L$=1.44 mS/cm$^2$    \\
 {\er  MR } & & cf Schild (1993)  \\
  
        {\er Carlin (2000)}  &    & \\
 {\er Spinal MN} &    &  \\
 {\er LVA \& HVA, soma} & 10mS/ cm$^2$  &  \\
  {\er LVA \& HVA, dend} & 0.3mS/ cm$^2$  &  \\

            {\er Donohoe (2000) }  & 16.8 pA/pF & at -10 mV \\
      {\er Rat VM}  & 18.3 pA/pF  & {\er RF} \\    
  
   {\er Joux  (2001)} & 19.3 pA/pF & \\
 {\er  SON} & &  \\
 
    {\er Fox  (2002)} & $\sim$ 1pA/pF, & \\
 {\er  VM} &  $\ov{P}_{Ca}=0.000026$ cm/ms  &  \\

     {\er Mangoni  (2003)} & 3.32 pA/pF  & 1.05 pA/pF\\
 {\er  Mouse SA PM} & (wild type) &  (1.3 knockout)\\
 
  {\er Poirazi  (2003)} &  7 mS / cm$^2$ &  \\
 {\er  CA1 pyramidal} & (soma) &\\

  \hline
  \end{tabular}
\newpage

\c{Table A5 continued}
\vskip .1 in  
\begin{tabular}{lll}
  \hline
     Source, cell type    & L-type magnitude & Remarks \\
  \hline

          {\er Tipparaju (2004)  } & 2.5 pA/pF  &  1.2 pA/pF, 2.6 pA/pF  \\
 {\er Human AM} & Young adults & Infants, Older Adults   \\

     {\er Durante (2004)  } & 27\% of ${\rm I_{Ca}}$ is L-type &   \\
 {\er Rat DA SNc} & &   \\
      
  {\er Komendantov  (2004)  } & 0.216 mS / cm$^2$  &  \\
 {\er MID  DA} & &    \\

   {\er Rhodes (2005)  }&  4 mS / cm$^2$ & \\
 {\er TR N soma} &    &  \\
 
  {\er Vignali (2006)  }& 60\% of ${\rm I_{Ca}}$ is L-type & \\
 {\er Pancreatic A,B} &    &  \\

 {\er  Komendantov (2007)}  & 0.4 mS / cm$^2$  & 0.22, 0.46 mS / cm$^2$  \\
 {\er Mag  Endocrine} & (soma)  & (prim., sec. dendrites) \\

   {\er  Marcantoni  (2007)}  & 10\% of ${\rm I_{Ca}}$  &  $\sim$ 15 pA \\
 {\er Adrenal CH cells} & -60 to -40 mV  & \\

     {\er  Kapela  (2008)}  & 100\% of ${\rm I_{Ca}}$ is \IL &   \\
{\er Mesenteric SMM} & & \\

  {\er  Xiang (2008)}  & 18.7 pA/pF  & ${\rm I_{Ca}}$ is mainly \IL \\
{\er CA1 pyramidal} & (control)  &  \\

  {\er  Fuller-Bicer (2009)}  & 5.76 pA/pF  &  3.17 pA/pF \\
{\er Mouse VM} &   {\er  WT}  & $\alpha_2/\delta$-1  {\er KO} \\

{\er Benitah  (2010)}  &   ${\rm I_{Ca}}$ is mainly  \IL &   \\
{\er VM}  & &  \\
{\er Empson  (2010)}  &  $\le$ 5\% of  ${\rm I_{Ca}}$  & Soma \& dendrites \\
{\er Purkinje neuron}  & &  \\
  \hline
  \end{tabular}

\newpage 
{Table A6}\\
Parameters for examples of  L-type channel subtype activation and VDI 
%\begin{center}
    \begin{tabular}{llllll}
  \hline
     Source      & Subtype &  $V_{m,\frac{1}{2}} $, Carrier  & $k_m$  &      $V_{h,\frac{1}{2}} $        & $k_h$ \\
  \hline

  {\er Koschak (2001)  } &  \ab   & -3.9  & 7.8  & -27.6  & 13.8 \\
 {\er  tsA201}  &  & 15 mM ${\rm Ba^{2+}}$  & &  &  \\

   &  \ac   & -17.5  & 8.9  & -42.7  & 6.6  \\
 &   & 15 mM ${\rm Ba^{2+}}$ & &  &  \\

  {\er Xu (2001)  } &  \ab$\alpha_1$   & 3.3   &  &  & \\
{\er Xenopus oocytes} &  & 5 mM \CA  & &  &  \\

 &  \ab$\alpha_1$   & -8.8   &  &  & \\
&  & 5 mM ${\rm Ba^{2+}}$   & &  &  \\

&  \ac$\alpha_1$   & -20  &  &  & \\
&  & 5 mM \CA  & &  &  \\

&  \ac$\alpha_1$   & -36.8  &  &  & \\
&  & 5 mM ${\rm Ba^{2+}}$ & &  &  \\

    {\er Schnee (2003)} &  \ac  &  -35 & 4.7  &  {\er high frequ} &    \\
 {\er Turtle AHC} &  & 2.8 mM \CA & &  &  \\
 & \ac & -43  &4.2 &{\er low frequ}  &  \\
 &  & 2.8 mM \CA & &  &  \\

  {\er  Lipscombe (2004)}  & \ab & -16.1  & 6$^*$& &  \\

    {\er tsA201  }   & \ac &  -40.4   & 5.5$^*$ & & \\
   &  & 2 mM \CA  &   & &  \\

  {\er  Hildebrand (2004)}  & \ab & -13.9  & &-33.7  &    {\er  control}  \\

   {\er tsA201,  2 mM ${\rm Ba^{2+}   }$ }  &   &    &  &-52.5  & {\er  allethrin}  \\

  {\er Helton (2005) }   & \ab &  -17.6    & & &  \\

    {\er tsA201  } & \ac & -39.4 & &  & \\
   &  & 2 mM \CA  &   & &  \\
   {\er Catterall  (2005)}  & \aa &  8 to 14 &  &  -8&\\
   &  & 10 mM \CA  &   & &  \\
  & \ab  &-17  & &-60 to -50 & \\
       {\er HEK}&  &    2mM \CA  &  &-60 to -50 &\\
  
  & \ac  & -37 & &-43 to -36 & \\
    {\er HEK}  &  &   2 mM \CA  &  &-43 to -36 &\\

  & \ad  & -2.5 to -12   &   &-27 to -9&\\
       {\er HEK}    & &  2 - 20 mM \CA  &  &-27 to -9&\\
  
  {\er Navedo  (2007)}  & \ab  & 31.9  & 9.3  & -2.7 & 5.2 \\
  &  & 110 mM ${\rm Ba^{2+}}$  &   &  &  \\
   {\er tsA201} & \ac   &  9.7     & 10.8  & -18.7 & 3.5  \\
   {\er Arterial SMM} &    &  33.8     & 8.0  & -4.9 & 5.4  \\

  {\er Shapiro  (2007)}  & \ab  & -20  & 6 & & \\
   {\er Model, MN} & \ac   & -41   & 6  & & \\

  \hline
   \end{tabular}
%   
%   
%\end{center} 
%
%
%}
%
%

\newpage
\c{Table A6 continued}
\vskip .1 in  
    \begin{tabular}{llllll}
  \hline
     Source      & Subtype &  $V_{m,\frac{1}{2}} $, Carrier  & $k_m$  &      $V_{h,\frac{1}{2}} $        & $k_h$ \\
  \hline

  {\er Luin (2008)}  &  \aa &8.73   {\er  young}   &7.72 & -5.77  &  7.30  \\

    {\er Cultured human SM } & {\er (Assumed)}   &  10 mM \CA    &  &  &   \\
    &  \aa & 14.47   {\er old}   &7.74 & 2.38  &  10.1  \\
 {\er Peloquin  (2008)}  & \ad   & 5 & 5.9 & & \\
          {\er tsA201  }  &    &  20 mM ${\rm Ba^{2+}}$ 23$^{\circ}$C &  & & \\
 &     & -13  & 3.9  & & \\
     &    &  20 mM ${\rm Ba^{2+}}$ 37$^{\circ}$C &  & & \\
  {\er Singh  (2008)}  & \ac (42, long) & -2.2   & 9.1 & & \\
 & \ac (42A, short) & -12.9   & 6.9  & & \\
   &  & 15 mM \CA  &   & &  \\
 & \ac (42, long) & -11.8   & 7.8 & & \\
  & \ac (42A, short) & -20.2   & 6.6  & & \\
&  & 15 mM ${\rm Ba^{2+}}$ & &  &  \\

   {\er Fuller-Bicer (2009)}  &  {\er WT}  & -3.5 & 4.87  & -15.15 & 5.54 \\
     {\er  Mouse VM}     &  $\alpha_2/\delta$-1  {\er KO} & 4.66  &  5.88 & -2.18 & 6.14  \\
  &  & 1.8 mM \CA  &   & &  \\

  {\er Putzier  (2009a)}  & \ac & -31.1  & 5.35 & & \\
 
    {\er Cordeiro  (2009)}  & \ab $\beta_2$  {\er WT} & 1.5 &  & -24.8 & 9.13 \\
          {\er tsA201  }      &  {\er B mutant}  & 3.5   &  & -30.0 & 6.25  \\
  &  & 15 mM \CA  &   & &  \\
 
   {\er Gebhart  (2010)}  & \ac $\beta_3$  & -0.9  & 9 & &  \\
           {\er tsA201  }   &      &  15mM \CA& &  &   \\
  \hline
   \end{tabular}

\end{center}

}
\newpage
\hoffset = .25  in 
% 
%\begin{thebibliography}{KW}
%\end{thebibliography}
\section{References}
\def\nh{\noindent\hangindent=1 true cm \hangafter = 1}

\nh Albrecht MA, Colegrove SL, Hongpaisan J, Pivovarova NB,
Andrews SB, and Friel DD (2001).  Multiple modes of calcium-induced
calcium release in sympathetic neurons. I. Attenuation of endoplasmic
reticulum \CA accumulation at low [\CAN]$_i$  during weak depolarization.
J Gen Physiol 118: 83-100.

\nh Amini B Clark JW Jr Canavier CC (1999) Calcium dynamics underlying pacemaker-like and burst
firing oscillations in midbrain dopaminergic neurons:
A computational study. J. Neurophysiol 82, 
2249-2261.

\nh Andrade A, Sandoval A, Ricardo Gonz\'alez-Ram\'irez R, Lipscombe D, Campbell K, Felix R (2009) 
The $\alpha_2\delta$ subunit augments functional expression and modifies the
pharmacology of ${\rm Ca_v1.3}$ L-type channels.  Cell Calcium 46: 282-292. 

\nh Anwyl R (1991) Modulation of vertebrate neuronal calcium channels
   by transmitters. Brain Res 
 16: 265-281. 

\nh Arikawa M, Takahashi N, Kira T, Hara M, Saikawa T, Sakata T (2002)
Enhanced inhibition of L-type calcium currents by troglitazone in
streptozotocin-induced diabetic rat cardiac ventricular myocytes.
Brit J Pharmacol 136: 803-810.

\nh Athanasiades A, Clark JW Jr, Ghorbel F,  Bidani A (2000) 
An ionic current model for medullary respiratory neurons. 
J Comp Neurosci 9, 237-257. 

\nh Augustine GJ, Santamaria F, Tanaka K (2003) Local calcium signaling in neurons.
          Neuron 40: 331-346.

\nh Avery RB, Johnston D (1996) 
Multiple channel types contribute to the low-voltage-activated
calcium current in hippocampal CA3 pyramidal neurons.
J Neurosci 16: 5567-5582.

\nh Baig S, Koschak A, Lieb A et al. (2010)
Loss of   \ac (CACNA1D) function in a human channelopathy with bradycardia and congenital deafness.
Nat Neurosci 14: 77-84.

\nh Barrett CF, Tsien RW (2008)
The Timothy syndrome mutation differentially affects
voltage- and calcium-dependent inactivation of
\ab L-type calcium channels. PNAS 105: 2157-2162.

\nh Bauer CS, Tran-Van-Minh A, Kadurin I,  Dolphin  AC (2010) 
A new look at calcium channel $\alpha_2\delta$ subunits. 
 Curr Opion Neurobiol 20: 563-71.

\nh Baumann L, Gerstner A, Zong X, Biel M, 
Wahl-Schott C (2004) Functional characterization of the L-type \CAN-channel
\d$\alpha_1$ from mouse retina. Invest Ophthal Vis Sci 45: 708-713. 
       
\nh Bazzazi H, Johnny MB, Yue DT (2010) 
Calmodulin release from the IQ Domain of \ac channels during
calcium dependent inactivation? Biophys J 98 Issue, Supp 1: 519a. 

\nh Bean BP (1989)  Classes of calcium channels in vertebrate cells.  Ann Rev Physiol 51:367-384.

\nh Belluzzi O, Sacchi O (1991) A five-conductance model of the action potential
in the rat sympathetic neurone. Prog Biophys Molec Biol 55: 1-30.

\nh Benitah J-P Alvarez JL, G\'omez AM (2010) L-type Ca2+ current in ventricular cardiomyocytes.
 J  Mol  Cel Cardiol 48:  26-36. 
 
 \nh Bernstein J (1902) Untersuchungen zur Thermodynamik der biolektrischen Str\"ome.
 Arch Ges Physiol Pfug 92: 521-562. 

\nh Bers DM (2008) Calcium cycling and signaling in cardiac myocytes. 
Ann Rev Physiol 70:23-49.

\nh Bidaud I,  Lory P (2011) 
Hallmarks of the channelopathies associated with L-type calcium channels :
A focus on the Timothy mutations in \ab channels. Biochimie xxx: 1-7.

\nh Blaustein MP (1988) Calcium transport and buffering in neurons. TINS 11: 438-443. 

\nh Blaustein MP, Lederer WJ (1999) Sodium/calcium exchange: its physiological
implications. Physiol Rev 79: 763-854.

\nh Bodi I, Mikala G, Koch SE, Akhter SA, Schwartz A (2005) 
The L-type calcium channel in the heart:
the beat goes on. J Clin Invest 115: 3306-3317.

\nh Booth VB, Rinzel J, Kiehn O (1997) 
Compartmental model of vertebrate motoneurons for Ca$^{2+}$-dependent
spiking and plateau potentials under pharmacological treatment. 
J Neurophysiol 78: 3371-3385. 

\nh Boyett MR, Clough A, Dekanski J, Holden AV (1997) Modelling cardiac excitation and excitability.
In:  Computational Biology of the Heart, pp 1-47. Eds Panfilov AV, Holden AV. Wiley,  New York.

\nh  Brasen JC, Olsen LF, Hallett MB (2010) Cell surface topology creates high Ca$^{2+}$ signalling
microdomains. Cell Calcium 47: 339-349. 

%\nh Bray JG, Mynlieff M (2011) 
%Involvement of protein kinase C and protein kinase A in
%the enhancement of L-type calcium current by GABA$_B$ 
%receptor activation in neonatal hippocampus.  Neurosci 179: 62-72.

\nh Brehm P, Eckert R (1978) Calcium entry leads to inactivation of calcium channel 
in {\it Paramecium}. Science 202: 1203-1206.

\nh    Brette F, Leroy J, Le Guennec JY,  Salle L (2006) \CA currents in 
cardiac myocytes: old story, new insights. Prog Biophys Mol Biol 91: 1-82.

\nh Brown TM, Piggins HD (2007) Electrophysiology of the suprachiasmatic circadian clock. 
Prog Neurobiol 82: 229-255. 

\nh Budde T, Meuth S, Pape H-C (2002) Calcium-dependent
inactivation of neuronal
calcium channels. Nat Rev
Neurosci 3: 873-883.

\nh Bui TV, Ter-Mikaelian M, Bedrossian D, Rose PK (2006)
Computational estimation of the distribution of L-type Ca$^{2+}$ channels in
motoneurons based on variable threshold of activation of persistent
inward currents. J Neurophysiol 95: 225-241. 

\nh Carlin KP, Jones KE, Jiang Z, Jordan LM, Brownstone, RM (2000)  Dendritic L-type calcium
currents in mouse spinal motoneurons: implications for bistability. Eur J Neurosci 12: 1635-1646.

\nh Catterall WA (1995) Structure and function of voltage-gated ion channels. 
Ann Rev Biochem 64:493-531.

\nh Catterall WA (2000) Structure and regulation of
voltage-gated Ca$^{2+}$ channels. Ann Rev Cell Dev Biol 16:521-555.

\nh Catterall WA (2010) Voltage-gated calcium channels.   Handbook of Cell Signaling, Ch 112, pp 897-909. 
Elsevier, Amsterdam. 

\nh  Catterall WA, Perez-Reyes E, Snutch TP, Striessnig J (2005) International Union of Pharmacology. XLVIII.
Nomenclature and structure-function relationships of
voltage-gated calcium channels. Pharmacol Rev 57:411-425.

\nh Cens T, Rousset M, Leyris J-P, Fesquet P, Charnet P (2006) 
Voltage- and calcium-dependent inactivation in high
voltage-gated Ca$^{2+}$  channels. Prog Biophys Mol  Biol 90: 104-117.

\nh Chameau P, Qin Y, Spijker S, Smit G, Jo\"els M (2007) Glucocorticoids specifically snhance L-type calcium current amplitude and
affect calcium channel subunit expression in the mouse hippocampus.  J Neurophysiol 97: 5-14.

\nh Chavis P, Fagni L, Lansman JB, Bockaert J (1996)  Functional coupling between ryanodine receptors and L-type
calcium channels in neurons. Nature 382: 719-722. 

\nh Collet C, Csernoch L, Jacquemond V (2003)
Intramembrane charge movement and l-type calcium current
in skeletal muscle fibers isolated from control and {\it mdx} mice. 
Biophys J 84: 251-265.

\nh Coombes S, Hinch R, Timofeeva Y (2004) Receptors, sparks and waves in a fire-diffuse-fire framework for 
calcium release.  Prog Biophys Mol  Biol  85: 197-216. 

\nh CooperJR, Bloom FE, Roth RH (2003) The Biochemical Basis of Neuropharmacology. Oxford University Press: Oxford. 

\nh Cordeiro JM, Marieb M, Pfeiffer R, Calloe K, Burashnikov E, Antzelevitch C (2009) 
Accelerated inactivation of the L-type calcium current due to a
mutation in CACNB2b underlies Brugada syndrome. 
J Mol Cell Cardiol 46: 695-703.

\nh Coulon P, Herr D, Kanyshkova T, Meuth P,
Budde T, Pape H-C (2009). Burst discharges in neurons of the thalamic reticular nucleus are
shaped by calcium-induced calcium release. 
Cell Calcium 46: 333-346.

\nh Cox DH, Dunlap K (1994)  Inactivation of N-type calcium current in
chick sensory neurons: calcium and
voltage dependence. J Gen Physiol 104: 311-366.

\nh Crump SM, Schroder EA, Sievert GA,
Andres DA, Satin J (2011)
Calmodulin interferes with \ac C-terminal regulation of L-type
channel current. Biophys J 100, Supp 1: 571a. 

\nh Cueni L, Canepari M,
Adelman JP, L\"uthi A (2009)  \CA signaling by T-type  \CA channels in neurons. 
Pflugers Arch 457:1161-1172. 

\nh   Davies A, Kadurin I, Alvarez-Laviada A, Douglas L, Nieto-Rostro M, Bauer CS, 
Pratt WS,  Dolphin AC (2010)
The $\alpha_2 \delta$ subunits of voltage-gated calcium channels
form GPI-anchored proteins, a posttranslational
modification essential for function.  PNAS 107: 1654-1659.

\nh Destexhe A, Sejnowski TJ (2001) Thalamocortical Assemblies. Oxford University Press: Oxford, UK.   

\nh De Waard M, Gurnett CA, Campbell KP (1996) Structural and functional diversity of
voltage-gated calcium channels. Ion Channels 4: 41-87. 

\nh DiFrancesco D, Noble D (1985) A model of cardiac electrical activity incorporating
ionic pumps and concentration changes.  Phil Trans R Soc Lond B 307: 353-398.

\nh Doan L (2010) Voltage-gated calcium channels and pain. 
Tech Reg Anesth Pain Manag 14: 42-47.

\nh Dolmetsch RE, Pajvani U, Fife K, Spotts JM, Greenberg ME (2001) 
Signaling to the nucleus by an
L-type Calcium channel-
calmodulin complex through
the MAP kinase pathway.  Science 294: 333-339. 

\nh Dolphin AC (2003)  $\beta$ subunits of voltage-gated calcium channels. 
J Bioenerg Biomembr 35:599-620.

\nh Dolphin AC (2006)  A short history of voltage-gated calcium channels. 
Br J Pharmacol 147(S1):  S56-S62. 

\nh Dolphin, AC (2009) Calcium channel diversity: multiple roles of calcium channel
subunits. Curr Opin Neurobiol 19:237-244.

\nh Dolphin AC (2010) Age of quantitative proteomics hits voltage-gated
calcium channels. PNAS 107: 14941-14942. 

\nh    Donohoe P,  McMahon AC, Walgama OV et al. (2000) L-type calcium current of isolated 
 rat cardiac myocytes in experimental uraemia. Nephrol Dial Transplant 15: 791-798. 

\nh  Durante P, Cardenas CG, Whittaker JA, Kitai ST, Scroggs RS (2004) 
Low-threshold L-type calcium channels in rat dopamine neurons. 
J Neurophysiol 91: 1450-1454.

\nh Empson RM, Kn\"opfel T (2010) 
Functional integration of calcium regulatory 
mechanisms at Purkinje neuron synapses.  Cerebellum 10: online.

\nh Erickson Mg, Liang H, Mori MX, Yue DT (2003) FRET two-hybrid mapping reveals function and
location of L-type Ca$^{2+}$ channel CaM preassociation. Neuron 39: 97-107.  

\nh Ertel EA, Campbell KP, Harpold MM, Hofmann F, Mori Y, Perez-Reyes E, Schwartz
A, Snutch TP, Tanabe T, Birnbaumer L, et al. (2000) Nomenclature of voltagegated
calcium channels. Neuron 25:533-535.

\nh Faber ESL (2010) Functional interplay between NMDA receptors, SK
channels and voltage-gated \CA channels regulates
synaptic excitability in the medial prefrontal cortex. 
J Physiol 588.8: 1281-1292.

\nh Faber GM, Silva J, Livshitz L, Rudy Y (2007) Kinetic properties of the cardiac L-Type Ca$^{2+}$  Channel and Its role in
myocyte electrophysiology: a theoretical investigation.  Biophys J 92: 1522-1543.

\nh  Findlay I, Suzuki S, Murakami S, Kurachi Y (2008) Physiological modulation of voltage-dependent inactivation in
the cardiac muscle L-type calcium channel: a modelling study.  Prog Biophys Mol  Biol 96: 482-498.   

\nh  Fink M, Niederer SA, Cherry EM et al.  (2011) Cardiac cell modelling: observations
from the heart of the cardiac
physiome project. Prog Biophys
Mol Biol  104: 2-21. 

\nh Fisher RE, Gray R, Johnston D (1990) 
Properties and distribution of single voltage-gated
calcium channels in adult hippocampal neurons. J Neurophysiol 64: 91-104.

\nh Forti L, Pietrobon D (1993) 
Functional diversity of L-type calcium channels
in rat cerebellar neurons. Neuron 10: 437-450.

\nh Fox AP  Nowycky MC Tsien RW (1987)  Kinetic and pharmacological properties distinguishing
three types of calcium currents in chick sensory
neurones. J Physiol 394: 149-172.  

\nh Fox JJ, McHarg JL, Gilmour RF Jr (2002) Ionic mechansim of electrical alternans. Am J Physiol Heart
Circ Physiol 282: H516-H530.

\nh Friel DD, Chiel HJ (2008) Calcium dynamics: analyzing the \CA
regulatory network in intact cells. 
TINS 31: 8-19. 

\nh Friel DD, Tsien RW (1992) 
A caffeine- and ryanodine-sensitive \CA store in bullfrog
sympathetic neurones modulates effects of \CA entry. 
J Physiol 450: 217-246.

\nh Fuller-Bicer GA, Varadi G, Koch  SE et al. (2009) 
Targeted disruption of the voltage-dependent calcium channel $\alpha_2/\delta$-1-subunit.
Am J Physiol Heart Circ Physiol 297: H117-H124. 

\nh Gebhart M, Juhasz-Vedresa G, Zuccotti A et al. (2010)  Modulation of \ac \CA channel gating by Rab3 interacting molecule.
Mol Cell Neurosci 44: 246-259.

\nh Graef IA, Mermelstein PG, Stankunas K, Neilson JR, Deisseroth K, Tsien RW, Crabtree GR (1999)
 L-type calcium channels and GSK-3
regulate the activity of NF-ATc4
in hippocampal neurons.  Nature 401: 703-708.

\nh Grandi E, Morotti S, Ginsburg KS, Severi  S, Bers DM (2010)  
Interplay of voltage and Ca-dependent inactivation of L-type Ca current.  Prog Biophys
Mol Biol 103: 44-50. 

\nh Greenstein JL, Hinch R, Winslow RL (2006) Mechanisms of excitation-contraction coupling in an integrative model
of the cardiac ventricular myocyte.  Biophys J 90: 77-91.

\nh Groff JR, Smith GD (2008) Calcium-dependent inactivation and the dynamics of
calcium puffs and sparks. J Theor Biol 252: 483-499. 

\nh Guyot A, Dupr\'e-Aucouturier S, Ojeda C, Rougier O, Bilbaut  A (2000)
Two types of pharmacologically distinct \CA currents with voltage-dependent
similarities in zona fasciculata cells isolated from bovine adrenal gland. 
J Membrane Biol 173: 149-163. 

\nh Habermann CJ, O'Brien BJ, W\"assle H, and Protti DA (2003)
 AII amacrine
 cells express L-type calcium channels at their output synapses.
 J Neurosci 17: 6904-6913.

\nh  Hamill OP, Marty A, Neher E, Sakmann B, Sigworth FJ (1981)
Improved patch-clamp techniques for high-resolution current recording
from cells and cell-free membrane patches.  Pfl\"ugers Arch 391: 85-100.

\nh Handrock R, Schr\"oder F, Hirt S, Haverich A, Mittmann C, Hrzig S (1998) 
Single-channel properties of L-type calcium channels from failing
human ventricle. Cardiovasc Res 37: 445-455. 

\nh Hardingham GE, Arnold FJL, Bading H (2001) Nuclear calcium signaling controls
CREB-mediated gene expression
triggered by synaptic activity. Nat Neurosci 4: 261-267.

\nh Hell JW, Westenbroek RE, Warner C, Ahlijanian MK, Prystay W,
Gilbert MM, Snutch TP, Catterall WA (1993) Identification and differential
subcellular localization of the neuronal class C and class D L-type calcium
channel alpha 1 subunits. J Cell Biol 123: 949-962.

\nh Helton TD, Xu W, Lipscombe D (2005) Neuronal L-Type calcium channels open quickly and are
inhibited slowly.  J Neurosci 25:10247-10251. 

\nh Higgins ER, Goel P, Puglisi JL, Bers DM, Cannell M, Sneyd J (2007) Modelling calcium 
microdomains using homogenisation. J Theor Biol 247: 623-644. 

\nh Hildebrand ME,  McRory JE, Snutch TP, Stea A (2004) 
Mammalian voltage-gated calcium channels are potently
blocked by the pyrethroid insecticide allethrin. 
J Pharmacol Exp Ther 308: 805-813.

\nh Hinch R, Greenstein JL, Tanskanen AJ, Xu L, Winslow RL (2004) A simplified local control model 
of calcium-induced calcium release
in cardiac ventricular myocytes. Biophys J 87: 3723-3736. 

\nh Hodgkin AL, Huxley AF (1952) A quantitative description of membrane
current and its application to conduction
and excitation in nerve.  J Physiol 117: 500-544.

\nh Hoesch RE, Weinreich D, Kao  JPY (2001) 
A novel \CA influx pathway in mammalian primary sensory
neurons is activated by caffeine. J Neurophysiol 86: 190-196. 

\nh H\"ofer GF, Hohenthanner K, Baumgartner W, Groschner K, Klugbauer N, Hofmann F, Romanin C (1997) 
intracellular Ca$^{2+}$ inactivates L-Type Ca$^{2+}$ channels with a Hill
coefficient of $\sim$1 and an inhibition constant of $\sim$4 $\mu$M by reducing
channel's open probability. Biophys J 73: 1857-1865. 

\nh Hofmann F,  Lacinov\'a, L, Klugbauer N (1999)  Voltage-dependent calcium
channels: from structure to function.  Rev Physiol Biochem Pharmacol
139: 33-87.

\nh Holmgaard K, Jensen K, Lambert JDC (2008) Imaging of Ca$^{2+}$  responses mediated by presynaptic
L-type channels on GABAergic boutons of cultured
hippocampal neurons.  Brain Res 1249: 79-90.

\nh Hutcheon B, Miura RM, Yarom Y, Puil E (1994) Low-threshold calcium 
current and resonance in thalamic
neurons: a model of frequency preference. J Neurophysiol 71: 583-594.

\nh Hu H, Marban E (1998) Isoform-specific inhibition of L-type calcium channels by
dihydropyridines is independent of isoform-specific gating
properties. Molec Pharmacol 53: 902-907.

\nh Huxley AF (1959)  Ion movements during nerve activity. 
Ann NY Acad Sci 81: 221-246.

\nh Imredy JP, Yue DT (1994) Mechanism of Ca$^{2+}$-sensitive inactivation
of L-type Ca$^{2+}$ channels. Neuron 12: 1301-1318. 

\nh Jackson AC, Yao GL, Bean BP (2004)  Mechanism of spontaneous f
iring in dorsomedial
suprachiasmatic nucleus neurons. J Neurosci 24: 7985-7998. 

\nh Jaffe DB, Ross WN, Lisman JE, Lasser-Ross N,
Miyakawa H, Johnston D (1994) 
A model for dendritic \CA  accumulation in hippocampal
pyramidal neurons based on fluorescence imaging measurements.
J Neurophysiol 71: 1065-1077.

\nh Jafri MS, Rice JJ, Winslow RL (1998)  Cardiac Ca$^{2+}$ dynamics: the roles of ryanodine
receptor adaptation and sarcoplasmic reticulum load.  Biophys J  74: 1149-1168.

\nh Johnston D, Wu S (1995) Foundations of Cellular Neurophysiology.  MIT Press: Cambridge, MA. 

\nh Jones SW (1998) Overview of voltage-dependent calcium channels. 
J Bioenerg  Biomemb 30: 299-312. 

\nh Joux N  Chevaleyre V  Alonso G et al.  (2001) 
High voltage-activated Ca$^{2+}$ currents in rat supraoptic neurones:
biophysical properties and expression of the various channel $\alpha_1$
subunits.  J Neuroendocrinol 13, 638-649.

  \nh Kager H, Wadman WJ, Somjen GG (2007)  Seizure-like afterdischarges simulated in a model neuron.
  J Comp Neurosci 22:105-128.

\nh Kamp TJ, Hell JW (2000)  Regulation of cardiac L-type calcium channels by protein kinase A and
protein kinase C.  Circ Res 87:1095-1102. 

\nh Kapela A, Bezerianos A, Tsoukias M (2008) 
A mathematical model of Ca$^{2+}$ dynamics in rat mesenteric smooth muscle
cell: agonist and NO stimulation. J Theor Biol 253: 238-260. 

\nh Kapur A, Yeckel MF, Gray R, Johnston D (1998) L-type calcium channels are required for one form of
hippocampal mossy fiber LTP. J Neurophysiol 79: 2181-2190.

\nh Kay AR (1991) Inactivation kinetics of calcium current of acutely dissociated CA1 pyramidal cells
of the mature guinea-pig hippocampus. J Physiol 437: 27-48.

\nh Kay AR, Wong RKS (1987) Calcium current activation kinetics in
isolated pyramidal neurons of the CA1 region of the mature guinea-pig
hippocampus. J Physiol 392: 603-616.

\nh Koch C (1999) Biophysics of Computation. Oxford University Press: Oxford UK. 

\nh Koh X, Srinivasan B, Ching HS, Levchenko A (2006) A 3D Monte Carlo analysis of the
role of dyadic space geometry
in spark generation. Biophys J  90: 1999-2014.

\nh  Komendantov AO, Komendantova OG, Johnson SW, Canavier CC (2004)
A modeling study suggests complementary roles for GABA$_A$ and NMDA
receptors and the SK channel in regulating the firing pattern in midbrain
dopamine neurons. J Neurophysiol 91: 346-357.

\nh Komendantov AO Trayanova NA Tasker JG (2007)  Somato-dendritic mechanisms underlying
the electrophysiological properties of hypothalamic
magnocellular neuroendocrine cells:
A multicompartmental model study. J Comp Neurosci 23,143-168. 

\nh Koschak A, Reimer D, Huber I et al. (2001) 
$\alpha_1D$ (\cc) subunits can form L-type \CA channels activating at
negative voltages.  J Biol Chem 276: 22100-22106.

\nh Koschak A, Reimer D, Walter D et al. (2003)  \d$\alpha_1$ 
 subunits can form slowly inactivating
dihydropyridine-sensitive L-type \CA  channels lacking
\CA-dependent inactivation. J Neurosci 23: 6041-6049. 

\nh Kotturi MF, Jefferies WA (2005). Molecular characterization
of L-type calcium channel splice variants expressed in
human T lymphocytes. Mol Immunol 42: 1461-1474.

\nh Kuryshev YA, Childs GV, Ritchie AK (1995) Three high threshold calcium channel
subtypes in rat corticotropes. Endocrinology 136: 3916-3924. 

\nh Lacinov\'a  L (2005) Voltage-dependent calcium channels. Gen
Physiol Biophys 24 (Suppl. 1): 1-78.

\nh Lacinov\'a  L, Hofmann F  (2005)  Ca$^{2+}$- and voltage-dependent inactivation of the expressed
L-type Ca$_v$1.2 calcium channel.  Arch Biochem Biophys 437: 42-50. 

\nh LeBeau  AP, Robson  AB, McKinnon AE, 
Donald RA, Sneyd J (1997) Generation of action potentials in a mathematical model of corticotrophs.
Biophys J 73: 1263-1275.

\nh  Lee A, Wong ST,  Gallagher D, Li B, Storm DR, Scheuer T, Catterall WA  (1999)  Ca$^{2+}$/calmodulin binds to
and modulates P/Q-type calcium channels. Nature 399: 155-159.

\nh Leitch B, Szostek A, Lin R, Shevtsova O (2009)  Subcellular distribution of L-type calcium channel
subtypes in rat hippocampal neurons.  Neuroscience 164: 641-657. 

\nh Levitan IB, Kaczmarek LK (1997) The Neuron. Oxford University Press: Oxford UK.  

\nh Li L, Bischofberger J, Jonas P (2007)   
Differential gating and recruitment of P/Q-, N-, and R-type
\CA  channels in hippocampal mossy fiber boutons. 
J Neurosci 27: 13420 -13429. 

\nh Li Y-X, Bertram R, Rinzel J (1996) 
Modeling N-methyl-D-aspartate-induced
bursting in dopamine neurons. Neuroscience 71: 397-410.

\nh Liebmann L, Karst H, Sidiropoulu K, van Gemert N, Meijer OC, Poirazi P, Jo\"els M (2008)  
Differential effects of corticosterone on the slow afterhyperpolarization
in the basolateral amygdala and CA1 region: possible role of calcium
channel subunits. J Neurophysiol 99: 958-968.

\nh Lindblad DS, Murphey CR, Clark JW, Giles WR (1996) 
A model of the action potential and underlying
membrane currents in a rabbit atrial cell. 
Am J Physiol
271:H1666-1696.

\nh Linz KW, Meyer R (1998)  Control of L-type calcium current during the action potential
of guinea-pig ventricular myocytes.  J Physiol  513: 425-442.

\nh Lipscombe D (2002) L-type calcium channels: highs and new lows. 
Circ Res 90: 933-935.

\nh Lipscombe D, Helton TD, Xu W (2004) L-type calcium channels: the low down.
J Neurophysiol 92: 2633-2641. 

\nh Liu X, Yang PS, Yang W, Yue DT (2010) 
Enzyme-inhibitor-like tuning of \CA channel
connectivity with calmodulin. Nature 463: 968-972.

\nh Liu Y,  Li  X, Ma C, Liu J, Lu H  (2005) Salicylate blocks L-type calcium channels in rat
inferior colliculus neurons. Hearing Res 205: 271-276. 

\nh Luin E, Lorenzon P, Wernig  A, Ruzzier F (2008)
Calcium current kinetics in young and aged human
cultured myotubes. Cell Calcium 44: 554-566.

\nh Luo C-H, Rudy Y (1994) A dynamic model of the cardiac
ventricular action potential. I. Simulations of ionic currents and concentration
changes. Circ Res 74:1071-1096.

\nh Magee JC, Johnston D (1995) 
Characterization of single voltage-gated Na$^+$ and Ca$^{2+}$ 
channels in apical dendrites of rat CA1 pyramidal neurons. 
J Physiol 487.1: 67-90.

\nh Mahajan A, Shiferaw Y,  Sato D et al. (2008) A rabbit ventricular action potential
model replicating cardiac dynamics
at rapid heart rates. Biophys J 94: 392-410.

\nh Mangoni  ME, Couette B, Bourinet E, Platzer J, Reimer D, Striessnig J, 
Nargeot J (2003) Functional role of L-type Cav1.3 Ca$^{2+}$ channels in
cardiac pacemaker activity. PNAS 100: 5543-5548.

\nh Marcantoni A, Baldelli P, Hernandez-Guijo JM, Comunanza V, Carabelli V, 
Carbone E (2007) L-type calcium channels in adrenal chromaffin cells:
role in pace-making and secretion. Cell Calcium 42: 397-408. 

\nh Marcantoni A, Carabelli V, Comunanza V, Hoddah H, Carbone E (2008)
Calcium channels in chromaffin cells: focus on L and T types. 
Acta Physiol 192: 233-246. 

\nh Marcantoni  A, Carabelli V, Vandael DH, Comunanza V, Carbone E (2009) 
PDE type-4 inhibition increases L-type Ca$^{2+}$ currents, action
potential firing, and quantal size of exocytosis in mouse
chromaffin cells. Eur J Physiol 457:1093-1110.

\nh Marcantoni A, Vandael DHF, Mahapatra S, Carabelli V, Sinnegger-Brauns MJ, 
Striessnig J, Carbone E (2010)  Loss of \ac channels reveals the critical role of L-type
and BK channel coupling in pacemaking mouse adrenal
chromaffin cells. J Neurosci 30:491-504.

\nh Martinez-Gomez J, Lopez-Garcia JA (2007) Simultaneous assessment of the effects of L-type current modulators
on sensory and motor pathways of the mouse spinal cord in vitro. Neuropharmacol 53: 464-471.  

\nh McConigle P, Molinoff PB (1989) Quantitative aspects of drug-receptor interactions. In: 
Basic Neurochemistry: Molecular, Cellular and Medical Aspects, pp 183-201. Eds Siegel GJ et al.. Raven Press: New York.

\nh McCormick DA, Huguenard JR (1992)  A model of the electrophysiological properties of 
thalamocortical relay neurons. J Neurophysiol 68: 1384-1400.

\nh Meir A, Ginsburg S,  Butkevich A, Kachalsky SG, Kaiserman I,
Ahdut  R, Demirgoren S, Rahamimoff R (1999)  Ion channels in presynaptic nerve terminals
and control of transmitter release.  Physiol Rev 79: 1019-1088.

\nh Mesirca P, Marger L, Torrente A, Striessnig J,
Nargeot J, Mangoni ME (2010)  Pacemaker cells of the atrioventricular node are \ac
 dependent
oscillators. 
  Biophys J 98, Supp 1: 339a.

\nh Meuth S, Budde T, Pape H-C (2001) Differential control of high-voltage activated Ca$^{2+}$ current components by
a Ca$^{2+}$-dependent inactivation mechanism in thalamic relay neurons. Thalamus Relat Syst 1: 31-38.

\nh Migliore M, Cook EP, Jaffe DB, Turner DA, Johnston D (1995)
Computer simulations of morphologically reconstructed CA3 hippocampal neurons.
J Neurophysiol 73: 1157-1168.

\nh Molitor SC, Manis PB (1999) 
Voltage-gated \CA conductances in acutely isolated guinea pig
dorsal cochlear nucleus neurons. J Neurophysiol 81: 985-998.

\nh Morad M, Soldatov N (2005) Calcium channel inactivation: possible
role in signal transduction and \CA signaling. Cell Calcium 38: 223-231.

\nh Mori MX, Erickson MG, Yue DT (2004) Functional stoichiometry and
local enrichment of calmodulin
interacting with Ca$^{2+}$ channels. Science 304: 432-5. 

\nh Muinuddin A, Kang Y, Gaisano HY, Diamant NE (2004)
Regional differences in L-type \CA channel expression in feline lower
esophageal sphincter. Am J Physiol Gastrointest Liver Physiol 287: G772–G781. 

\nh Navedo MF, Amberg GC, Westenbroek RE, Sinnegger-Brauns MJ, Catterall
WA, Striessnig J, Santana LF (2007) Cav1.3 channels produce persistent
calcium sparklets, but Cav1.2 channels are responsible for sparklets in
mouse arterial smooth muscle. Am J Physiol Heart Circ Physiol
293: H1359-H1370.

\nh Neher E, Sakaba T (2008) Multiple roles of calcium ions in the regulation of
neurotransmitter release. Neuron 59: 861-872.

\nh Nernst W (1889) Die elektromotorische Wirksamkeit der Ionen. Z. Phys Chem Leipzig 4: 129-181. 

\nh Newcomb R, Szoke B, Palma A, Wang G, Chen X, Hopkins W, Cong R, Miller
J, Urge L, Tarczy-Hornoch K, Loo JA, Dooley DJ, Nadasdi L, Tsien RW,
Lemos J, Miljanich G (1998) Selective peptide antagonist of the class E
calcium channel from the venom of the tarantula Hysterocrates gigas.
Biochemistry 37:15353-15362.

\nh N'Gouemo P, Faingold CL, Morad M (2009) 
Calcium channel dysfunction in inferior colliculus neurons
 of the genetically
epilepsy-prone rat. Neuropharmacology 56: 665-675. 

\nh Noble D (1995)  The development of mathematical models of the heart. 
Chaos, Solitons \& Fractals 5:321-333.

\nh Norris CM, Blalock EM, Chen K-C, Poter NM, Thibault O, Kraner SD, Landfield PW (2010)
Hippocampal 'zipper' slice studies reveal a necessary role for calcineurin
in the increased activity of L-type Ca2+ channels with aging. Neurobiol Aging 31: 328-338.

\nh Nowycky MC,  Fox AP, Tsien RW (1985) Three types of neuronal calcium channel
with different calcium agonist sensitivity. Nature 316: 440-443. 

\nh Ono K, Iijima T (2010)  Cardiac T-type Ca2+ channels in the heart. 
 J Mol Cell Card 48: 65-70. 

\nh Ouyang K, Wu C, Cheng H (2005) \CAN-induced \CA release in sensory neurons.
J Biol Chem 280: 15898-15902.

\nh Peloquin JB, Doering CJ,  Rehak R, McRory JE (2008)
Temperature dependence of \ad calcium
channel gating. Neurosci 151:1066-1083.

\nh Penington NJ,  Kelly JS,  Fox AP (1991) A study of the mechanism of Ca$^{2+}$ current 
inhibition produced by
serotonin in rat dorsal raphe neurons.  J Neurosci I7: 3594-3609.

\nh Perez-Reyes, E (2003) Molecular physiology of low-voltage-activated
T-type calcium channels. Physiol Rev 83: 117-161.

\nh Peterson BZ, DeMaria CD, Yue DT (1999) Calmodulin is the Ca$^{2+}$ sensor for
 Ca$^{2+}$-dependent
inactivation of L-type calcium channels.  Neuron 22: 549-558.  

\nh Piedras-Renter\'ia ES, Barrett CF, CaoY-Q, Tsien RW (2007)
Voltage-gated calcium channels, calcium signaling,
and channelopathies. In, Calcium: A Matter of Life or Death, 
Krebs J, Michalak M (Eds). Elsevier BV, Amsterdam.

\nh Planck M (1890) \"Uber die Erregung von Electricit\"at und Warme in Electrolyten.
    Ann Phys Chem 39: 161-186.

\nh Poirazi P, Brannon T, Mel BW (2003)  Arithmetic of subthreshold synaptic
summation in a model CA1 pyramidal cell. Neuron 37: 977-987.

\nh Pospischil M, Toledo-Rodriguez M, Monier C,  Piwkowska Z, Bal T, Fr\'egnac  Y, Markram H, Destexhe A (2008)
Minimal Hodgkin-Huxley type models for different classes
of cortical and thalamic neurons. Biol Cybern 99: 427-441.

\nh Power JM, Sah P (2005)
Intracellular calcium store filling by an L-type calcium current in the 
basolateral amygdala at subthreshold membrane potentials.
J Physiol 562: 439-453.

\nh Puglisi JL, Wang F, Bers DM (2004) Modeling the isolated cardiac myocyte.
Prog  Biophys Mol Biol 85:163-178.

\nh Putzier I, Kullmann PHM, Horn JP, Levitan ES (2009a) 
Ca$_v$1.3 channel voltage dependence, not Ca$^{2+}$ selectivity,
drives pacemaker activity and amplifies bursts in nigral
dopamine neurons. J Neurosci 29: 15414-15419. 

\nh Putzier I, Kullmann PHM, Horn JP, Levitan ES (2009b) 
Dopamine neuron responses depend exponentially on pacemaker interval.
J Neurophysiol  101: 926-933.

\nh  Qin N, Olcese R, Bransby M, Lin T, Birnbaumer L (1999)
Ca$^{2+}$-induced inhibition of the cardiac Ca$^{2+}$ channel depends
on calmodulin. PNAS 96: 2435-2438.

\nh Ravindran A, Lao QZ, Harry JB, Abrahimi P, Kobrinsky E, Soldatov NM 
(2008) Calmodulin-dependent gating of \ab calcium
channels in the absence of Ca${\rm _v\beta}$ subunits. PNAS 105: 8154-8159.

\nh Ravindran A,  Kobrinsky E, Lao QZ, Soldatov NM (2009)
Functional properties of the \ab calcium channel activated by 
calmodulin in the absence of ${\rm \alpha_2\delta}$ subunits. Channels (Austin) 3: 25-31

\nh Rhodes PA Llin\'as R (2001)  Apical tuft input efficacy in layer 5 pyramidal cells from
rat visual cortex.
J Physiol 536.1, 167-187. 

\nh Rhodes PA Llin\'as R (2005)  A model of thalamocortical relay cells. 
J Physiol 565, 765-781. 

\nh  Romanin C, Gamsjaeger R, Kahr H, Schaufler D, Carlson O, Abernethy DR, Soldatov NM (2000)
Ca$^{2+}$ sensors of L-type Ca$^{2+}$ channel.  FEBS Lett 487: 301-306.  

\nh Roussel C, Erneux T, Schiffmann SN, Gall D (2006) 
Modulation of neuronal excitability by intracellular
calcium buffering: from spiking to bursting. Cell Calcium 39: 455-466.

\nh Sala F (1991)  Activation kinetics of calcium currents in bull-frog
sympathetic neurones. J Physiol 437: 221-238.

\nh  Santos SF, Pierrot N, Morel N, Gailly P, 
Sindic C, Octave J-N (2009) Expression of human 
amyloid precursor protein in rat
cortical neurons inhibits calcium oscillations. 
J Neurosci 29: 4708-4718.

\nh  Satin J, Schroder EA, Crump  SM ( 2011)
L-type calcium channel auto-regulation of transcription.
Cell Calcium 49: 306-313.

\nh  Schild JH, Khushalani S, Clark JW, Andresen MC, Kunzei DL, Yang M (1993)
  An ionic current model for neurons in the rat medial
nucleus tractus solitarii receiving sensory afferent input. J Physiol 469: 341-363. 

\nh Schlick B, Flucher BE, Obermair GJ (2010) 
Voltage-activated calcium channel expression profiles
in mouse brain and cultured hippocampal neurons. Neurosci 167: 786-798. 

\nh Schnee ME, Ricci AJ (2003) 
Biophysical and pharmacological characterization of voltage-gated
calcium currents in turtle auditory hair cells. 
J Physiol 549: 697-717. 

\nh Schr\"oder F, Handrock R, Beuckelmann DJ, Hirt S, Hullin R, 
Priebe L, Schwinger RHG,  Weil J, Herzig S (1998) Increased availability and open probability of 
single L-type calcium channels from failing compared with nonfailing human ventricle. 
Circulation 98: 969-976.

\nh Scriven DRL, Asghari P, Schulson MN, Moore EDW (2010) 
Analysis of \ad and ryanodine receptor clusters in rat ventricular
myocytes. Biophys J 99: 3923-3929. 

\nh  Shannon TR, Wang F,  Puglisi J, Weber C, Bers DM (2004) 
A mathematical treatment of integrated Ca dynamics within the
ventricular myocyte. Biophys J 87: 3351-3371. 

\nh Shapiro NP, Lee RH (2007) 
Synaptic amplification versus bistability in motoneuron dendritic
processing: a top-down modeling approach. 
J Neurophysiol 97: 3948-3960.

\nh Sherman A,  Keizer J, Rinzel J (1990) Domain model for \CAN-inactivation of \CA channels at low
channel density. Biophys J 58: 985-995.

\nh Shiferaw Y, Watanabe MA, Garfinkel A, Weiss JN, Karma A (2003) 
Model of intracellular calcium cycling in ventricular myocytes.  
Biophys J 85: 3666-3686. 

\nh  Shorten PR, Wall DJN ( 2000) A Hodgkin-Huxley model exhibiting bursting
oscillations.  Bull Math Biol 62: 695-715.

\nh Singh A , Gebhart M, Fritsch R et al. (2008) Modulation of voltage- and \CAN-dependent gating of
\ac L-type calcium channels by alternative splicing of a
C-terminal regulatory domain. J Biol Chem 283: 20733-20744. 

\nh Singh A, Hamedinger D, Hoda J-C et al. (2006) 
C-terminal modulator controls \CA-dependent
gating of \ad L-type \CA channels. Nat Neurosci 9: 1108-1116.

\nh Soeller C, Cannell MB (2004) Analysing cardiac excitation-contraction coupling with
mathematical models of local control. Prog  Biophys Mol Biol 85: 141-162. 

\nh Soldatov, NM (2003) \CA channel moving tail: 
link between
\CAN-induced inactivation and \CA 
signal transduction. Trends Pharm Sci 24: 167-171.

\nh Splawski I, Timothy KW, Sharpe LM, Decher N, Kumar P,
Bloise R, Napolitano C, Schwartz PJ, Joseph RM, Condouris
K, Tager-Flusberg H, Priori SG, Sanguinetti MC,  Keating
MT (2004) \ab calcium channel dysfunction causes a
multisystem disorder including arrhythmia and autism. Cell
119: 19-31.

\nh Standen NB, Stanfield PR (1982) A binding-site model for calcium channel inactivation that
depends on calcium entry. Proc R Soc Lond B 217: 101-110.

\nh Stokes DL, Green NM (2003) Structure and function of the calcium pump. 
Ann Rev Biophys Biomol Struct 32:445-468.

\nh  Striessnig J, Hoda J-C, Koschak A et al. (2004)  L-type \CA channels in \CA channelopathies.
Biochem Biophys Res Comm 322: 1341-1346.

\nh Striessnig J, Koschak A (2008)
Exploring the function and pharmacotherapeutic potential
of voltage-gated \CA channels with gene knockout models.
Channels 2, 1-19.

\nh Striessnig J, Koschak A, Sinnegger-Brauns MJ, Hetzenauer A, 
Nguyen NK, Busquet  P, Pelster G, Singewald N (2006)
Role of voltage-gated L-type \CA channel
isoforms for brain function. 
Biochem Soc Trans 34: 903-909.

\nh Striessnig J (2007)  C-terminal tailoring of L-type
calcium channel function. J Physiol 585:  643-644. 

\nh  Sun L, Fan J-S, Clark JW, Palade PT (2000) A model of the Ltype
Ca$^{2+}$  channel in rat ventricular
myocytes: ion selectivity and inactivation mechanisms.  J Physiol 529: 139-158. 

\nh Suzuki Y, Inoue T, Ra C (2010) L-type Ca$^{2+}$  channels: a new player in the regulation of \CA  signaling, cell
activation and cell survival in immune cells. Mol Immunol 47: 640-648.

\nh Tadross MR, Yue DT (2010) 
Systematic mapping of the state dependence of voltage- and \CAN-
dependent inactivation using simple open-channel measurements. 
J Gen  Physiol 135: 217-227.

\nh Tanabe M,  G\"ahwiler BH,  Gerber U (1998) 
L-Type  \CA channels mediate the slow \CAN-dependent
afterhyperpolarization current in rat CA3 pyramidal cells in vitro.
J Neurophysiol 80: 2268-2273.

\nh Tank DW, Fiegehr WG, Delaney KR (1995) A quantitative analysis of presynaptic
calcium dynamics that
contribute to short-term enhancement. J Neurosci 15: 7940-7952. 

\nh Tanskanen A, Greenstein JL, O’Rourke B, Winslow RL
(2005)  The role of stochastic and modal gating of cardiac L-type \CA
channels on early after-depolarizations. 
Biophys J 88: 85-95.

\nh Thibault O, Landfield PW (1996) Increase in single L-type calcium channels in
hippocampal neurons during aging. Science 272: 1017-1020. 

\nh  Thibault O, Hadley R,  Landfield PW (2001) Elevated postsynaptic ${\rm[Ca^{2+}]_i}$  and L-type calcium channel activity
in aged hippocampal neurons: relationship to impaired
synaptic plasticity.   J Neurosci 21: 9744-9756.  

\nh  Thibault O, Gant JC,  Landfield PW (2007)  Expansion of the calcium hypothesis of brain aging and
Alzheimer's disease: minding the store. Aging Cell 6: 307-317.

\nh Thompson SM, Wong RKS (1991) Development of calcium current subtypes in isolated rat
hippocampal pyramidal cells. J Physiol 439: 671-689.

\nh Tillotson D (I979) Inactivation of Ca conductance dependent on entry of 
Ca ions in molluscan neurones. PNAS 76: 1497-1500.

\nh Tipparaju SM, Kumar R, Wang Y, Joyner RW, Wagner MB (2004) 
Developmental differences in L-type calcium current of human atrial
myocytes. Am J Physiol Heart Circ Physiol 286: H1963-H1969.

\nh Torrente A, Mesirca P, Neco P et al. (2011)
\ac L-type calcium channels-mediated ryanodine receptor
dependent calcium release controls heart rate. 
 Biophys J 100, Supp 1: 567a. 

\nh Traub RD, Wong RKS, Miles R, Michelson H (1991) 
A model of a CA3 hippocampal pyramidal neuron incorporating
voltage-clamp data on intrinsic conductances. J Neurophysiol 66: 635-650.

\nh Traub  RD, Jeffery JGR, Miles R,
Miles A. Whittington MA, T\'oth  K (1994)
A branching dendritic model of a rodent CA3
pyramidal neurone. J Physiol 481: 79-95. 

\nh Tsien RW, Barrett CF (2005) Brief history of calcium channel discovery. In, 
Voltage-Gated Calcium Channels, pp 27-47.  Ed Zamponi GW. 
  Kluwer Academic/Plenum:  New York.

\nh Tsien RW, Tsien RY (1990) Calcium channels, stores and oscillations. 
 Ann Rev Cell Biol 6: 715-60.

\nh Tsien RW, Lipscombe D, Madison  DV, Bley  KR, Fox AP (1988)
Multiple types of neuronal calcium channels and their
selective modulation. TINS 11: 432-438.  

\nh Tuckwell HC (1988) Introduction to Theoretical Neurobiology. Cambridge University Press:  Cambridge. 

\nh Vandael DH, Marcantoni A, Mahapatra S, Caro A, Ruth P, Zuccotti A,
 Knipper M, Carbone E (2010) Ca${_{\rm v}}$1.3 and BK channels for timing and regulating
cell firing. Mol Neurobiol 42:185-198.

\nh Van Wagoner DR, Pond A, Lamorgese M, Rossie SS,  McCarthy PM, Nerbonne JM 
(1999)  Atrial L-Type \CA currents and human atrial fibrillation.  Circ Res 85: 428-436.

\nh Verkhratsky A (2005) Physiology and pathophysiology of the calcium store
in the endoplasmic reticulum of neurons. Physiol Rev 85: 201-279. 

\nh Vignali S, Leiss V, Karl R, Hofmann F, Welling A (2006) 
Characterization of voltage-dependent sodium and
calcium channels in mouse pancreatic A- and B-cells.
J Physiol 572: 691-706.

\nh Wahl-Schott C, Baumann L, Cuny H, Eckert C, Griessmeier K, Biel M (2006) 
Switching off calcium-dependent inactivation in
L-type calcium channels by an autoinhibitory domain. PNAS 103: 15657-15662.

\nh Wallenstein G (1994) A model of the electrophysiological properties of nucleus reticularis
thalami neurons. Biophys J 66: 978-988.

\nh Wang X-J (1998) Calcium coding and adaptive temporal computation in cortical
pyramidal neurons. J Neurophysiol 79: 1549-1566.

\nh Weiergr\"aber M, Stephani U, K\"ohling R (2010) Voltage-gated calcium channels in the etiopathogenesis and
treatment of absence epilepsy. Brain Res Rev 62: 245-271. 

\nh Weiss JN (1997) The Hill equation revisited: uses and misuses. FASEB J 11: 835-841.  

\nh Wilders R (2007) Computer modelling of the sinoatrial node. 
Med Bio Eng Comput 45:189-207. 

\nh Williams GSB, Huertas MA, Sobie EA, Jafri MS,Smith GD (2007)
A probability density approach to modeling local control of calcium-induced calcium release in cardiac myocytes.
  	Biophys J  92: 2311-2328.

\nh Williams GSB, Smith  GD, Sobie EA, Jafri MS (2010) 
Models of cardiac excitation-contraction coupling in ventricular myocytes.
Math Biosci 226: 1-15. 
  
\nh Wu WW, Chan CS, Surmeier DJ, Disterhoft JF (2008) 
Coupling of L-Type  \CA channels to K$_v$7/KCNQ channels creates a novel,
activity-dependent, homeostatic intrinsic plasticity. 
J Neurophysiol 100: 1897-1908. 

  \nh  Xiang K, Earl DE, Davis KM, Giovannucci DR, Greenfield LJ Jr, Tietz EI (2008) 
  Chronic benzodiazepine administration potentiates
high voltage-activated calcium currents
in hippocampal CA1 neurons. J Pharm Exptal Therapeutics 327: 872-883.

     \nh Xu W, Lipscombe D (2001) Neuronal ${\rm Ca_v1.3\alpha_1}$ L-type channels activate at relatively
hyperpolarized membrane potentials and are incompletely
inhibited by dihydropyridines. J Neurosci  21: 5944-5951. 

\nh Yarotskyy V, Gao G, Peterson BZ, Elmslie KS (2009) 
The Timothy syndrome mutation of cardiac \ab
(L-type) channels: multiple altered gating mechanisms
and pharmacological restoration of inactivation. J Physiol 587: 551-565.

\nh Zalk R, Lehnart SE, Marks AR (2007) Modulation of the
ryanodine receptor
and intracellular calcium. Ann Rev Biochem 76:367-85.

\nh Zamponi GW, Ed  (2005) Voltage-gated calcium channels. Kluwer/Plenum: New York. 

\nh Zhang H, Holden AV, Kodama I, Honjo H, Lei M, Varghese T, Boyett MR (2000)
Mathematical models of action potentials in the
periphery and center of the rabbit sinoatrial node. Am J Physiol Heart Circ Physiol
279: H397-H421.

\nh  Zhang M,  Sukiasyan N,   M$\phi$ller M,  Bezprozvanny I,  Zhang H,
Wienecke J,  Hultborn H (2006)   Localization of L-type calcium channel \ac in cat lumbar spinal
cord - with emphasis on motoneurons.  Neurosci Lett 407: 42-47. 

\nh Zhang Q, Timofeyev V, Qiu H et al. (2011)
Expression and roles of \ac  ($\alpha_{1D}$) L-Type \CA channel in
 atrioventricular
node automaticity. J Mol Cell Cardiol 50: 194-202. 

\nh Zhuravleva SO, Kostyuk PG, Shuba YM (2001) 
Subtypes of low voltage-activated \CA channels in laterodorsal
thalamic neurons: possible localization and physiological roles.
Pflügers Arch 441: 832-839.

\nh Zuccotti A, Clementi S, Reinbothe T, Torrente A,
 Vandael DH, Pirone A (2011) Structural and functional differences
between L-type calcium channels:
crucial issues for future selective
targeting. TIPS 868: avail online.

\nh Z\"uhlke RD, Pitt GS, Deisseroth K, Tsien RW, 
Reuter H (1999)  Calmodulin supports both inactivation and
facilitation of L-type calcium channels. Nature 399: 159-162.

 \end{document}